%% file: ms.tex
\documentclass[11pt]{article}

\RequirePackage{amsthm,amsmath,amsfonts,amssymb}
\RequirePackage[numbers]{natbib}
\RequirePackage[colorlinks,citecolor=blue,urlcolor=blue,backref=page,backref=page]{hyperref}
\RequirePackage{graphicx}
\RequirePackage{float}
\usepackage{tikz}
\usepackage[ruled,vlined]{algorithm2e}
\usetikzlibrary{3d, calc,decorations.markings}
\usepackage[margin=1in]{geometry}
\usepackage{setspace}

\newtheorem{proposition}{Proposition}[section]

\theoremstyle{plain}

\newtheorem{theorem}{Theorem}[section]
\newtheorem{lemma}[theorem]{Lemma}
\newtheorem{definition}{Definition}

\newcommand{\supplementaryproof}[2]{%
    \vspace{1cm} 
    \section*{Proof of Proposition #1} 
    \addcontentsline{toc}{section}{Proof of Proposition #1} 
    \begin{proof}[\textit{Proof}] 
    #2 
    \end{proof}
}

\newcommand{\supplementarylemma}[2]{%
    \vspace{1cm} 
    \section*{Proof of  Lemma #1} 
    \addcontentsline{toc}{section}{Proof of  Lemma #1} 
    \begin{proof}[\textit{Proof}] 
    #2 
    \end{proof}
}

\title{Separable Geodesic Lagrangian Monte Carlo for Inference in 2-Way Covariance Models}
\author{
    Quinn Simonis and Martin T. Wells \thanks{Department of Statistics and Data Science, Cornell University; qas3@cornell.edu, mtw1@cornell.edu}
}
\date{Cornell University\\December 2024}

\begin{document}

\maketitle

\begin{abstract}
Matrix normal models have an associated 4-tensor for their covariance representation. The covariance array associated with a matrix normal model is naturally represented as a Kronecker-product structured covariance associated with the vector normal, also known as separable covariance matrices. Separable covariance matrices have been studied extensively in the context of multiway data, but little work has been done within the scope of MCMC beyond Gibbs sampling. This paper aims to fill this gap by considering the pullback geometry induced from the Kronecker structure of the parameter space to develop a geodesic Hamiltonian Monte Carlo sampler.
\end{abstract}
\noindent
\begin{small}
\end{small}

\section*{Introduction}
A covariance matrix $\Sigma \in \mathcal{P}(d)$ is separable if for $d = d_{1}\times d_{2}$, we can decompose $\Sigma$ into $\Sigma_{1} \in \mathcal{P}(d_{1})$ and  $\Sigma_{2} \in \mathcal{P}(d_{2})$ as
\begin{equation} \label{eq:kp model}
    \Sigma = \Sigma_{1} \otimes \Sigma_{2},
\end{equation}
where for $A \in \mathbb{R}^{m \times m}$, $B \in \mathbb{R}^{n \times n}$, their Kronecker product, $A \otimes B \in \mathbb{R}^{mn \times mn}$ is defined as
\begin{equation}
    A \otimes B = \begin{pmatrix}
        a_{11}B & a_{12} B & \cdots & a_{1m}B \\
        a_{21}B & \ddots & & \vdots \\
        \vdots & & \ddots & \vdots \\
        a_{m1}B & \cdots & \cdots & a_{mm}B
    \end{pmatrix}.
\end{equation}

Such covariance structures have been studied in the literature on separable covariance \cite{fosdick2014separable, genton2007separable, leorato2021bayesian}, with tests of separability being studied in \cite{dutilleul2018estimation}. Within the scope of MCMC, Hoff \cite{hoff2011separable} studied tensors with a separable covariance structure under the Tucker decomposition and provides a Gibbs sampling algorithm. Within the scope of geometry, \cite{mccormack2023information} investigated separable covariance matrices from an information geometry perspective, and \cite{wiesel2012geodesic} discussed the geodesic convexity of multiway covariances in a multivariate normal assuming a product manifold structure under the affine-invariant metric.

Hamiltonian Monte Carlo (HMC)  \cite{neal2011mcmc} is a flexible MCMC methodology that takes advantage of Hamiltonian dynamics to build a random walk-like sampler based on gradients.  HMC may be accelerated by replacing the static mass matrix tuning parameter with the posterior Fisher information matrix \cite{girolami2011riemann}. HMC has, however, also been shown to be a powerful tool for sampling parameter spaces that possess geometric structure in the form of a Riemannian manifold \cite{byrne2013geodesic}. Algorithms for the MCMC sampling of parameter spaces valued by Riemann manifolds have been extensively studied for manifolds that may be isometrically embedded, such as the Stiefel manifold \cite{jauch2020random, jauch2021monte, pmlr-v97-nirwan19a, pourzanjani2020general} and the sphere \cite{byrne2013geodesic, lan2014spherical}. Regarding covariance estimation, probabilistic PCA methods can be developed using Stiefel sampling methods as was done in \cite{NirwanBerchtold2019}, however, quantifying a suitable prior distribution for the eigenvalues can be challenging in anything but the simplest cases.

Our work directly builds on \cite{holbrook2018geodesic}, which investigates a methodology for direct MCMC simulation on the manifold of symmetric positive definite and Hermitian matrices by leveraging Lagrangian dynamics \cite{lan2015markov}. However, the methodology in \cite{lan2015markov} only pays attention to the estimation of unstructured matrices and has limited scalability without considering additional structure. To this end, we extend their method to covariance matrices which admit a Kronecker structure, such as those found in the vectorization of 2-way data. 

The package {\it stan} \cite{carpenter2017stan} is a probabilistic programming tool for the easy implementation of Hamiltonian Monte Carlo models. However, the Kronecker structure of the separable covariance imposes limitations on the scalability of its implementation. This is due to the nature of {\it stan}'s reverse-mode automatic differentiation techniques and Kronecker products not being optimized in its numerical libraries. 

The organization of the paper is as follows. Section \ref{sec: Mathematical Background} details the matrix normal model, some useful reshaping arguments for efficient interpretation of SGLMC, an introduction to Bayesian inference, and relevant general concepts in Riemannian geometry.  Details the construction of four variants of the affine-invariant metric for separable covariances. In particular, the degeneracy of the metric is investigated when pulled from the Kronecker space to the Cartesian product, and resolutions to this degeneracy are discussed in Section \ref{sec: separable SPD matrices}.  Section \ref{sec: Van-Loan} describes the Pitsianis - Van Loan (PVL) decomposition and how it can be used to straightforwardly derive a Gibbs sampler and compute gradients with respect to the components of the Kronecker structured covariance. Section \ref{sec: HMC} details HMC, Riemann manifold HMC, Lagrangian Monte Carlo, Separable Geodesic Lagrangian Monte Carlo (SGLMC), and adaptation techniques for SGLMC. 
 Empirical comparisons between {\textit{stan}}, Gibbs, and SGLMC using the Riemannian metrics from Section \ref{sec: separable SPD matrices} are given in Section \ref{sec: Empirical Comparisons}. Section \ref{sec: real data} gives an analysis of the Wisconsin breast cancer dataset \cite{misc_breast_cancer_wisconsin_(diagnostic)_17}. All proofs, full empirical comparisons, and additional discussion are included in Section \ref{sec: supplement}.

\section{Mathematical Background} \label{sec: Mathematical Background}
\subsection{The Matrix Normal and Separable Normal Model} \label{sec: Matrix Normal}
The matrix normal likelihood of classical multivariate statistical analysis has recently been applied to capture nuanced relationships in the covariance and mean structure of tabular data, as is frequently found in image analysis \cite{kamm1998kronecker} and spatiotemporal data \cite{greenewald2015robust}. Beyond capturing subtle relationships induced from the tabular format, it provides a useful dimensionality reduction property because of the structure provided by the covariance relationship. Explicitly, if we were to naively vectorize the matrix observations and model them as coming from a normal distribution with unstructured covariance, we would need to estimate $\frac{d(d+1)}{2}$ parameters associated with our covariance if the vector observations were $Y_{i} \sim \mathcal{N}_{d}(0, \Sigma)$. However, in the case where $Y_{i} \sim \mathcal{M} \mathcal{N}_{d_{2}, d_{1}}(0, \Sigma_{2}, \Sigma_{1})$, the number of parameters to be estimated from our covariance structure in (\ref{eq:kp model}) reduces to $\frac{d_{1}(d_{1} + 1)}{2} + \frac{d_{2}(d_{2} + 1)}{2} << \frac{d(d+1)}{2}$ when $d$ is large. This is characterized by the following Theorem.
\begin{theorem} \label{theorem: matrix normal}
    Let $Y_{i} \sim \mathcal{M}\mathcal{N}(0, \Sigma_{2}, \Sigma_{1})$, then the matrix normal is defined as $y_{i} = vec(Y_{i}) \sim \mathcal{N}(0, \Sigma_{1} \otimes \Sigma_{2})$.
\end{theorem}
See \cite{gupta2018matrix} for the corresponding proof of Theorem \ref{theorem: matrix normal}.

Moving from the unstructured covariance produced by naively modeling vectorized matrix observations to vectorizing matrix observations, which are themselves modeled as coming from a matrix normal, reduces our need to model $\Sigma \in \mathcal{P}(d)$ to modeling only the independent pieces $(\Sigma_{1}, \Sigma_{2}) \in \mathcal{P}(d_{1}) \times \mathcal{P}(d_{2})$, which then structurally define $\Sigma$ through the Kronecker product.

\subsection{Bayesian Inference}
Let $y_{1} ,\ldots, y_{n} \in \mathbb{R}^{k}$ denote our observed data generated under the likelihood function $\mathcal{L}(y | q)$. Let $q \in \mathcal{M}$ be a $k$-dimensional vector parameter defined on the manifold $\mathcal{M}$ that parameterize our likelihood. Giving $q$ the prior distribution $P(q)$ produces the posterior distribution
\[
\pi(q) = P(q | y) = \frac{\mathcal{L}(y | q) P(q)}{\int \mathcal{L}(y | q) P(q) dq}.
\]
The integral in the denominator is often intractable in complex problems. MCMC methods are a frequent resolution to such intractability. Generally, such MCMC methods may be classified as Gibbs or random-walk samplers. Gibbs sampling restricts our prior distribution choices to yield conjugacy with the likelihood, and random walk methods are often inefficient. HMC has been a recent development in the MCMC literature which has been shown to be an efficient random walk type methodology which is competitive in performance with Gibbs samplers.

\subsection{Riemannian Manifolds and Geodesics}
A topological manifold $\mathcal{M}$ is a second-countable Hausdorff space such that for every point $q \in \mathcal{M}$, there exists a neighborhood around $q$, $N_{q} \subseteq \mathcal{M}$, such that $N_{q}$ is homeomorphic to Euclidean space. That is, for each $q \in \mathcal{M}$, there exists a continuous bijective mapping $\phi$ from an open set $\mathcal{O}_{q} \subset \mathcal{M}$ into $\mathbb{R}^{k}$ with $q \in \mathcal{O}_{q}$. This implies that topologically $\mathcal{M}$ locally acts like $\mathbb{R}^{k}$. This property allows gradient-based algorithms defined on Euclidean space such as Hamiltonian Monte Carlo to be naturally extended to Riemannian manifolds.

Let $\Gamma = \{\gamma(t) \in \mathcal{M}: \gamma(0) = q, \quad t\in \mathbb{R}^{\geq 0}\}$ be the collection of curves in $\mathcal{M}$ paramterized by $t$ starting at $\gamma(0) = q \in \mathcal{M}$. Then the tangent space at $q$, $T_{q}\mathcal{M}$, will be defined as the equivalence class:
\begin{equation}
    T_{q} \mathcal{M} := \{\gamma'(0): \gamma \in \Gamma\}.
\end{equation}

A Riemannian manifold incorporates more structure on a topological manifold $\mathcal{M}$ as a pair $(\mathcal{M},g)$, where $g$ is the Riemannian metric, or in matrix form, $G$, which we will refer to as the metric tensor. The Riemannian metric is a smooth-varying, bilinear, positive definite inner product defined on the tangent space of $\mathcal{M}$. This Riemannian metric gives the ability to compute inner products between tangent vectors $V_{1}, V_{2} \in \mathcal{T}_{q}\mathcal{M}$ as
\[
<V_{1}, V_{2}>_{q} = V_{1}^{T} G(q) V_{2}.
\]
The matrix form $G$ will give a natural way to extend the first-order gradient updates from Euclidean Hamiltonian Monte Carlo to curved surfaces with a Riemann manifold structure. 

If there exists a smooth map from $\mathcal{M}$ to $\mathbb{R}^{k}$ such that the Riemannian inner product is equivalent to the Euclidean inner product, then such a map is an isometric embedding. \cite{byrne2013geodesic} show how to leverage isometric embedding manifolds to do fast geodesic calculations for MCMC inference. Although such an embedding must exist for any Riemannian manifold by the Nash Embedding Theorem \cite{nash1956imbedding}, the embedding is not known for symmetric positive definite matrices.

The affine connection of a manifold defines the relationship between tangent spaces of distinct points on a manifold. Informally, it gives "rules" for how vector fields are differentiated along a path on a manifold. For a vector field $V(t) \in \mathcal{T}_{\gamma(t)}$, the derivative of $V(t)$ under the affine connection is called the covariant derivative, and is how we measure the change between distinct tangent spaces. The time derivative itself, $\dot{\gamma}(t) = \frac{d \gamma(t)}{dt}$ is a vector field, and when the covariant derivative of $\gamma$ is 0, $\gamma$ is then a geodesic. This property is expressible via the geodesic equation
\begin{equation} \label{eq: Geodesic Equation}
    \dot{\dot{\gamma_{i}}}(t) + \sum_{j,k} \Gamma_{jk}^{i}(\gamma(t)) \dot{\gamma_{j}}(t) \dot{\gamma_{k}}(t) = 0,
\end{equation}
where $\Gamma_{jk}^{i}(x)$ are the Christoffel symbols. Riemannian manifolds induce a unique affine connection named the Levi-Civita connection.

On a Riemannian manifold, geodesics are local extremal paths of the integrated path length
\[
\int_{a}^{b} \| \dot{\gamma}(t) \|_{G} dt, \;\; \text{ such that \; } \|V\|_{G}^{2} = V^{T} G V.
\]

For a geodesic $\gamma: [a,b] \rightarrow \mathcal{M}$, the geodesic flow describes the pair $(\gamma(t), \dot{\gamma}(t))$ which is unique to the initial conditions $(q,p) = (\gamma(0), \dot{\gamma}(0))$. Geodesics are simply the notion of straight lines on a curved surface.
\subsection{Distributions on Manifolds}
The Lebesgue measure is insufficient for sampling from distributions defined on curved surfaces such as the manifold of SPD matrices. In Riemann manifold sampling methods, the Hausdorff measure is a common alternative probability measure. Let $\mathcal{H}^{k}$ denote the $k-$dimensional Hausdorff measure, and $\lambda^{k}$ denote the Lebesgue measure on $\mathbb{R}^{k}$. The relationship between these measures is given by the \textit{area formula} \cite{federer2014geometric}

\begin{equation} \label{eq:area formula}
    \int_{A} g(f(u)) J_{m}f(u) \lambda^{m}(du) = \int_{\mathbb{R}^{k}} g(x) \vert \{u \in A: f(u) = x\} \vert \mathcal{H}^{m}(dx),
\end{equation}
where $J_{m}f(u)$ is the $m-$dimensional Jacobian of f. For $\Sigma \in \mathcal{P}(d)$. Note that (\ref{eq:area formula}) can be interpreted in the context of the Riemannian manifold of SPD matrices as
\begin{equation}
    \mathcal{H}^{k}(d\Sigma) = \sqrt{\vert G(\Sigma) \vert} \lambda^{k}(d\Sigma).
\end{equation}

\section{Geometry of Separable SPD Matrices} \label{sec: separable SPD matrices}
A covariance matrix $\Sigma \in \mathcal{P}(d_{1} \times d_{2})$ is separable if it may be decomposed as $\Sigma = \Sigma_{1} \otimes \Sigma_{2}$, 
where $\Sigma_{1} \in \mathcal{P}(d_{1})$ and $\Sigma_{2} \in \mathcal{P}(d_{2})$. Note the differential of a separable covariance matrix follows a product rule
 $d(\Sigma_{1} \otimes \Sigma_{2}) = (d\Sigma_{1})\otimes \Sigma_{2} + \Sigma_{1} \otimes (d\Sigma_{2})$.
Hence, the tangent space is expressible as
\[
\mathcal{T}_{\Sigma_{1} \otimes \Sigma_{2}}^{kr}(d_{1} \times d_{2}) =  \{ V_{1} \otimes \Sigma_{2} + \Sigma_{1} \otimes V_{2}: V_{1} \in \mathcal{T}_{\Sigma_{1}}(d_{1}), V_{2} \in \mathcal{T}_{\Sigma_{2}}\}.
\]
Let $\Sigma \in \mathcal{P}(d)$, and $S_{1},S_{2} \in \mathcal{T}_{\Sigma}\mathcal{P}(d)$. Define $\psi(\Sigma)$ to be the Boltzmann entropy of $\Sigma$
\begin{equation} \label{eq: Boltzmann entropy}
\psi(\Sigma) = \log(\vert \Sigma \vert) = tr(\log (\Sigma)).
\end{equation}
The Riemannian inner product at $\Sigma \in \mathcal{P}(d)$ with $S_{1}, S_{2} \in \mathcal{T}_{\Sigma} \mathcal{P}(d)$ under the affine-invariant metric was derived in \cite{moakher2011riemannian} through the Hessian of the Boltzmann Entropy
\begin{align} 
    g_{\Sigma}(S_{1}, S_{2}) &= - Hess \psi(\Sigma)(S_{1}, S_{2}) = -\frac{\partial^{2}}{\partial s \partial t} \psi(P + tS_{1} + sS_{2})\vert_{t = s = 0} \nonumber \quad t,s \in \mathbb{R}\\
    &= tr(\Sigma^{-1} S_{1} \Sigma^{-1} S_{2}). \label{eq: Riemannian metric full}
\end{align}
Let $\nabla_{\Sigma}^{E} f(\Sigma)$ denote the euclidean gradient of a $f$ at $\Sigma \in \mathcal{P}(d)$. We express the Riemannian gradient, $\nabla_{\Sigma}^{R} f(\Sigma) \in \mathcal{T}_{\Sigma} P(d)$ as
\[
vec(\nabla_{\Sigma}^{R}) = G(\Sigma)^{-1} vec(\nabla_{\Sigma}^{E}),
\]
where the vec operator is defined as
\begin{equation} \label{eq: vec operator}
    vec(A) = [A_{11}, A_{21}, \ldots, A_{d1}, A_{12}, \ldots, A_{dd}]
\end{equation}
and $G(\Sigma)$ is the metric $g_{\Sigma}$ in matrix form when vectorizing $S_{1}$ and $S_{2}$ and rewriting as a quadratic form. Under the affine-invariant metric, the solution to (\ref{eq: Geodesic Equation}) has a convenient analytic expression given an initial tangent vector:
\[
\Sigma(t) = \Sigma(0)^{-\frac{1}{2}} \exp(t \Sigma(0)^{\frac{1}{2}} V(0) \Sigma(0)^{\frac{1}{2}}) \Sigma(0)^{-\frac{1}{2}} \quad V(0)\in \mathcal{T}_{\Sigma}\mathcal{P}(d).
\]

Given the Kronecker product is a smooth map from two SPD matrix spaces to a higher dimensional SPD matrix space, observing that the Hessian of the Boltzmann entropy would follow a product rule acting on itself, and given that Kronecker products themselves follow a convenient product rule, we would imagine there is some possibility of considering a local geometric structure on the Kronecker components being induced from a global geometric structure on the entire Kronecker structured SPD matrix.
\newline
With the geometric formulation above, we can generalize the affine-invariant metric to Kronecker-structured SPD matrices in the following proposition:
\begin{proposition} \label{prop: Separable Metric Tensor}
    Let $\Sigma_{1} \in \mathcal{P}(d_{1})$, $\Sigma_{2} \in \mathcal{P}(d_{2})$, and $V_{1} \in \mathcal{T}\mathcal{P}(d_{1})$, $V_{2} \in \mathcal{T} \mathcal{P}(d_{2})$. For $V = V_{1} \otimes \Sigma_{2} + \Sigma_{1} \otimes V_{2} \in T_{\Sigma_{1} \otimes \Sigma_2}$, the corresponding norm under the affine invariant metric is
    \[
    \|V\|_{\Sigma_{1} \otimes \Sigma_{2}} = d_{2} tr(\Sigma_{1}^{-1} V_{1} \Sigma_{1}^{-1} V_{1}) + d_{1} tr(\Sigma_{2}^{-1} V_{2} \Sigma_{2}^{-1} V_{2}) + 2 tr(\Sigma_{1}^{-1} V_{1}) tr(\Sigma_{2}^{-1} V_{2})
    \]
    or in matrix form, with 
    \[
    v = \begin{pmatrix}
        vec(V_{1}) \\
        vec(V_{2})
    \end{pmatrix}.
    \]
    Then $\| V \|_{\Sigma_{1} \otimes \Sigma_{2}} = v^{t} G^{\otimes}(\Sigma_{1}, \Sigma_{2}) v$ where
    \[
    G^{\otimes}(\Sigma_{1}, \Sigma_{2}) = \begin{pmatrix}
        d_{2} \Sigma_{1}^{-1} \otimes \Sigma_{1}^{-1} & vec(\Sigma_{1}^{-1}) vec(\Sigma_{2}^{-1})^{T} \\
        vec(\Sigma_{2}^{-1}) vec(\Sigma_{1}^{-1})^{T} & d_{1} \Sigma_{2}^{-1} \otimes \Sigma_{2}^{-1}
    \end{pmatrix}.
    \]
\end{proposition}
Note that this norm is not conducive to HMC sampling as it is degenerate:
\begin{proposition} \label{prop: degenerate determinant}
    Let 
    \[
    G^{\otimes}(\Sigma_{1}, \Sigma_{2}) = \begin{pmatrix}
         d_{2} \Sigma_{1}^{-1}(\tau) \otimes \Sigma_{1}^{-1}(\tau)  &  vec(\Sigma_{1}^{-1}(\tau))vec(\Sigma_{2}^{-1}(\tau))^{T} \\
         vec(\Sigma_{2}^{-1}(\tau))vec(\Sigma_{2}^{-1}(\tau))^{T} &  d_{1} \Sigma_{2}^{-1}(\tau) \otimes \Sigma_{2}^{-1}(\tau) 
    \end{pmatrix},
    \]
then $|G^{\otimes}(\Sigma_{1}, \Sigma_{2})| = 0$.
\end{proposition}
Proposition \ref{prop: degenerate determinant} provides an induced metric known as a "pullback metric" and highlights that this metric is degenerate in the sense that it is not a valid Riemannian metric. We refrain from any of the formality regarding its construction or degeneracy, and defer to Section \ref{sec: supplement} for the associated discussion. In summary, however, Proposition \ref{prop: degenerate determinant} highlights that the induced inner product between tangent vectors on $\mathcal{P}(d_{1}) \times \mathcal{P}(d_{2})$ is not conducive to HMC sampling as it is not positive definite. This posits three possibilities for what we can do to deal with this degeneracy: regularization of the metric tensor, orthogonalization of the metric tensor, or simplifying the geometry from its natural (pullback) structure to its canonical (product) structure.

Regularization will be the first to be treated in the following lemma.
\begin{lemma} \label{lemma: vec metric tensor}
    Let
    \begin{equation} \label{eq: half vec metric tensor}
     G_{\alpha}^{\otimes}(\Sigma_{1}, \Sigma_{2}) = \begin{pmatrix}
         d_{2} \Sigma_{1}^{-1}(\tau) \otimes \Sigma_{1}^{-1}(\tau)  &  \alpha \times vec(\Sigma_{1}^{-1}(\tau))vec(\Sigma_{2}^{-1}(\tau))^{T} \\
         \alpha \times vec(\Sigma_{2}^{-1}(\tau))vec(\Sigma_{1}^{-1}(\tau))^{T} &  d_{1} \Sigma_{2}^{-1}(\tau) \otimes \Sigma_{2}^{-1}(\tau) 
    \end{pmatrix}.
    \end{equation}
Then $ G_{\alpha}^{\otimes}(\Sigma_{1}, \Sigma_{2})$ is positive definite for any $0 \leq \alpha < 1$.
\end{lemma}
Hence, Lemma \ref{lemma: vec metric tensor} shows that regularizing the interaction terms of the tensor satisfies the necessary property of producing a positive definite metric tensor. Regularization can be a compelling approach from a sampling perspective, as it guarantees that we do not need to enforce any constraints during our sampling procedure and simultaneously allows for interaction between Kronecker components within the geodesic paths. Geometrically, 
when $\alpha$ is close to 1 we can view this metric as the one that generates a geodesic path with the closest inner product to when we endow the separable space naively with the affine-invariant metric.

Orthognalization of the metric tensor may, however, be more compelling computationally to deal with the degeneracy of the manifold. The degeneracy of the geometry of the manifolds can be intuitively understood by observing the volume indeterminacy of the space
\[
\vert \Sigma_{1} \otimes \Sigma_{2} \vert = \vert c \Sigma_{1} \vert \vert \frac{1}{c} \Sigma_{2} \vert
\]
for any $c \in \mathbb{R}$. Dealing with this degeneracy, as highlighted in \cite{bouchard2021line} and \cite{mccormack2023information}, means imposing the constraint $\vert \Sigma_{2} \vert = 1$. Orthogonalization is treated in the following lemma.
\begin{lemma}
    Let $V_{2} \in T_{\Sigma_{2}} \mathcal{P}(d_{2})$, and define the orthogonal map
    \[
    P_{\Sigma_{2}}(V_{2}) = V_{2} - \frac{tr(V_{2} \Sigma_{2}^{-1})}{d_{2}} V_{2}.
    \]
    Then $\|V_{1} \vert P_{\Sigma_{2}}(V_{2})\| g_{\Sigma_{1} \otimes \Sigma_{2}} = d_{1} tr(\Sigma_{1}^{-1}V_{1} \Sigma_{1}^{-1}V_{1}) + d_{2} tr(\Sigma_{2}^{-1}V_{2} \Sigma_{2}^{-1}V_{2})$. Or in matrix form
    \begin{equation} \label{eq: orthogonalized metric tensor}
        G^{O}(\Sigma_{1}, \Sigma_{2}) = \begin{pmatrix}
            d_{1} \Sigma_{1}^{-1} \otimes \Sigma_{1}^{-1} & 0 \\
            0 & d_{2} \Sigma_{2}^{-1} \otimes \Sigma_{2}^{-1}
        \end{pmatrix}.
    \end{equation}
\end{lemma}
Notice that we can just as easily rewrite the momentum in vector form as the idempotent projection matrix:
\[
P_{\Sigma_{2}} (vec(V_{2})) = I_{d_{2}^{2}} - \frac{vec(\Sigma_{2}) \otimes vec(\Sigma_{2})^{t}}{d_{2}}.
\]
Such a projection can be viewed analogously in the work of \cite{byrne2013geodesic}.

Neither of these first two metrics gives any notion to the discrepancy between the likelihood and prior geometry (with the former being pullback of the Kronecker product and the latter being a standard product manifold geometry by assumption of independence), and we found empirically both of these choices to work well in terms of expected sample size per iteration (ESS/it). However, assuming the orthogonality conditions of metric (\ref{eq: orthogonalized metric tensor}) are satisfied, to account for the prior geometry in the computation of the metric tensor, we could consider constructing a weighted metric where for $\omega \in (0,1)$
\begin{equation} \label{eq: weighted metric}
G^{W}(\Sigma_{1}, \Sigma_{2} \vert \omega) = \begin{pmatrix}
    (\omega d_{2} + (1-\omega)) \Sigma_{1}^{-1} \otimes \Sigma_{1}^{-1} & 0 \\
    0 & (\omega d_{1} + (1-\omega)) \Sigma_{2}^{-1} \otimes \Sigma_{2}^{-1}
\end{pmatrix}.
\end{equation}

Lastly, we may consider a product manifold geometry choice:
\begin{equation} \label{eq: Product Manifold Metric}
G^{\times}(\Sigma_{1}, \Sigma_{2}) = \begin{pmatrix}
    \Sigma_{1}^{-1} \otimes \Sigma_{1}^{-1} & 0 \\
    0 & \Sigma_{2}^{-1} \otimes \Sigma_{2}^{-1}
\end{pmatrix}.
\end{equation}

In any case of using regularization or orthognalization of the pullback metric, the use of an independent product manifold metric, or the weighted metric, as shown in Section \ref{sec: supplement}, the geodesic solutions are identical to the geodesic solutions of the product manifold.
\begin{proposition} \label{prop: Separable flows}
    Let $\Sigma(0) = \Sigma_{1}(0) \otimes \Sigma_{2}(0)$, be a path on $\mathcal{P}^{k_{1} \otimes k_{2}}(k_{1} \times k_{2})$. Given initial velocities $V_{1}(0) \in \mathcal{T}_{\Sigma_{1}(0)} \mathcal{P}(k_{1})$, $V_{2}(0) \in \mathcal{T}_{\Sigma_{2}(0)} \mathcal{P}(k_{2})$, then the co-geodesic flow on $\Sigma(t)$ under either metric (\ref{eq: half vec metric tensor}), (\ref{eq: Product Manifold Metric}), (\ref{eq: weighted metric}), or (\ref{eq: orthogonalized metric tensor}) (the latter two with additional constraint $\vert \Sigma_{2} \vert = 1$)
    \begin{align}
    \Sigma(\tau) = \Sigma_{1}(\tau) \otimes \Sigma_{2}(\tau),
    \end{align}
    with corresponding velocity
    \begin{align}
        V(\tau) = V_{1}(\tau) \otimes \Sigma_{2}(\tau) + \Sigma_{2}(\tau) \otimes V_{2}(\tau),
    \end{align}
    where
    \begin{equation}
    \Sigma_{i}(t) = \exp_{\Sigma_{i}}tV(0) = \Sigma_{i}(0)^{\frac{1}{2}} \exp (t \Sigma_{i}(0)^{-\frac{1}{2}} V_{i}(0) \Sigma_{i}(0)^{-\frac{1}{2}}) \Sigma_{i}(0)^{\frac{1}{2}}.
\end{equation}
The corresponding flow on the tangent bundle is obtained by taking the derivative with respect to $t$ is
\begin{equation} \label{eq: velocity flow}
    V_{i}(t) = \dot{\Sigma_{i}}(t) = \frac{d}{dt} \exp_{\Sigma_{i}} tV(0) = V_{i}(0) \Sigma_{i}(0)^{-\frac{1}{2}} \exp (t \Sigma_{i}(0)^{-\frac{1}{2}} V_{i}(0) \Sigma_{i}(0)^{-\frac{1}{2}}) \Sigma_{i}(0)^{\frac{1}{2}}.
\end{equation}
\end{proposition}
The only analytic differences being in efficiency and the difference in the Hausdorff measure augmentation is highlighted in the next result.
\begin{lemma} \label{lemma: half vec metric tensor}
    Under metrics (\ref{eq: half vec metric tensor}) or (\ref{eq: Product Manifold Metric}), the determinant is given as
    \begin{equation}
        \vert G(\Sigma_{1}, \Sigma_{2}) \vert \propto \vert \Sigma_{1} \vert^{-(d_{1} + 1)} \vert \Sigma_{2} \vert^{-(d_{2} + 1)}.
    \end{equation}
    With metrics (\ref{eq: orthogonalized metric tensor}) or (\ref{eq: weighted metric}), the determinant is
    \begin{equation}
        \vert G(\Sigma_{1}, \Sigma_{2}) \vert \propto \vert \Sigma_{1} \vert^{-(d_{1} + 1)}.
    \end{equation}
\end{lemma}

As a matter of efficiency, computing with the regularized metric (\ref{eq: half vec metric tensor}) requires explicit construction and inversion of the metric tensor, being at least $\mathcal{O}((d_{1}^{2} + d_{2}^{2})^{3})$ operation when done naively (although the Schur complement \cite{zhang2006schur} could give a more efficient formula for inversion). This can be prohibitively expensive when either $d_{1}$ or $d_{2}$ are large. However, under the orthognalized metric, we can instead see that
\begin{align*}
    G^{O}(\Sigma_{1}, \Sigma_{2})^{-1} vec(\nabla_{\Sigma_{1}}, \nabla_{\Sigma_{2}}) = vec(\frac{1}{d_{2}} \Sigma_{1} \otimes \Sigma_{1} vec(\nabla_{\Sigma_{1}}), P_{\Sigma_{2}}(\frac{1}{d_{1}} \Sigma_{2} \otimes \Sigma_{2} vec(\Sigma_{2}))).
\end{align*}

With an abuse of notation, using $G^{O}(\Sigma_{i})$ to refer to the component of the metric tensor corresponding to $\Sigma_{i}$, the Riemannian gradients can instead be written efficiently in matrix form (bypassing the explicit construction of the metric tensor)
\begin{align*}
    \nabla_{\Sigma_{1}}^{R} = mat(G^{O}(\Sigma_{1})^{-1} vec(\nabla_{\Sigma_{1}})) &= \frac{1}{d_{2}} \Sigma_{1} \nabla_{\Sigma_{1}} \Sigma_{1} \\
    \nabla_{\Sigma_{2}}^{R} = mat(G^{O}(\Sigma_{2})^{-1} vec(\nabla_{\Sigma_{2}})) &= \frac{1}{d_{1}} P_{\Sigma_{2}}(\Sigma_{2} \nabla_{\Sigma_{2}} \Sigma_{2}).
\end{align*}

The critical distinction between these two metrics can be viewed as under the regularized metric, the corresponding posterior samples will be representative of the unconstrained posterior. That is, they will be identical to the unconstrained Gibbs samples. The $\vert \Sigma_{2} \vert = 1$ constraint clearly imposes some bias on the posterior samples, although such a bias is redundant in the Kronecker model due to scale indeterminacy. However, this computational gain is substantial, and the orthogonal metric often provides a much more stable sampler in regards to both acceptance probability choice during dynamic tuning and the choice of prior distribution, which can be crucial in the performance of the resulting algorithm. This bias is demonstrated in Figure \ref{fig: bias comparisons} below. Note, however, that this bias is subtle, and equivalent to the bias introduced when normalizing samples of $\Sigma_{2}$ with the Gibbs sampler.

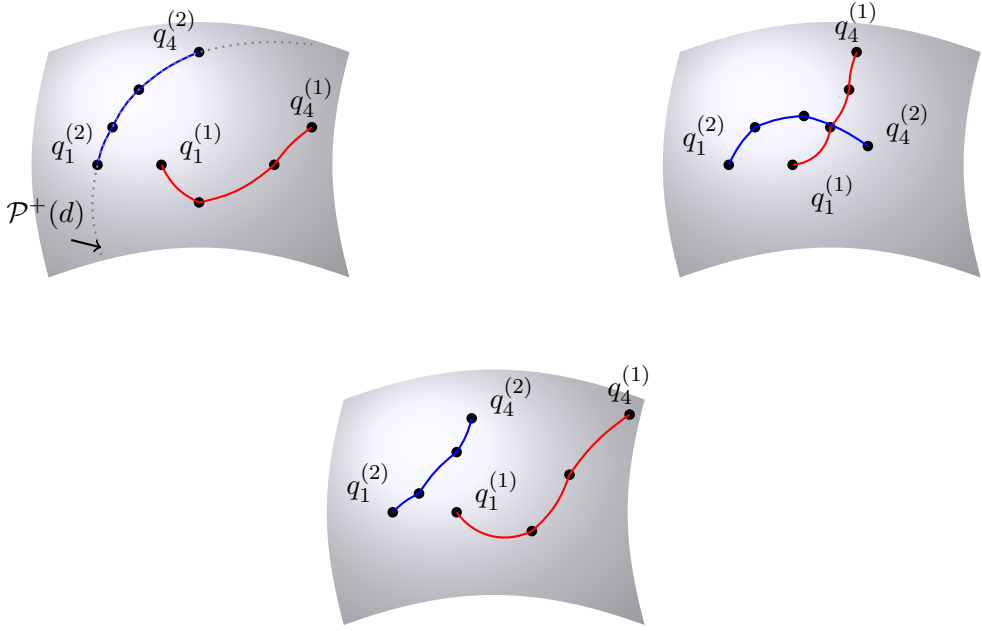
\begin{figure}[H] 
    \centering
    \begin{minipage}[t]{0.48\textwidth}
        \centering
        \begin{tikzpicture}
            \shade [ball color=blue!10!white, opacity=0.50] (0,0) to [bend left=20] (4,0) to [bend left=15] (4,3) to [bend right=20] (0,3) to [bend right=15] cycle;
            
            \coordinate (P1) at (1.5, 1.5);
            \coordinate (P2) at (2, 1);
            \coordinate (P3) at (3, 1.5);
            \coordinate (P4) at (3.5, 2);

            \fill (P1) circle (2pt) node[above right, xshift=3pt, yshift=-3pt] {$q_{1}^{(1)}$};
            \fill (P2) circle (2pt);
            \fill (P3) circle (2pt);
            \fill (P4) circle (2pt) node[above] {$q_{4}^{(1)}$};

            \draw [red, thick] (P1) to [bend right=20] (P2) to [bend right=15] (P3) to [bend left=10] (P4);

            \coordinate (B1) at (.65, 1.5);
            \coordinate (B2) at (.85, 2);
            \coordinate (B3) at (1.2, 2.5);
            \coordinate (B4) at (2, 3);

            \fill (B1) circle (2pt) node[above left, xshift=3pt, yshift=-3pt] {$q_{1}^{(2)}$};
            \fill (B2) circle (2pt);
            \fill (B3) circle (2pt);
            \fill (B4) circle (2pt) node[above left, xshift=3pt, yshift=-3pt] {$q_{4}^{(2)}$};
            \draw [blue, thick] (B1) to [bend left=10] (B2) to [bend left=10] (B3) to [bend left=10] (B4);
            
            \draw [black!50, dotted, thick] (.7,.3) to [bend left=15] (B1) to [bend left=10] (B2) to [bend left=10] (B3) to [bend left=10] (B4) to [bend left=10] (3.5,3.1);

            \draw [->, thick] (.3,.5) -- (.7,.4) node[above left, yshift = 3pt,xshift = -2pt] {$\mathcal{P}^{+}(d)$};
        \end{tikzpicture}
    \end{minipage}
    \hfill
    \begin{minipage}[t]{0.48\textwidth}
        \centering
        \begin{tikzpicture}
            \shade [ball color=blue!10!white, opacity=0.50] (0,0) to [bend left=20] (4,0) to [bend left=15] (4,3) to [bend right=20] (0,3) to [bend right=15] cycle;
            
            \coordinate (P1) at (1.5, 1.5);
            \coordinate (P2) at (2, 2);
            \coordinate (P3) at (2.25, 2.5);
            \coordinate (P4) at (2.35, 3);

            \fill (P1) circle (2pt) node[below right, xshift=3pt] {$q_{1}^{(1)}$};
            \fill (P2) circle (2pt);
            \fill (P3) circle (2pt);
            \fill (P4) circle (2pt) node[above] {$q_{4}^{(1)}$};

            \draw [red, thick] (P1) to [bend right=40] (P2) to [bend right=15] (P3) to [bend left=10] (P4);

            \coordinate (B1) at (.65, 1.5);
            \coordinate (B2) at (1, 2);
            \coordinate (B3) at (1.65, 2.15);
            \coordinate (B4) at (2.5, 1.75);

            \fill (B1) circle (2pt) node[above left, xshift=3pt] {$q_{1}^{(2)}$};
            \fill (B2) circle (2pt);
            \fill (B3) circle (2pt);
            \fill (B4) circle (2pt) node[above right, xshift=3pt, yshift=-3pt] {$q_{4}^{(2)}$};
            \draw [blue, thick] (B1) to [bend left=10] (B2) to [bend left=10] (B3) to [bend left=10] (B4);
        \end{tikzpicture}
    \end{minipage}

    \vspace{1cm}

    \begin{minipage}[t]{0.98\textwidth}
        \centering
        \begin{tikzpicture}
            \shade [ball color=blue!10!white, opacity=0.50] (0,0) to [bend left=20] (4,0) to [bend left=15] (4,3) to [bend right=20] (0,3) to [bend right=15] cycle;
            
            \coordinate (P1) at (1.5, 1.5);
            \coordinate (P2) at (2.5, 1.25);
            \coordinate (P3) at (3, 2);
            \coordinate (P4) at (3.8, 2.8);

            \fill (P1) circle (2pt) node[above right, xshift=3pt, yshift=-3pt] {$q_{1}^{(1)}$};
            \fill (P2) circle (2pt);
            \fill (P3) circle (2pt);
            \fill (P4) circle (2pt) node[above] {$q_{4}^{(1)}$};

            \draw [red, thick] (P1) to [bend right=40] (P2) to [bend right=15] (P3) to [bend left=10] (P4);

            \coordinate (B1) at (.65, 1.5);
            \coordinate (B2) at (1, 1.75);
            \coordinate (B3) at (1.5, 2.3);
            \coordinate (B4) at (1.7, 2.75);

            \fill (B1) circle (2pt) node[above left, xshift=3pt] {$q_{1}^{(2)}$};
            \fill (B2) circle (2pt);
            \fill (B3) circle (2pt);
            \fill (B4) circle (2pt) node[above right, xshift=3pt, yshift=-3pt] {$q_{4}^{(2)}$};
            \draw [blue, thick] (B1) to [bend left=10] (B2) to [bend left=10] (B3) to [bend right=10] (B4);
        \end{tikzpicture}
    \end{minipage}
    \caption{Illustration of HMC trajectories when $\Sigma_{1}, \Sigma_{2}$ are both $d\times d$ SPD matrices under metrics $G^{O}$ (top left), $G^{\otimes}_{\alpha}$ (top right), and $G^{\times}$ (bottom).}
    \label{fig: HMC trajectories}
\end{figure}

Figure \ref{fig: HMC trajectories} illustrate the HMC trajectories intuitively when $\Sigma_{1}, \Sigma_{2}$ are both $d\times d$ SPD matrices under metrics $G^{O}$ , $G^{\otimes}_{\alpha}$, and $G^{\times}$. $G^{O}$ follows a constrained path determined by $\vert \Sigma_{1} \vert$, $G^{\otimes}_{\alpha}$ follows a coupled geodesic path, where the positions in $\Sigma_{1}$ are determined by the positions in $\Sigma_{2}$ and vise versa. $G^{\times}$ ignores any implications of the Kronecker structure in the construction of geodesic paths.

\section{The Pitsianis-Van Loan Decomposition and Posterior Sampling} \label{sec: Van-Loan}
Given $\Sigma \in \mathcal{P}(d)$ such that $d$ is non-prime with $d = d_{1}d_{2}$, $\Sigma$ may be decomposed into $\Sigma = \sum_{i = 1}^{r} \Sigma_{1}^{(i)} \otimes \Sigma_{2}^{(i)}$ where $\Sigma_{1}^{(i)} \in \mathbb{R}^{d_{1}}$ and $\Sigma_{2}^{(i)} \in \mathbb{R}^{d_{2}}$. The result of this follows from Sections $2.2$ and $2.4$ of \cite{pitsanis1997kronecker}, which we refer to as the Pitsianis - Van Loan (P-VL) decomposition \cite{VanLoan1993}. Ultimately, we can view this decomposition as a restructured SVD decomposition, hence it's full summand is exact for representing $\Sigma$.

For Hamiltonian Monte Carlo, we need to compute derivatives with respect to $\Sigma_{1}$ and $\Sigma_{2}$. It is straightforward but expensive to do this for the matrix normal model, and inexpensive but difficult for the vector normal model. The vector normal model is naturally defined on the manifold of Kronecker-product structured matrices $\Sigma \in \mathcal{P}(d_{1}) \otimes \mathcal{P}(d_{2}) \not = \mathcal{P}(d_{1}) \times \mathcal{P}(d_{2})$.\\
To deal with this, for $Y_{1}, \ldots, Y_{n} \sim \mathcal{N}(0, \Sigma_{1} \otimes \Sigma_{2})$, and through the Van Loan decomposition, let $\sum_{i = 1}^{n} Y_{i} = \sum_{k = 1}^{d_{min}^{2}} A_{k} \otimes B_{k}$. Then
\begin{align}
    \mathcal{L}(\Sigma_{1}, \Sigma_{2}|y_{1},\ldots, y_{n}) &\propto - \frac{nd_{2}}{2}\log \vert \Sigma_{1} \vert - \frac{nd_{1}}{2} \log \vert \Sigma_{2} \vert - \frac{1}{2} tr(\Sigma_{1}^{-1} \otimes \Sigma_{2}^{-1} (\sum_{k = 1}^{d_{min}^{2}} A_{k} \otimes B_{k})) \nonumber \\
    &= - \frac{nd_{2}}{2}\log \vert \Sigma_{1}^{-1} \vert - \frac{nd_{1}}{2} \log \vert \Sigma_{2}^{-1} \vert - \frac{1}{2} tr(\sum_{k = 1}^{d_{min}^{2}} [\Sigma_{1}A_{k} \otimes \Sigma_{2}  B_{k}]) \label{eq: mixed product likelihood}\\
    &= - \frac{nd_{2}}{2}\log \vert \Sigma_{1}^{-1} \vert - \frac{nd_{1}}{2} \log \vert \Sigma_{2}^{-1} \vert - \frac{1}{2} \sum_{k = 1}^{d_{min}^{2}}tr(\Sigma_{1} A_{k})tr( \Sigma_{2} B_{k}) \label{eq: mixed trace likelihood}
\end{align}
where (\ref{eq: mixed product likelihood}) and (\ref{eq: mixed trace likelihood}) follow from the mixed product (\ref{eq: mixed product}) and mixed trace (\ref{eq: mixed trace}) properties of Kronecker products, respectively.

If $Q \in \mathcal{P}(d)$, the Inverse-Wishart density is defined as:
\[
Q \sim IW(\nu, T) \implies f_{IW}(Q) = \frac{\vert T \vert^{\nu/2}}{2^{\nu d/2} \Gamma_{d}(\frac{\nu}{2})} \vert Q \vert^{-(\nu + d + 1)/2}e^{-\frac{1}{2} tr(TQ^{-1})}
\]

Expressing the likelihood through the P-VL decomposition and placing independent Inverse Wishart priors on each of $\Sigma_{1},\Sigma_{2}$ leads to the following full conditionals.
\begin{proposition} \label{prop: Gibbs Lemma} Let $Y_{i} \sim \mathcal{N}(0, \Sigma_{1} \otimes \Sigma_{2})$ with P-VL decomposition given by $\sum_{i = 1}^{n} Y_{i} Y_{i}^{T} = \sum_{k = 1}^{r} A_{k} \otimes B_{k}$. Imposing the following priors on $\Sigma_{1}$ and $\Sigma_{2}$
\begin{equation}
    \Sigma_{j} \sim \mathcal{I}\mathcal{W}(\nu_{0j} = d_{j} + 2, S_{0j} = \frac{\gamma}{d_{j}} I_{d_{j}})  \quad j=1, 2.
\end{equation}
The full conditionals posteriors are expressible as
\begin{align}
    \Sigma_{1} &\vert Y, \Sigma_{2} \sim \mathcal{I} \mathcal{W} (\nu_{01} = d_{1} + 2 + d_{2}n, S_{01} + \sum_{k = 1}^{r} A_{k} tr(\Sigma_{2}^{-1}B_{k})) \\
    \Sigma_{2} &\vert Y, \Sigma_{1} \sim \mathcal{I} \mathcal{W} (\nu_{02} = d_{2} + 2 + d_{1}n, S_{02} + \sum_{k = 1}^{r} B_{k} tr(\Sigma_{1}^{-1}A_{k})).
\end{align}
\end{proposition}
Likewise, the likelihood gradients can be calculated in a straightforward way using the P-VL decomposition.
\begin{proposition} \label{Prop: HMC Lemma}
For a separable covariance model of the form $Y_{i} \sim \mathcal{N}(0, \Sigma_{1} \otimes \Sigma_{2})$ with corresponding P-VL decomposition given by $\sum_{i = 1}^{n} Y_{i} Y_{i}^{T} = \sum_{k = 1}^{r} A_{k} \otimes B_{k}$, the gradient of the negative log-likelihood for $Y \in (\mathbb{R}^{m})^{n}$ with respect to $\Sigma_{1}$ and $\Sigma_{2}$ are given by
\begin{align} \label{eq: Seperable likelihood gradient fixed 1}
    -\nabla_{\Sigma_1} \mathcal{L}(\Sigma_{1}, \Sigma_{2}; Y) &= -\frac{m_{2} + n}{2} \Sigma_{1}^{-1} + \frac{1}{2}\sum_{k = 1}^{r}[tr(\Sigma_{2}^{-1}B_{k}) \Sigma_{1}^{-1} A_{k} \Sigma_{1}^{-1}]  \\
    \label{eq: Seperable likelihood gradient fixed 2} -\nabla_{\Sigma_2} \mathcal{L}(\Sigma_{1}, \Sigma_{2}; Y) &= -\frac{m_{1} + n}{2} \Sigma_{2}^{-1} + \frac{1}{2}\sum_{k = 1}^{r}[tr(\Sigma_{1}^{-1}A_{k}) \Sigma_{2}^{-1} B_{k} \Sigma_{2}^{-1}].
\end{align}
\end{proposition}

For $\Sigma \in \mathcal{P}(d)$, computation of the exponential map is $\mathcal{O}(d^{3})$. For $\Sigma_{1} \otimes \Sigma_{2} \in \mathcal{P}(d_{1}) \times \mathcal{P}_(d_{2})$, this scales as $\mathcal{O}(d_{1}^{3} + d_{2}^{3})$. Note these operations can be run in parallel to result in $\mathcal{O}(d_{max}^{3})$, where $d_{\max} = \max \{d_{1}, d_{2}\}$. The most expensive operation of the gradient, the inversion of the larger of $\Sigma_{1}$ and $\Sigma_{2}$, results in a computational complexity of the gradient of $\mathcal{O}(r \cdot d_{max}^{3})$. Comparatively for gradients in the case of a matrix normal is limited to $\mathcal{O}(n \cdot d_{max}^{3})$. Hence, gradients are much more efficient in the case of treating a matrix normal as a vector normal in the context of Hamiltonian or Lagrangian Monte Carlo.

\subsection{Other Priors for Covariance Matrices} \label{sec: priors}
In all the examples considered, we impose independent priors on the components $\Sigma_{1}$ and $\Sigma_{2}$. In each dimensionality experiment and regularization parameter experiment, we impose IW priors. In Section \ref{sec: Empirical Comparisons}, we consider the following prior choices with corresponding log density and gradient up to proportionality.

The covariance reference prior derived in \cite{yang1994estimation} has been applied in various settings, and is notably interesting in applications within an HMC context \cite{holbrook2018geodesic} due to its non-conjugacy with the multivariate normal likelihood. Compared to the Jeffreys or inverse Wishart prior, the reference
prior places considerably more mass near the region of equality of the
eigenvalues.  It is plausible that the reference prior produces a covariance matrix estimator with better eigenstructure shrinkage.  The reference prior density is given by
    \begin{align}
        \log f^{(R)}(\Sigma) &\propto - \log \vert \Sigma \vert - \sum_{k < j} \log (\lambda_{k} - \lambda_{j}) \nonumber \\
        \nabla_{\Sigma} \log f^{(R)}(\Sigma) &= - \Sigma^{-1} - \sum_{k < j}(\sum_{i = 1}^{d}(V_{ki}^{-1} - V_{ji}^{-1} ) \Sigma^{i - 1})/(\lambda_{k} - \lambda_{j}). \nonumber
    \end{align}
    Here, $V$ is the Vandermonde matrix
    \[
    V^{T} = \begin{pmatrix}
        1 & \lambda_{1} & \lambda_{1}^{2} & \cdots & \lambda_{1}^{n - 2} & \lambda_{1}^{n - 1} \\
        1 & \lambda_{2} & \lambda_{2}^{2} & \cdots & \lambda_{2}^{n - 2} & \lambda_{2}^{n - 1} \\
        \vdots & \vdots & \vdots & \ddots & \vdots & \vdots \\
        1 & \lambda_{n} & \lambda_{n}^{2} & \cdots & \lambda_{n}^{n - 2} & \lambda_{n}^{n-1}
    \end{pmatrix}.
    \]
    The reference prior gradient was derived in \cite{magnus2019matrix} when the eigenvalues $\{\lambda_{i}\}_{i = 1}^{d}$ are distinct.
   
Recently, \cite{berger2020bayesian} proposed the Shrinkage Inverse Wishart (SIW) prior and show that it has excellent decision-theoretic estimation properties as well as good eigenstructure shrinkage.  The prior density is given by
    \begin{align}
        \log f^{(SIW)}(\Sigma | a, H = c I_{d})) &\propto - \frac{1}{2} tr(\Sigma^{-1} c) - a \log \vert \Sigma \vert - \sum_{k < j} \log (\lambda_{k} - \lambda_{j}) \nonumber \\
        \nabla_{\Sigma} \log f^{(SIW)}(\Sigma | a, H = c I_{d})) &= \frac{1}{2} \Sigma^{-1} c \Sigma^{-1} - a \Sigma^{-1} - \sum_{k < j}(\sum_{i = 1}^{d}(V_{ki}^{-1} - V_{ji}^{-1} ) \Sigma^{i - 1})/(\lambda_{k} - \lambda_{j}),
    \end{align}
    where 
    \begin{align}
    a &= 2 + \frac{(2(d+2) - d - 2)(d + 2 - d - 1)}{(d+2)(d+1) - d(d + 2)} = 3 \label{eq: moment match 1}\\
    c &= \frac{\frac{\sqrt{\gamma}}{d}(2(d + 2) - d - 2)}{(d+2)(d + 1) - d(d+2)} = \frac{\sqrt{\gamma}}{d} \label{eq: moment match 2}.
    \end{align}
    These constants were chosen to moment-match the SIW prior with the inverse Wishart prior, as discussed in Lemma 2 of \cite{berger2020bayesian}.

\section{Hamiltonian Monte Carlo, Geodesic Lagrangian Monte Carlo, and Separable Geodesic Lagrangian Monte Carlo} \label{sec: HMC}

Hamiltonian Monte Carlo is an MCMC methodology which leverages the mathematical insights underlying conservation of energy to develop a clever system for effective sampler by only sampling with the intent of preserving the total energy of our posterior with an auxiliary momentum variable. More specifically, Hamiltonian Monte Carlo works by constructing a Hamiltonian system that is composed of a parameter vector, $q$, and an auxiliary "momentum" variable $p$ of the same dimension. Taking the negative logarithm of the posterior density turns the density into a potential energy function, $U(q) = - \log \pi (q)$. The kinetic energy $K(p)$ is defined entirely by the auxiliary momentum variable. The Hamiltonian is defined by the sum of these two quantities
\[
\mathcal{H}(q,p) = U(q) + K(p) = -\log \pi(q) + \frac{1}{2} p^{t}M^{-1} p
\]
where $p \sim N(0,M)$.

Hamiltonian dynamics are defined by the differential equations
\begin{align}
    \frac{dq}{dt} &= \frac{\partial \mathcal{H}}{dp} = M^{-1}p \\
    \frac{dp}{dt} &= - \frac{\partial \mathcal{H}}{\partial q} = \nabla_{q} \log \pi (q). \nonumber
\end{align}
Such dynamics are volume preserving, reversible, and we can ensure detailed balance is satisfied via the Metropolis correction. Hence, HMC yields a valid MCMC scheme.

However, these dynamics are described through a continuous-time dynamical system, solutions cannot be analytically derived except in simple cases. Instead, solutions are numerically simulated through the Leapfrog integrator; a numerical integration scheme that approximately preserves the symplecticity of Hamiltonian dynamics \cite{betancourt2013generalizing}.

\subsection{Geodesic Lagrangian Monte Carlo}
Geodesic Lagrangian Monte Carlo \cite{holbrook2018geodesic, lan2015markov} augments HMC by replacing the Hamiltonian $\mathcal{H}(q,p)$ with total energy $E(q,v)$, the static mass matrix $M$ with a dynamic mass matrix $G(q)$, and momentum $p$ with velocity $v$
\begin{align}
E(q,v) &= - \log \pi(q) - \frac{1}{2} \log \vert G(q) \vert + \frac{1}{2}v^{T} G(q)v =  -\pi_{\mathcal{H}}(q)  + \frac{1}{2}v^{T} G(q)v   \label{eq:Lagrangian energy}
\end{align}
where $v \sim N(0, G^{-1}(q))$. Here, the Euler-Lagrange equations of the first kind (those associated with Hamiltonian dynamics) are reframed as those of the second kind (Lagrangian dynamics). The velocity and position from (\ref{eq:Lagrangian energy}) are not separable \cite{neal2011mcmc}, and so the energy is split  into the potential and kinetic components
\begin{align}
E^{(1)}(q,v) &= - \log \pi(q) - \frac{1}{2} \log \vert G(q) \vert \label{eq: energy 1}\\
E^{(2)}(q,v) &= \frac{1}{2}v^{T} G(q)v \label{eq: energy 2}
\end{align}
Dynamics are then simulated iteratively between (\ref{eq: energy 1}) and (\ref{eq: energy 2}).
Various vec, matricization, and duplication operators are used to deal with the symmetries of tangent vectors in Lagrangian Monte Carlo.
The half vectorization operator is defined analogously to (\ref{eq: vec operator}) except for only the upper triangular components:
\begin{equation}
    vech(A) = [A_{11}, A_{21}, \ldots, A_{d1}, A_{22}, \ldots, A_{dd}]
\end{equation}
that is, the $vech(\cdot)$ operator is the vectorization of the lower triangular elements. The vectorization and half vectorizations are connected through the duplication matrix and it's pseudo-inverse:
\[
vec(A) = D_{d} vech(A) \quad vech(A) = D_{d}^{+} vec(A)
\]
where $D_{d}^{+} = (D_{d}^{T} D_{d})^{-1} D_{d}^{T}$.

Using the affine invariant metric tensor derived in \cite{moakher2011riemannian}, the full Lagrangian Monte Carlo algorithm is given in Algorithm 1 below.\\
\begin{algorithm}[H] \label{alg: Lagrangian Monte Carlo}
\SetAlgoLined
\textbf{Input:} Initial position $q^{(1)} \in \mathcal{P}(k)$, number of iterations $N$, step size $\epsilon$, number of leapfrog steps $L$ \\
\For{$i \leftarrow 1$ : $N$}{
    Sample velocity $v^{(1)} \sim N(0, D_{d}^{+}[G(q^{(1)})]^{-1} [D_{d}^{+}]^{t})$\;
    $h \leftarrow -\pi_{\mathcal{H}}(q^{(1)}) + \frac{1}{2} (v^{(1)})^{t} (D_{d}^{t} G(q^{(1)}) D_{d}) v^{(1)}$ \;
    \For{$j \leftarrow 1$ : $L$}{
        $v^{(j + \frac{1}{2})} \leftarrow v^{(j)} + \frac{\epsilon}{2}*D_{d}^{+}[G(q^{(1)})]^{-1} [D_{d}^{+}]^{t} \nabla_{vech(q^{(j)})} \pi_{\mathcal{H}}(q^{j})$\;
        Update $q^{(j)} \rightarrow q^{(j + 1)}$ via the geodesic flow for a step of size $\epsilon$\;
        $v^{(j + 1)} \leftarrow v^{(j + \frac{1}{2})} + \frac{\epsilon}{2}*D_{d}^{+}[G(q^{(1)})]^{-1} [D_{d}^{+}]^{t} \nabla_{vech(q^{(j + \frac{1}{2})})} \pi_{\mathcal{H}}(q^{j + 1})$\;
    }
    $h^{*} \leftarrow -\pi_{\mathcal{H}}(q^{(L + 1)}) + \frac{1}{2} (v^{(L + 1)})^{t} (D_{d}^{t} G(q^{(L + 1)}) D_{d}) v^{(L + 1)}$ \;
    Compute $\Delta E = h^{*} - h$\;
    Accept or reject the new state $(q, p)$ based on the Metropolis criterion with probability $\min(1, e^{\Delta E})$\;
    \eIf{accept}{
         $q \leftarrow q^{(L)}$\;
       }{
         $q \leftarrow q^{(0)}$\;
    }
}
\caption{Lagrangian Monte Carlo via Generalized Leapfrog Integrator}
\end{algorithm}

\subsection{Separable Geodesic Lagrangian Monte Carlo (SGLMC)}
Letting $q = (\Sigma_{1}, \Sigma_{2}) \in \mathcal{P}(k_{1}) \times \mathcal{P}(k_{2})$, and replacing $G(q)$ with $G_{\alpha}^{\otimes}(\Sigma_{1}, \Sigma_{2})$, the algorithm of \ref{alg: Lagrangian Monte Carlo} is extending in a straightforward way. However, we emphasize that, under the metrics $G^{O}(\Sigma_{1}, \Sigma_{2})$, $G^{W}(\Sigma_{1}, \Sigma_{2})$, and $G^{\times}(\Sigma_{1}, \Sigma_{2})$, we can make some straightforward manipulations to provide substantial computational savings. First note, according to the matrix normal model, if $A \in \mathbb{R}^{d \times d}$ such that $A_{i,j} \sim N(0,1)$ for all $(i,j)$. Then $
\Sigma^{\frac{1}{2}} A \Sigma^{\frac{1}{2}} \sim MN(0, \Sigma, \Sigma)$
and consequently $vec(\Sigma^{\frac{1}{2}} A \Sigma^{\frac{1}{2}}) \sim N(0, \Sigma \otimes \Sigma)$.  Moreover, it was shown in \cite{seber2008matrix} that for an appropriately sized matrix $A$,  reshaping a half vectorization back into the full vectorization can be computed as
\[
D_{n} D_{n}^{+} vec(A) = vec(\frac{A + A^{t}}{2}),
\]
in other words, 
\[
mat(D_{n} D_{n}^{+} vec(A)) =  \frac{A + A^{t}}{2} = symm(A).
\]

The key point is to note that Algorithm \ref{alg: Lagrangian Monte Carlo} works by generating half vectorizations of the random velocities and reshaping into symmetric matrices to compute random tangent vectors (and for consistency, this is also done when computing the Riemannian gradients as was done in \cite{moakher2011riemannian}). However, we can just as easily work directly with full matrices rather than half-vectorizations by appropriately symmetrizing. By doing this, we can avoid the expensive matrix products between duplication matrices and explicit construction of the metric tensor. Under either of the orthogonalized metrics, we can independently simulate velocities on $\Sigma_{1}$ and $\Sigma_{2}$ according to
\begin{align*}
    V_{1}^{(0)} &\sim \frac{1}{d_{2}}symm(\Sigma_{1}^{\frac{1}{2}} A_{1} \Sigma_{1}^{\frac{1}{2}})\\
    V_{2}^{(0)} &\sim \frac{1}{d_{1}} P_{\Sigma_{2}}(symm(\Sigma_{2}^{\frac{1}{2}} A_{2} \Sigma_{2}^{\frac{1}{2}}))
\end{align*}
where each $A_{i}$ is independent with iid standard normal entries. Moreover, we can write the kinetic term of the Lagrangian as
\begin{align*}
    &(v^{(k)})^{t} (D_{d}^{t} G(q^{(k)}) D_{d}) v^{(k)} \\
    &= d_{2} tr(V_{1}^{(k)} [\Sigma_{1}^{(k)}]^{-1} V_{1}^{(k)} [\Sigma_{1}^{(k)}]^{-1}) + d_{1} tr(V_{2}^{(k)} [\Sigma_{2}^{(k)}]^{-1} V_{2}^{(k)} [\Sigma_{2}^{(k)}]^{-1}).
\end{align*}

Lastly, we can write the Riemannian gradients of the velocity updates
\begin{align*}
    D_{d}^{+}[G(q^{(k)})]^{-1} [D_{d}^{+}]^{t} \nabla_{vech(q^{(j)})} \pi_{\mathcal{H}(q^{(j)})}
\end{align*}
as two separate Riemannian gradients:
\begin{align*}
    \nabla_{\Sigma_{1}^{(k)}}^{R} \pi_{\mathcal{H}(q^{(k)})} &= \frac{1}{d_{2}} \Sigma_{1}^{(k)} \nabla_{\Sigma_{1}^{(k)}} \pi_{\mathcal{H}(q^{(j)})} \Sigma_{1}^{(k)} \\
    \nabla_{\Sigma_{2}^{(k)}}^{R} \pi_{\mathcal{H}(q^{(k)})} &= \frac{1}{d_{1}} \Sigma_{2}^{(k)} \nabla_{\Sigma_{2}^{(k)}} \pi_{\mathcal{H}(q^{(j)})} \Sigma_{2}^{(k)}.
\end{align*}

The explicit construction of $G(q)$ at each step during an iteration would be expensive to compute. While a necessity for the regularized metric in the current implementation, for the orthogonalized and weighted metrics, these equations show that we can significantly reduce the complexity of the computations by breaking the operations up into smaller sequential matrix products. These also hold for the product manifold metric by simply omitting any of the dimensional multiplications (such as those found in the kinetic energy computation or Riemannian gradients). These equations are formalized in an SGLMC context in Algorithm \ref{alg: orthogonal SGLMC}.

\begin{algorithm}[H] \caption{Orthogonal Separable Geodesic Lagrangian Monte Carlo (SGLMC-O)}
\label{alg: orthogonal SGLMC}
\SetAlgoLined
\textbf{Input:} Initial position $q^{(1)} = (\Sigma_{1}^{(1)}, \Sigma_{2}^{(1)}) \in \mathcal{P}(k_{1}) \times \mathcal{P}(k_{2})$, number of iterations $N$, step size $\epsilon$, number of leapfrog steps $L$, regularization parameter $\alpha$ \\
\For{$i \leftarrow 1$ : $N$}{
    \For{$k \leftarrow$ $1:2$}{
    Sample matrices $A_{k}$ such that $A_{k}[m,n] \sim N(0,1)$\;
    Set $M_{k} \leftarrow d_{-k}^{\frac{1}{2}} \Sigma_{k}^{\frac{1}{2}} A_{k} \Sigma_{k}^{\frac{1}{2}}$\;
    \If{$k = 2$}{
    $M_{2} \leftarrow P_{\Sigma_{2}}(M_{2})$\;
    }
    Compute velocity $V_{k}^{(1)} \leftarrow symm(M_{k})$
    }
    $h \leftarrow -\pi_{\mathcal{H}}(q^{(1)}) + d_{2}tr(\Sigma_{1}^{-1} V_{1} \Sigma_{1}^{-1} V_{1}) + d_{1} tr(\Sigma_{2}^{-1} V_{2} \Sigma_{2}^{-1} V_{2})$ \;
    \For{$j \leftarrow 1$ : $L$}{
        $V_{1}^{(j + \frac{1}{2})} \leftarrow V_{1}^{(j)} + \frac{\epsilon}{2}*[\frac{1}{d_{2}}\Sigma_{1} \nabla_{\Sigma_{1}} \pi_{\mathcal{H}}(q^{j}) \Sigma_{1}]$\;
        $V_{2}^{(j + \frac{1}{2})} \leftarrow V_{2}^{(j)} + \frac{\epsilon}{2}*P_{\Sigma_{1}}( \frac{1}{d_{1}}\Sigma_{2} \nabla_{\Sigma_{2}} \pi_{\mathcal{H}}(q^{j}) \Sigma_{2})$\;
        Update $q^{(j)} \rightarrow q^{(j + 1)}$ via the geodesic flow for a step of size $\epsilon$\;
        $V_{1}^{(j + \frac{1}{2})} \leftarrow V_{1}^{(j)} + \frac{\epsilon}{2}*[\frac{1}{d_{2}}\Sigma_{1} \nabla_{\Sigma_{1}} \pi_{\mathcal{H}}(q^{j}) \Sigma_{1}]$\;
        $V_{2}^{(j + \frac{1}{2})} \leftarrow V_{2}^{(j)} + \frac{\epsilon}{2}*P_{\Sigma_{1}}( \frac{1}{d_{1}}\Sigma_{2} \nabla_{\Sigma_{2}} \pi_{\mathcal{H}}(q^{j}) \Sigma_{2})$\;
    }
    $h^{*} \leftarrow -\pi_{\mathcal{H}}(q^{(L + 1)}) + d_{2}tr(\Sigma_{1}^{-1} V_{1} \Sigma_{1}^{-1} V_{1}) + d_{1} tr(\Sigma_{2}^{-1} V_{2} \Sigma_{2}^{-1} V_{2})$ \;
    Compute $\Delta E = h^{*} - h$\;
    Accept or reject the new state $(q, p)$ based on the Metropolis criterion with probability $\min(1, e^{\Delta E})$\;
    \eIf{accept}{
         $q \leftarrow (\Sigma_{1}^{(L + 1)}, \Sigma_{2}^{(L + 1)})$\;
       }{
         $q \leftarrow q^{(1)}$\;
    }
}
\end{algorithm}

\subsection{Implementation and Adaptation of SGLMC} \label{sec: Implementation and Adaptation}
The performance of Hamiltonian Monte Carlo is sensitive to each of its tuning parameters $\epsilon$ (the step size)
and  $L$ (the number of steps).  The parameter $\epsilon$ was tuned via the Nesterov dual averaging algorithm from the no U-turn sampler \cite{carpenter2017stan}. The hyperparameters associated with dual averaging tuning were the same as those used in \cite{carpenter2017stan}. Except for $a_{0}$ was set to $.8$ to avoid numerical overflow and underflow. Tuning of $L$ is more challenging on a Riemannian manifold, and implementation of a dynamic choice, while investigated in \cite{betancourt2013generalizing}, we refrain from using and instead use a static choice of $L= 10$ for all comparisons on different data dimensions and comparisons of regularization parameters. However, the use of the log map can be used to give a simple dynamic termination criterion, as discussed in Section \ref{sec: supplement}.

\section{Empirical Comparisons} \label{sec: Empirical Comparisons}

In this section, we demonstrate the empirical effectiveness of SGLMC. We generate $\Sigma_{1}$ and $\Sigma_{2}$ as $Y_{1}, \ldots, Y_{n} \sim \mathcal{N}(0, \Sigma_{1} \otimes \Sigma_{2})$ with $\Sigma_{i} \sim \mathcal{I}\mathcal{W}(\nu = d_{i}$$ + 10, \frac{\sqrt{\gamma}}{d_{i}} I_{d_{i}})$ for $\gamma = 5$, $n = 300$ and $d_{1} = 15, d_{2} = 6$. 

Comparisons of global matrix summaries, such as traces, determinants and posterior densities, as well as summaries of the properties of the components of the Kronecker product such as their eigenvectors, and condition numbers between Gibbs, SGLMC, and {\it stan}.  Each sampling method was initialized in the iterative MLE from the algorithm of \cite{dutilleul1999mle}.

$5000$ samples were drawn from the Gibbs sampler, and $5000$ samples were drawn from {\it stan}. The Gibbs sampler was given a burn-in of $30000$,and {\it stan}'s "warm-up" parameter was set to $1000$ with $2000$ samples drawn afterwards. The only tuning parameter adjusted for {\it stan} was {\it int time}, which was set to $.05$. SGLMC was given $500$ adaptation and $500$ burn-in iterations, with $2000$ samples generated afterward. We set the regularization parameter at $\alpha = .95$, with the acceptance probability for the dual averaging of $\epsilon$ set to $a_{0} = .8$. The results of these samples can be found in Figure \ref{fig: d1= 15, d2 = 6, alpha = .95}.  For brevity, we do not include our results for the robustness to dimensionality on sampling efficiency and the effects of dynamically tuning $L$. These particular results may be found in Section \ref{sec: supplement}. 

\begin{figure}[ht]
    \centering
    \begin{minipage}{0.45\textwidth}
        \centering
        \includegraphics[width=\textwidth]{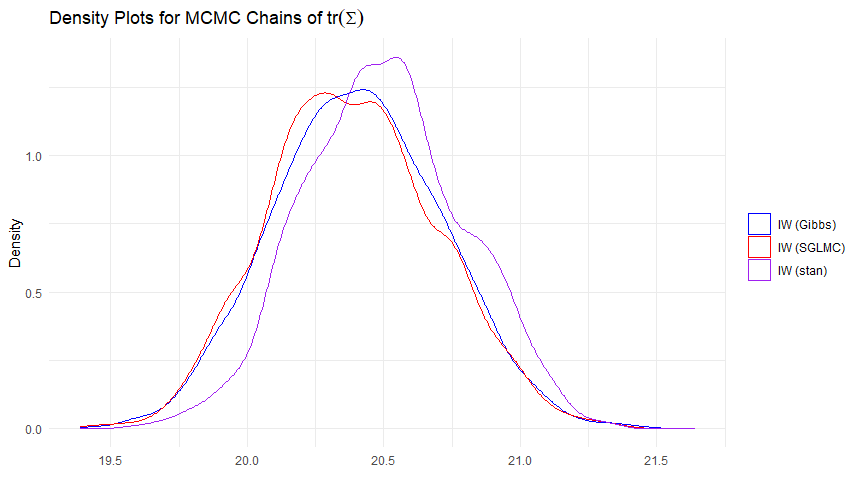} 
    \end{minipage}\hfill
    \begin{minipage}{0.45\textwidth}
        \centering
        \includegraphics[width=\textwidth]{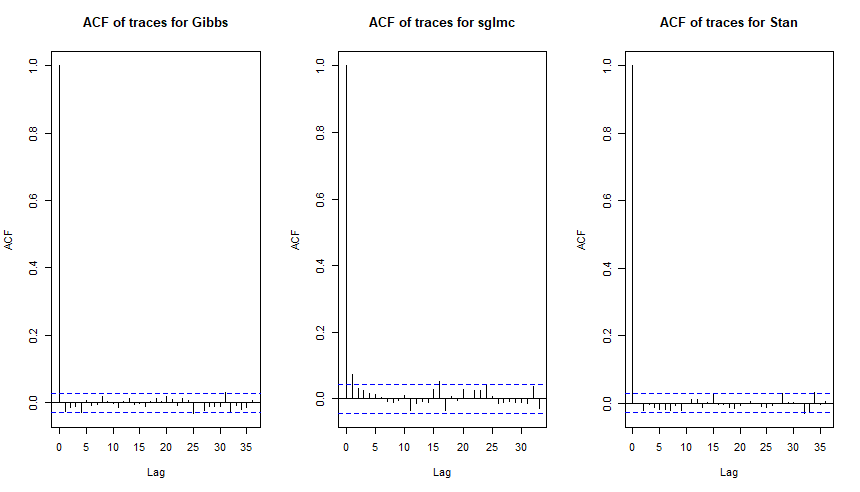} 
    \end{minipage}
           \centering
        \begin{minipage}{0.45\textwidth}
        \centering
        \includegraphics[width=\textwidth]{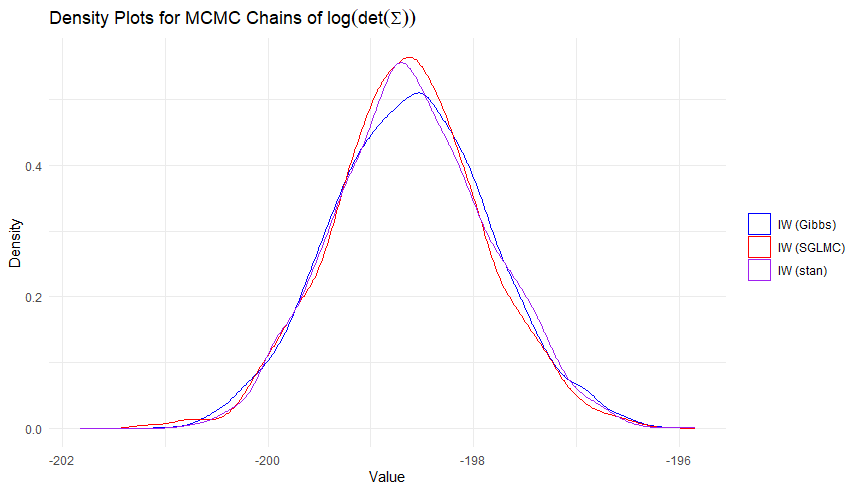} 
    \end{minipage}\hfill
    \begin{minipage}{0.45\textwidth}
        \centering
        \includegraphics[width=\textwidth]{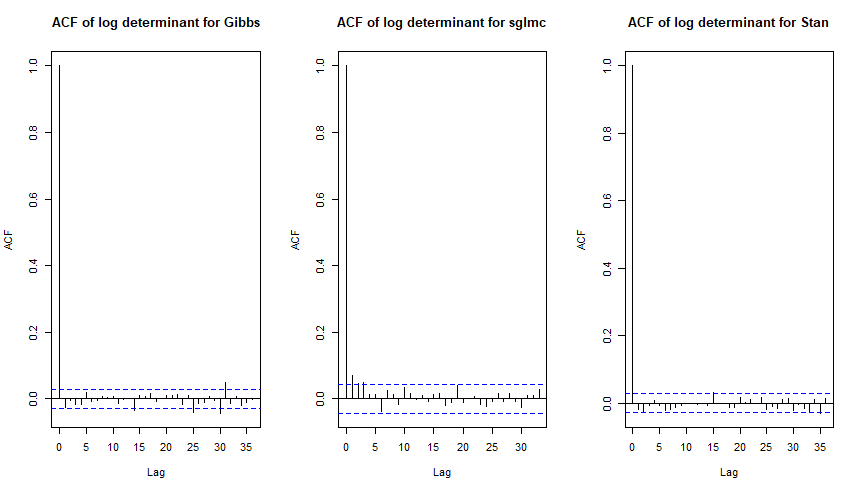} 
    \end{minipage}
    \centering
    \begin{minipage}{0.45\textwidth}
        \centering
        \includegraphics[width=\textwidth]{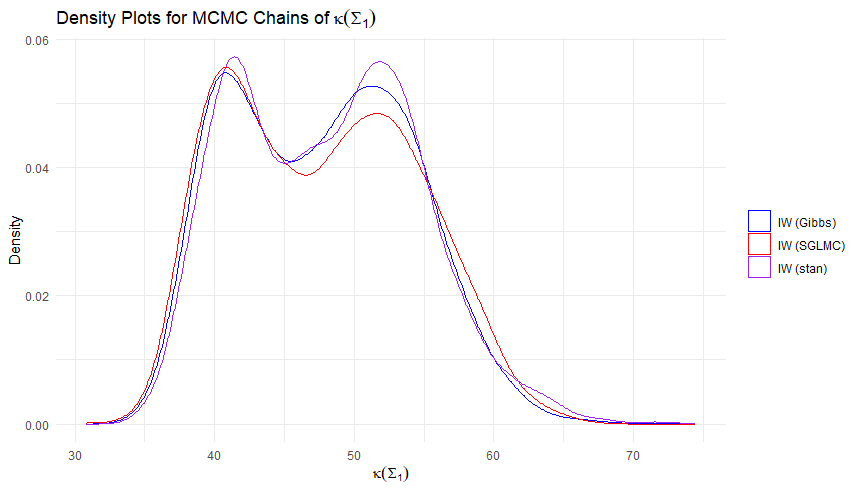} 
    \end{minipage}\hfill
    \begin{minipage}{0.45\textwidth}
        \centering
        \includegraphics[width=\textwidth]{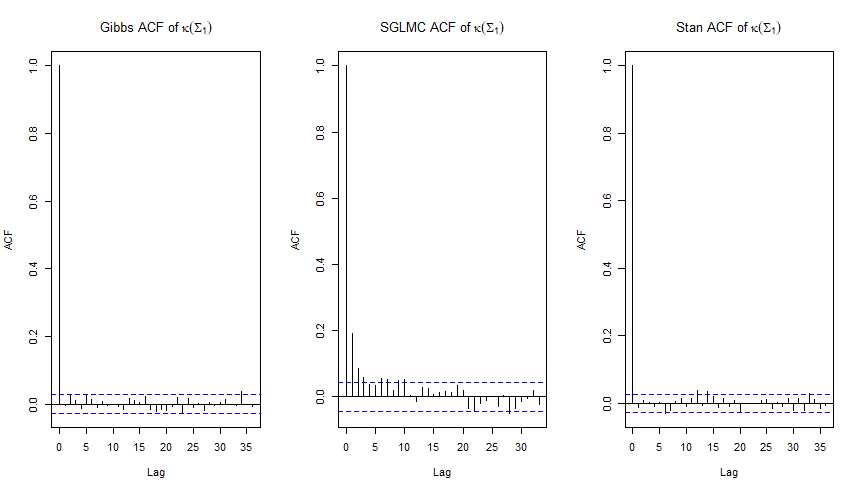} 
    \end{minipage}
           \centering
        \begin{minipage}{0.45\textwidth}
        \centering
        \includegraphics[width=\textwidth]{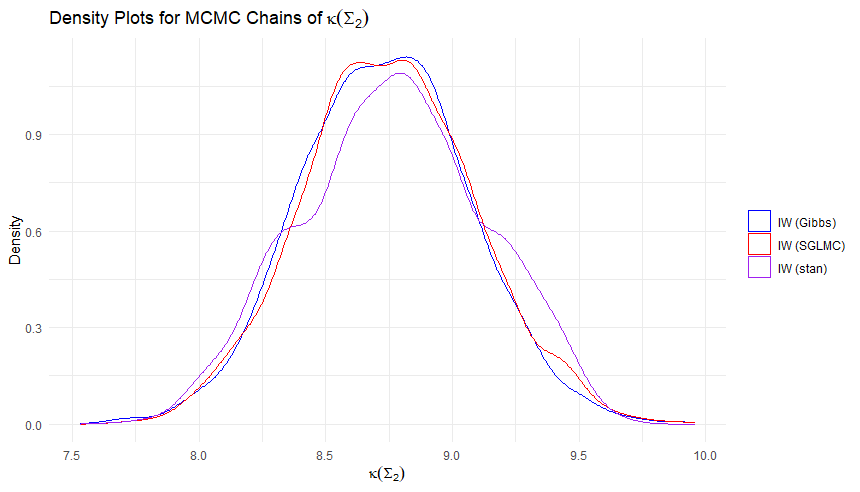} 
    \end{minipage}\hfill
    \begin{minipage}{0.45\textwidth}
        \centering
        \includegraphics[width=\textwidth]{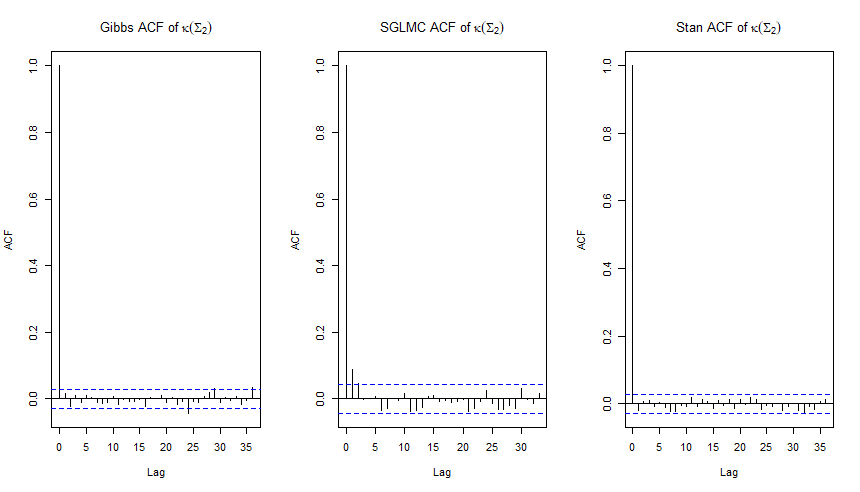} 
    \end{minipage}
    \caption{Density and corresponding ACFs plots of the traces, determinants for samples of $\Sigma$, and the condition numbers for samples of $\Sigma_{1}, \Sigma_{2}$ from SGLMC, Gibbs, and {\it stan} ($d_{1} = 15$, $d_{2} = 6$, $\alpha = .95$) with Inverse-Wishart priors on $\Sigma_{1}$ and $\Sigma_{2}$. From each of these figures, we can see that regularized SGLMC aligns with the Gibbs sampler in distribution of the corresponding global matrix statistics. SGLMC seems to have a heavier autocorrelation in the first and second lags, although {\it stan} appears to provide a worse fit for $tr(\Sigma)$ and $\kappa(\Sigma_{2})$.}    
    \label{fig: d1= 15, d2 = 6, alpha = .95}
\end{figure}

 In Figures \ref{fig: d1= 3, d2 = 8, alpha comparison Sig 1 and 2} and \ref{fig: d1= 3, d2 = 8, alpha comparison Sig}, we compare the effects of different choices of the regularization parameter for $\alpha \in \{0,.1,.25,.5,.75,.9,.95\}$. In general, we found that any choice of $\alpha$ had little effect on the density estimates of $\Sigma_{1}, \Sigma_{2}$ or $\Sigma$, however, the autocorrelation patterns grew significantly as $\alpha \rightarrow 0$. We note, however, that using $\alpha > .95$ may give diminishing returns in the autocorrelation, potentially due to computational instability. Thus, we recommend using $\alpha = .95$ as a robust choice. To ensure comparability between samples, we ran each choice of $\alpha$ in parallel, with each parallel chain sharing the same seed. We also did not use the dual average algorithm and used a static choice of $\epsilon = .7$ to improve comparability.

\begin{figure}[ht]
    \centering
    \begin{minipage}{0.45\textwidth}
        \centering
        \includegraphics[width=\textwidth]{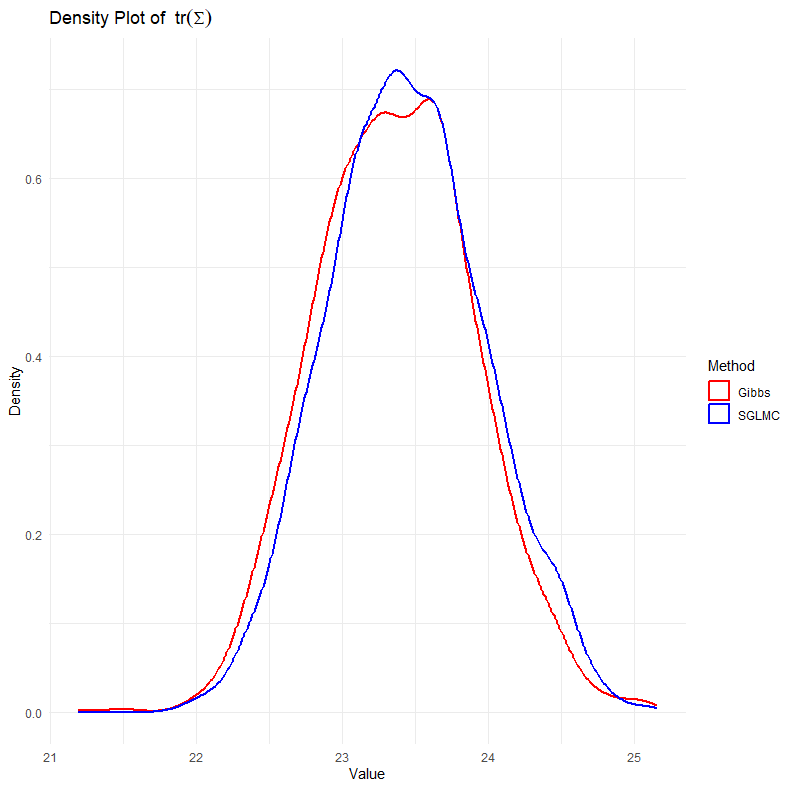} 
    \end{minipage}\hfill
    \begin{minipage}{0.45\textwidth}
        \centering
        \includegraphics[width=\textwidth]{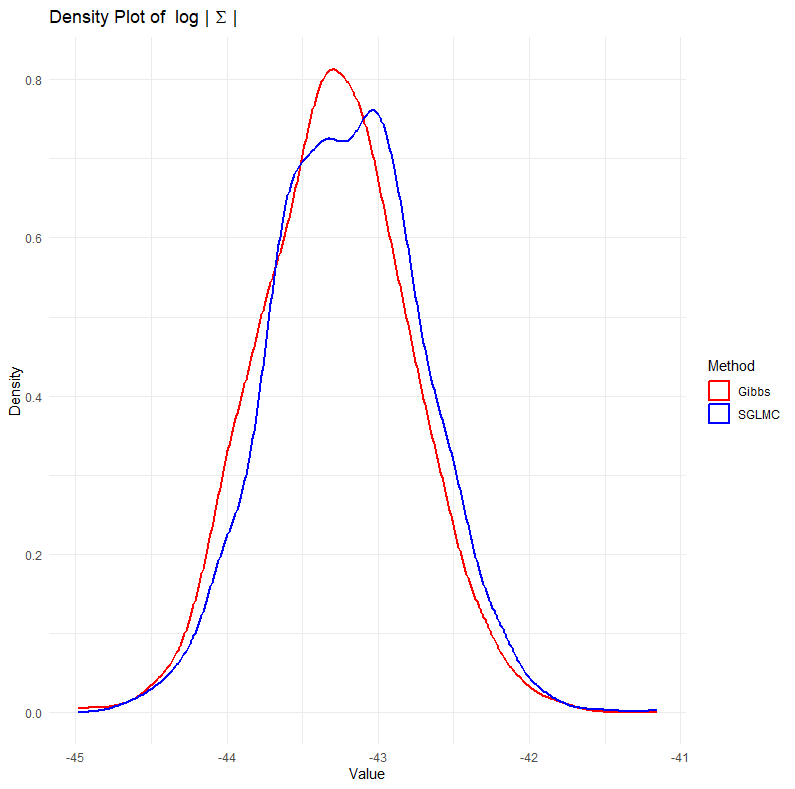} 
    \end{minipage}
           \centering
        \begin{minipage}{0.45\textwidth}
        \centering
        \includegraphics[width=\textwidth]{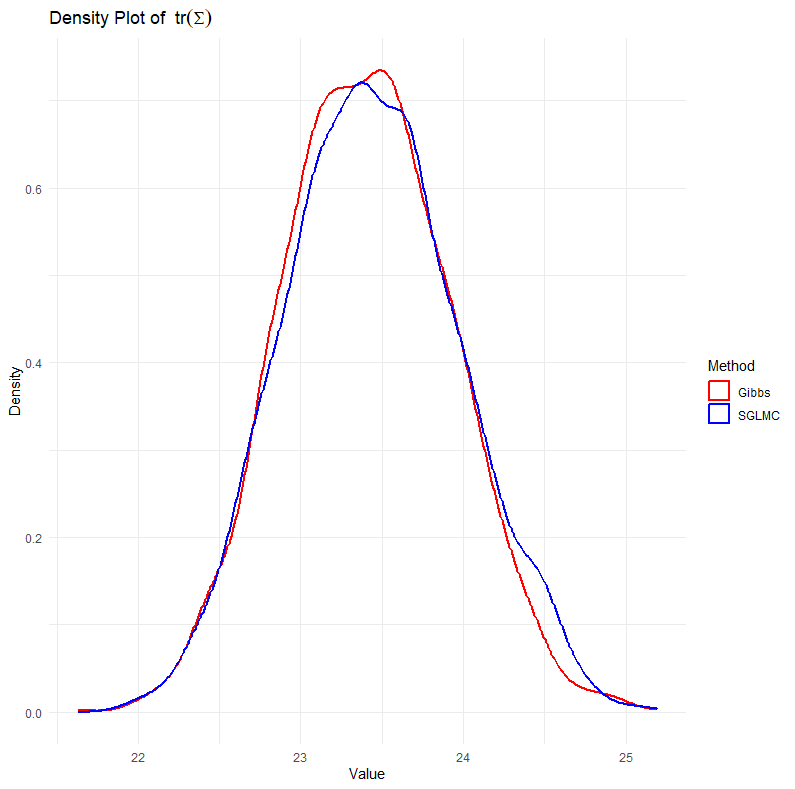} 
    \end{minipage}\hfill
    \begin{minipage}{0.45\textwidth}
        \centering
        \includegraphics[width=\textwidth]{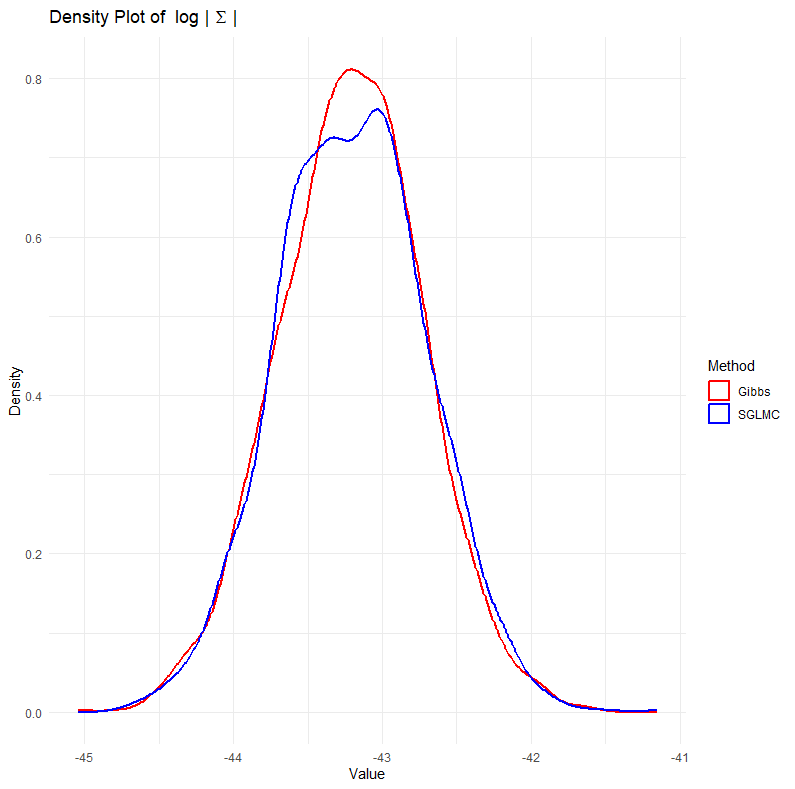} 
    \end{minipage}
      \caption{Comparisons between un-normalized and normalized Gibbs samples for $tr(\Sigma_{1} \otimes \Sigma_{2})$ and $\log \vert \Sigma_{1} \otimes \Sigma_{2}\vert$. First row is unnormalized, second row is normalized. The first row illustrates a bias introduced by the normalization of $\Sigma_{2}$ in the SGLMC sampler when the Gibbs samples are unnormalized, but the bias vanishes when we normalize the $\Sigma_{2}$ samples at each iteration.}
    \label{fig: bias comparisons}
\end{figure}

 We illustrate the bias introduced by the constrained sampling under the orthogonalized metric in Figure \ref{fig: bias comparisons}, but also demonstrate the densities match up exactly to the Gibbs sampler when normalizing the Gibbs samples as $\Sigma_{2} \leftarrow \frac{\Sigma_{2}}{\vert \Sigma_{2} \vert^{1/d_{2}}}$.

\begin{figure}[ht]
    \centering
    \begin{minipage}{0.425\textwidth}
        \centering
        \includegraphics[width=\textwidth]{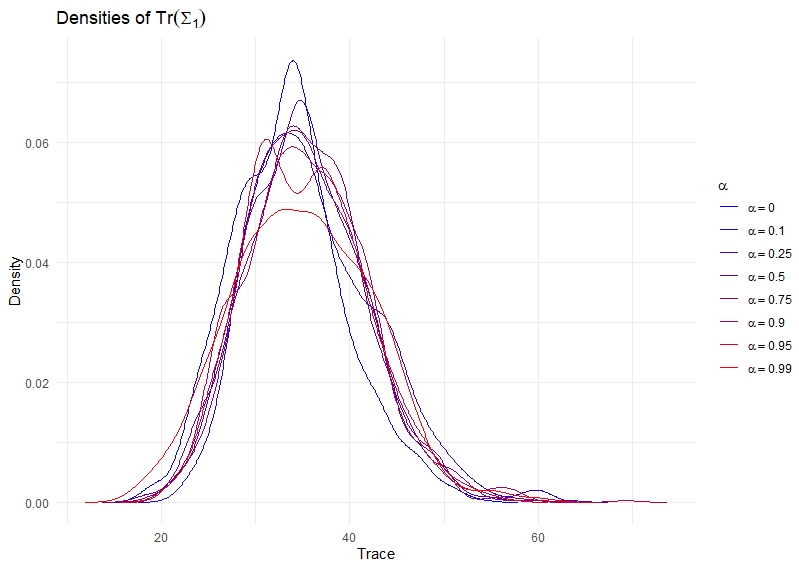} 
    \end{minipage}\hfill
    \begin{minipage}{0.425\textwidth}
        \centering
        \includegraphics[width=\textwidth]{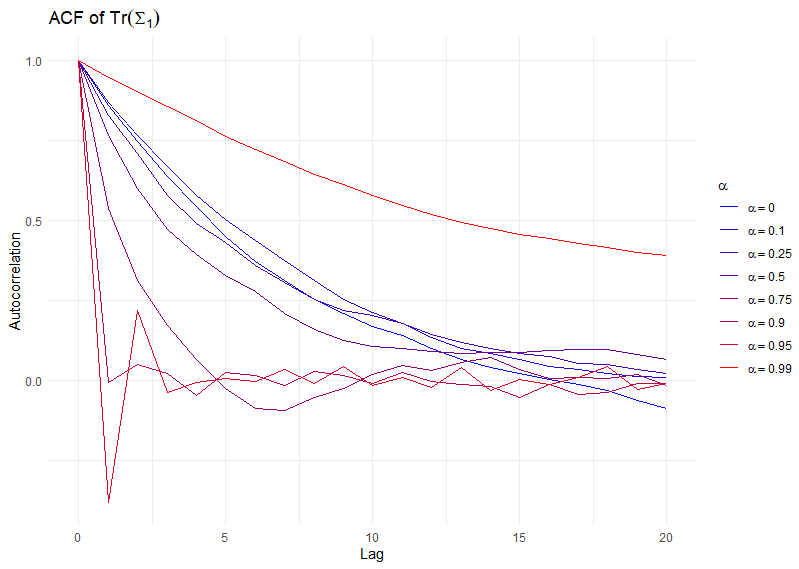} 
    \end{minipage}
           \centering
        \begin{minipage}{0.425\textwidth}
        \centering
        \includegraphics[width=\textwidth]{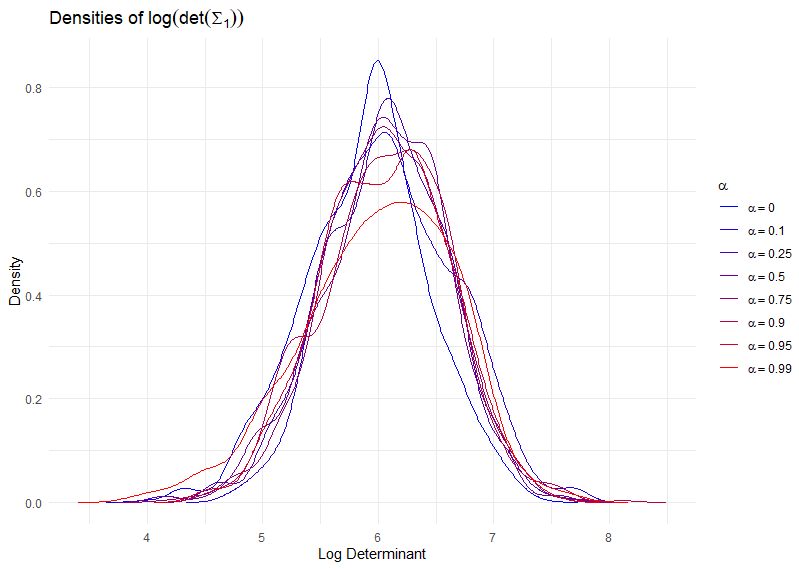} 
    \end{minipage}\hfill
    \begin{minipage}{0.425\textwidth}
        \centering
        \includegraphics[width=\textwidth]{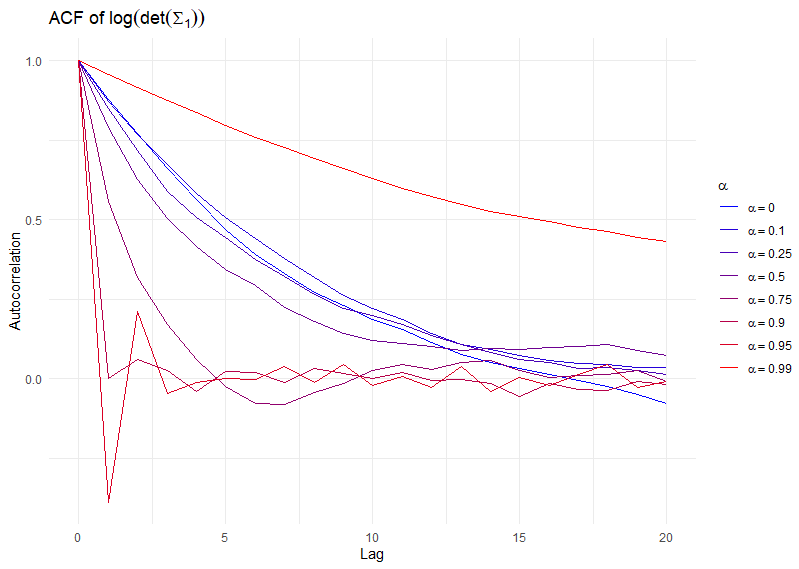} 
    \end{minipage}
    \centering
    \begin{minipage}{0.425\textwidth}
        \centering
        \includegraphics[width=\textwidth]{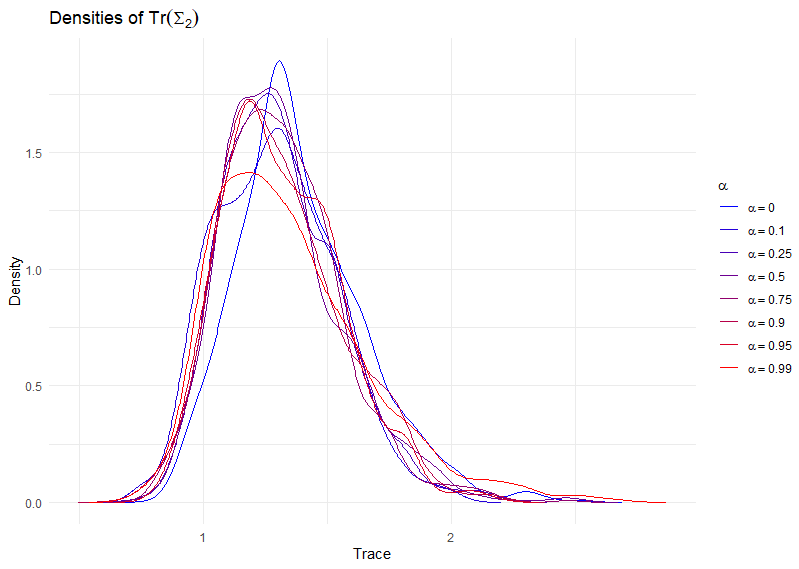} 
    \end{minipage}\hfill
    \begin{minipage}{0.425\textwidth}
        \centering
        \includegraphics[width=\textwidth]{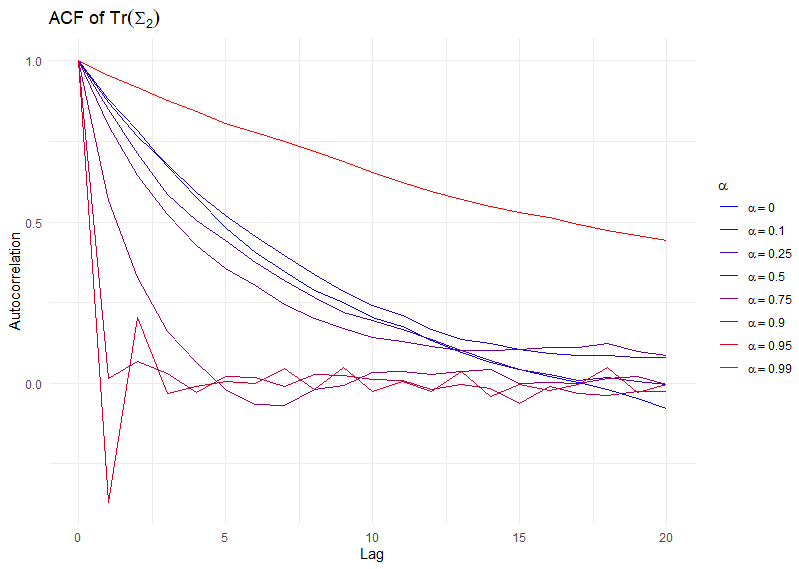} 
    \end{minipage}
           \centering
        \begin{minipage}{0.425\textwidth}
        \centering
        \includegraphics[width=\textwidth]{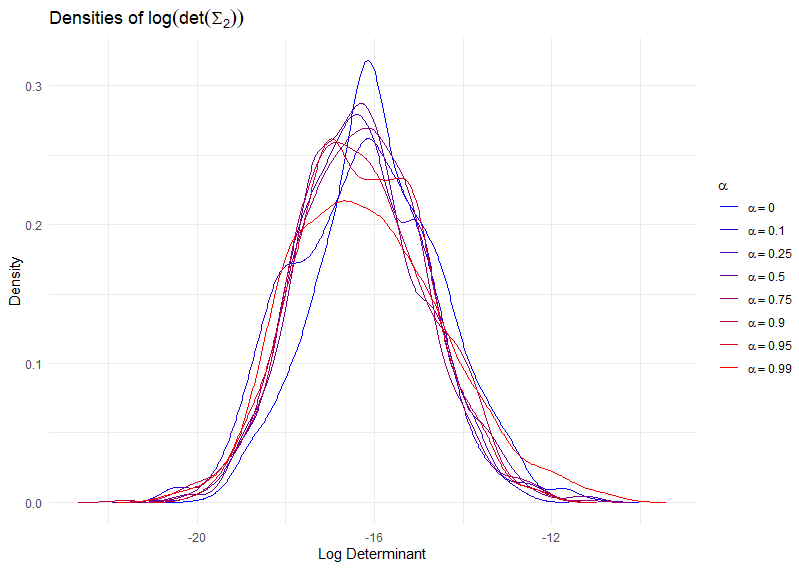} 
    \end{minipage}\hfill
    \begin{minipage}{0.425\textwidth}
        \centering
        \includegraphics[width=\textwidth]{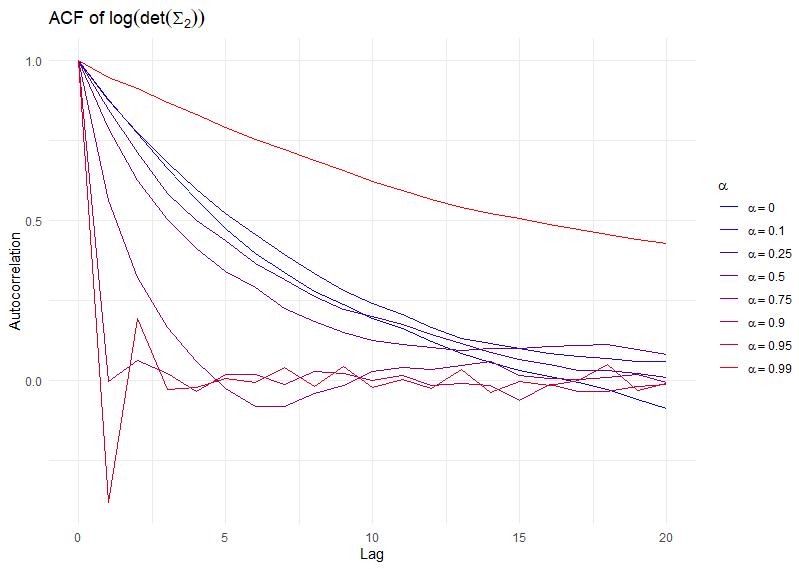} 
    \end{minipage}
               \centering
    \caption{Comparison of density and ACFs of the trace and log determinant of $\Sigma_{1}$, $\Sigma_{2}$ across different $\alpha$ ($d_{1} = 3$, $d_{2} = 8$). Note that the autocorrelations appear to diminish as $\alpha$ grows large, but the benefit diminishes when $\alpha \approx 1$. This may be attributed to the instability of the metric in this region. There appears to be anticorrelation in the first couple of lags for $\alpha = .95$, while minimal auto or anticorrelation for $\alpha = .9$ for any lags.}  
    \label{fig: d1= 3, d2 = 8, alpha comparison Sig 1 and 2}
\end{figure}

\begin{figure}[ht]
               \centering
        \begin{minipage}{0.45\textwidth}
        \centering
        \includegraphics[width=\textwidth]{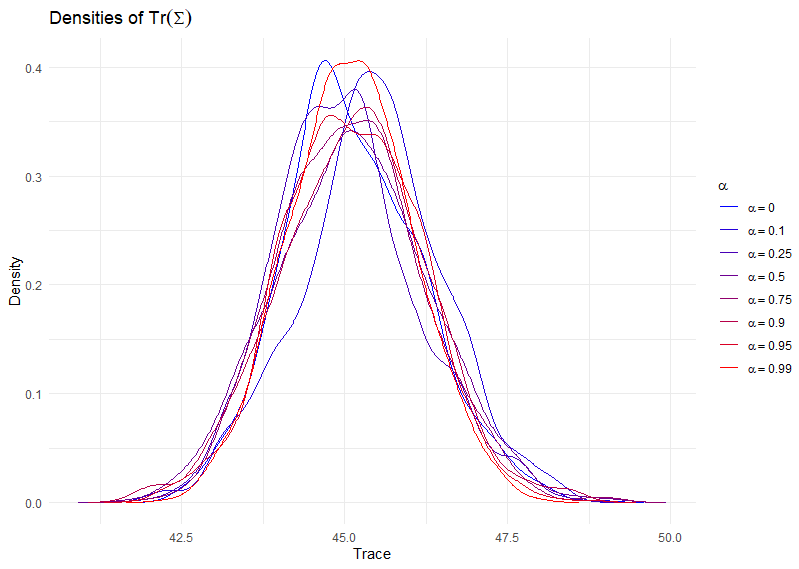} 
    \end{minipage}\hfill
    \begin{minipage}{0.45\textwidth}
        \centering
        \includegraphics[width=\textwidth]{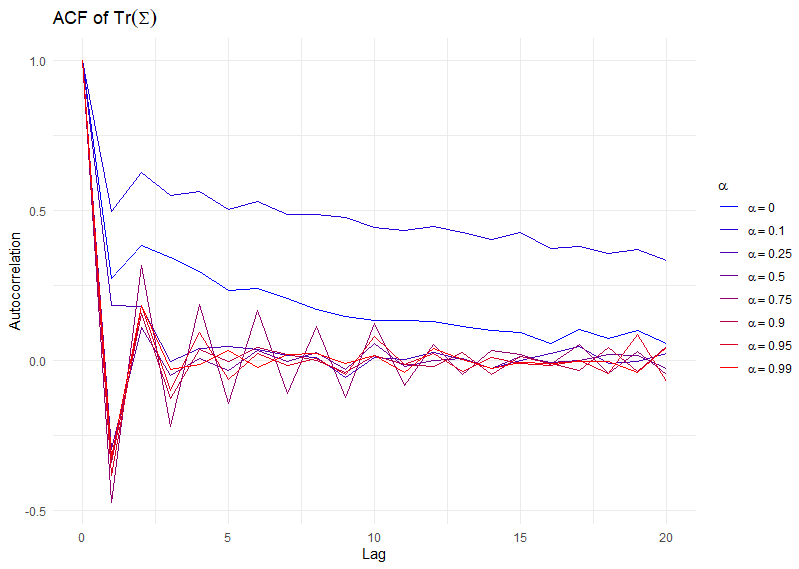} 
    \end{minipage}
                   \centering
        \begin{minipage}{0.45\textwidth}
        \centering
        \includegraphics[width=\textwidth]{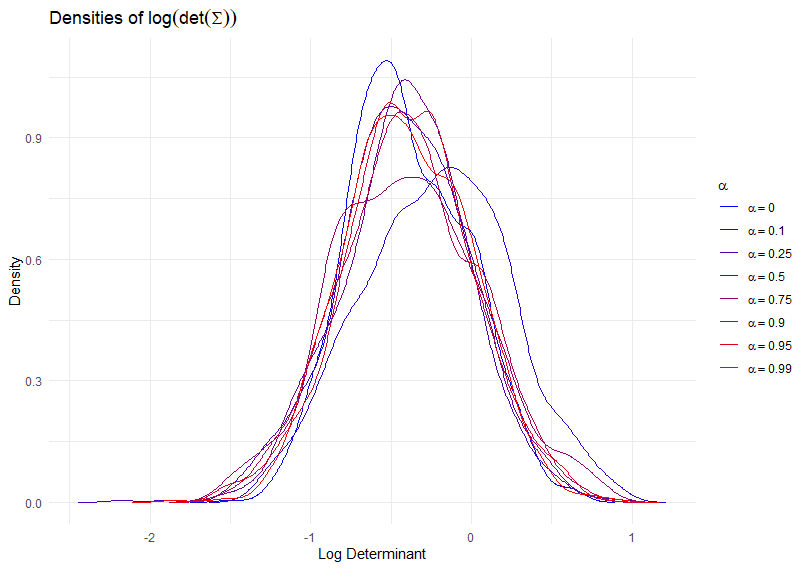} 
    \end{minipage}\hfill
    \begin{minipage}{0.45\textwidth}
        \centering
        \includegraphics[width=\textwidth]{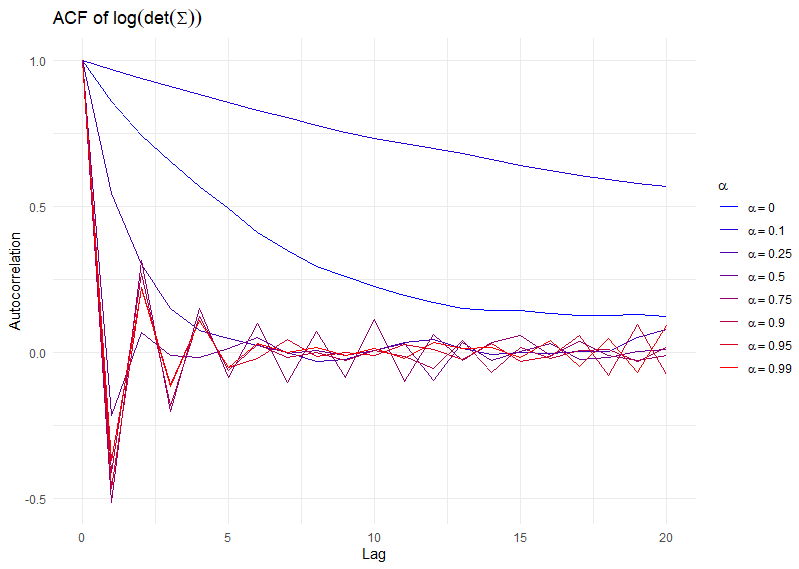} 
    \end{minipage}
    \caption{Comparison of density and ACFs of the trace and log determinant of $\Sigma$ across different $\alpha$ ($d_{1} = 3$, $d_{2} = 8$). We see for any choice of $\alpha \geq .5$, there appears to be substantial anticorrelation across many lags. Regarding the density estimates, most share a shape and centering, although this structure appears to deviate for $\alpha \leq .25$.}    
\label{fig: d1= 3, d2 = 8, alpha comparison Sig}
\end{figure}

We give comparisons with each prior considered in Section \ref{sec: priors} with the regularization parameter $\alpha = .95$ in the supplementary file for the generated data. However, we include the comparison between the IW and SIW priors for the real data example in the next section. We would like to highlight that while the propriety of the IW and SIW priors is guaranteed, propriety of the reference prior for this problem was not investigated in this paper and remains questionable. In our experiments, we found its stability to be very sensitive to the choice of $a_{0}$.

\begin{figure}[ht] \label{fig: eigenvalue comparisons}
  \centering
  \begin{minipage}[t]{0.4\textwidth}
    \centering
    \includegraphics[width=\linewidth]{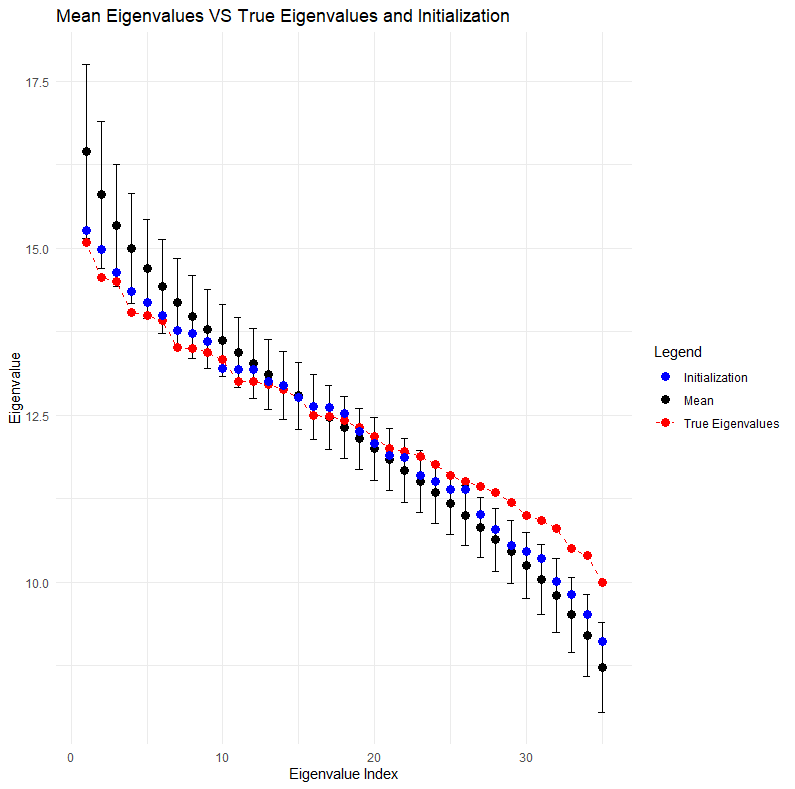} 
    \caption{IW priors}
  \end{minipage}%
  \hfill
  \begin{minipage}[t]{0.4\textwidth}
    \centering
    \includegraphics[width=\linewidth]{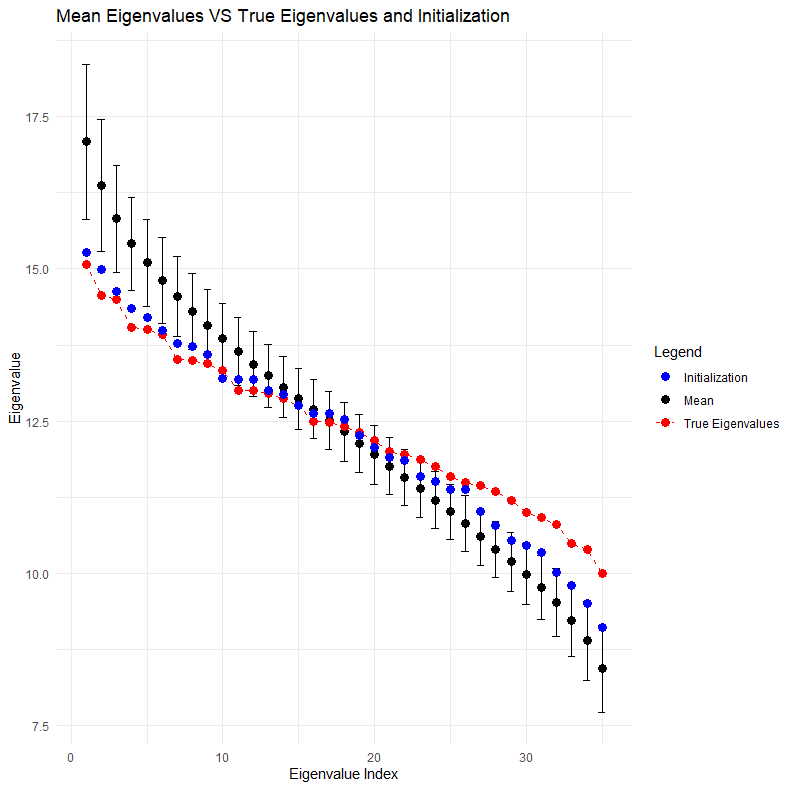} 
    \caption{IW and Ref priors}
  \end{minipage}

  \vspace{0.5cm} 
  \begin{minipage}[t]{0.6\textwidth} 
    \centering
    \includegraphics[width=\linewidth]{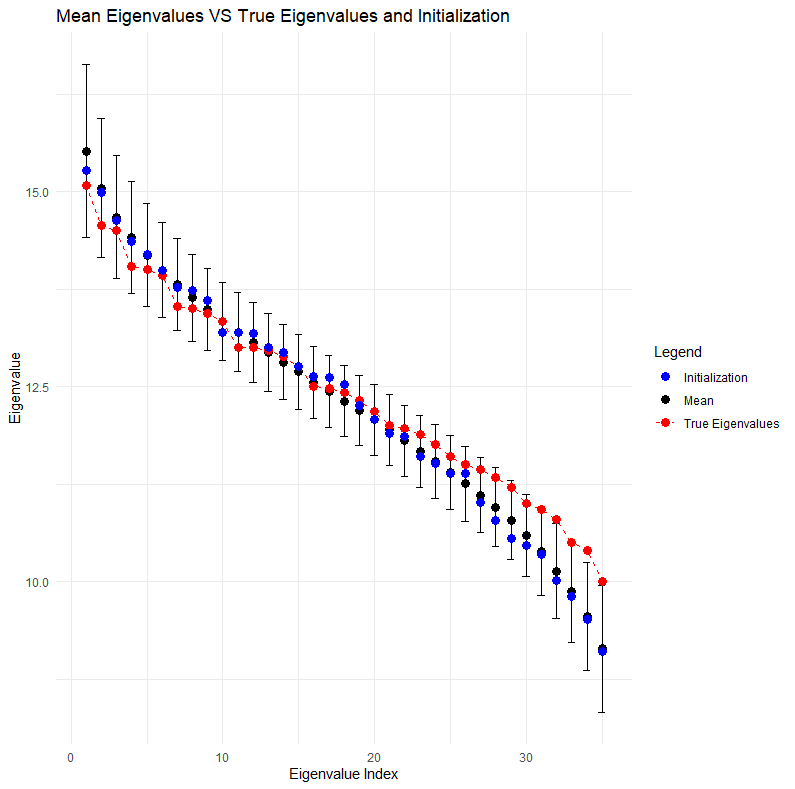} 
    \caption{IW and SIW priors}
  \end{minipage}
  \caption{Eigenvalue comparisons between posterior samples of $\hat{\Sigma}_{1} \otimes \hat{\Sigma}_{2}$ with an IW prior on $\Sigma_{1}$ and IW, Reference, and SIW priors on $\Sigma_{2}$ in respective figures 1,2, and 3. Note the SIW prior under the linear eigenvalue decay places much more support to the appropriate region than either of the other two prior choices.}
\end{figure}

In Figure \ref{fig: eigenvalue comparisons}, we give comparisons of posterior samples under a data generation with a small linear decay in the eigenvalues, namely, we generated $\Sigma_{1}$ and $\Sigma_{2}$ with eigenspacings of $.2$ and $.1$ such that $\lambda_{\max}(\Sigma_{1}) = 5$, $\lambda_{\max}(\Sigma_{2}) = 2$, the Inverse Wishart is known to space eigenvalues apart. Hence, such data generation is known to be problematic for the posterior sampling of an Inverse Wishart. In this data generation, we see that with Inverse Wishart priors on both $\Sigma_{1}$ and $\Sigma_{2}$, the posterior fails to provide adequate support on both the smallest and largest true eigenvalues. We see the same relationship when we impose a reference prior on $\Sigma_{2}$, however, when applying the moment-matched Shrinkage Inverse Wishart (see equations (\ref{eq: moment match 1}, \ref{eq: moment match 2})), we see most of the eigenvalues are adequately captured by the posterior density.
\newline
Lastly, we generated $5$ separate data generation processes with corresponding dimensions $d_{1}, d_{2} \in \{(2,3), (5,4), (8,10), (15,2), (20,5)\}$, we ran each of the metrics (\ref{eq: half vec metric tensor}), (\ref{eq: orthogonalized metric tensor}),  (\ref{eq: weighted metric}), and (\ref{eq: Product Manifold Metric}) in parallel with $300$ adaptation steps, $300$ burn-in, and total samples generated after burn-in to 1000. In each example, we used an acceptance rate of $a_{0} = .85$. Table \ref{table: ESS} illustrates the average effective sample size per iteration across the five data-generating processes with each of the 4 metrics.  Note that the regularized metric appears to provide a robust compromise between the regularized metrics in terms of ESS/it for global matrix statistics and identifiable individual matrix statistics such as condition number. Note that in any of the statistics evaluated, the product manifold appears to under perform, indicating a deficiency in the metric choice for this sampling problem.
\begin{table}[ht]
    \centering
    \begin{tabular}{|c|c|c|c|c|c|c|c|c|}
        \hline
        Metric & $tr(\Sigma_{1})$ & $tr(\Sigma_{2})$ & $tr(\Sigma_{1} \otimes \Sigma_{2})$ & $\log \vert \Sigma_{1} \vert$ & $\log \vert \Sigma_{2} \vert$ & $\log \vert \Sigma_{1} \otimes \Sigma_{2} \vert$ & $\kappa(\Sigma_{1})$ & $\kappa(\Sigma_{2})$ \\
        \hline
        $G^{\times}$   & .027  & .028 & .346 & .0235 & .0264 & 2.13 & .98 & .58  \\
        \hline
        $G^{O}$  & 1.41  & 4.602 & 1.56 & 6.97 & - & 6.97 & .874 & 4.65   \\
        \hline
        $G^{W}$  & 2.52  & 4.85 & 2.84 & 4.30 & - & 4.30 & .596 & .74   \\
        \hline
        $G^{\otimes}_{.95}$ & .64  & .61 & 2.13 & .62 & .62 & 2.57 & .79 & 2.05 \\
        \hline
    \end{tabular}
    \caption{$ESS/it$ for each of the 4 metrics.}
    \label{table: ESS}
\end{table}

\section{Real Data Example} \label{sec: real data}

\begin{figure}[ht]
    \centering
    \begin{minipage}{0.45\textwidth}
        \centering
        \includegraphics[width=\textwidth]{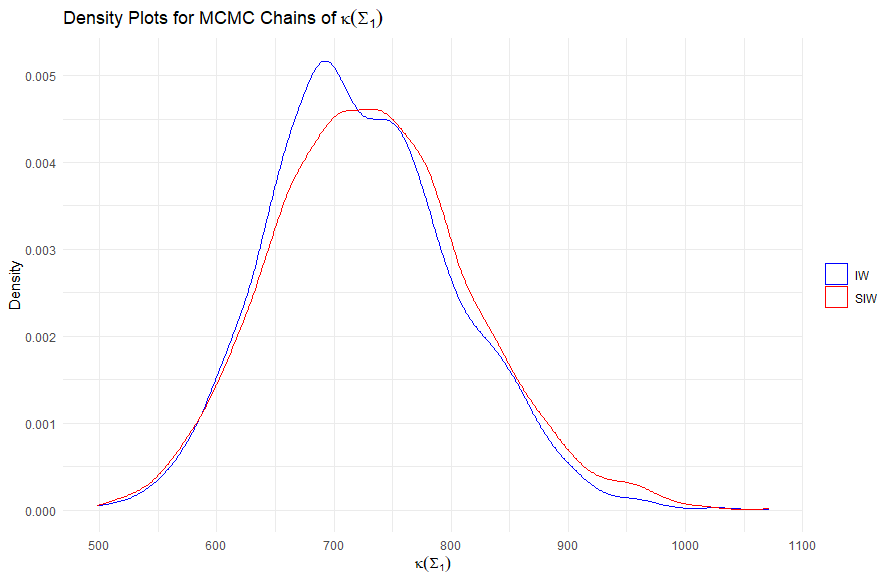} 
    \end{minipage}\hfill
    \begin{minipage}{0.45\textwidth}
        \centering
        \includegraphics[width=\textwidth]{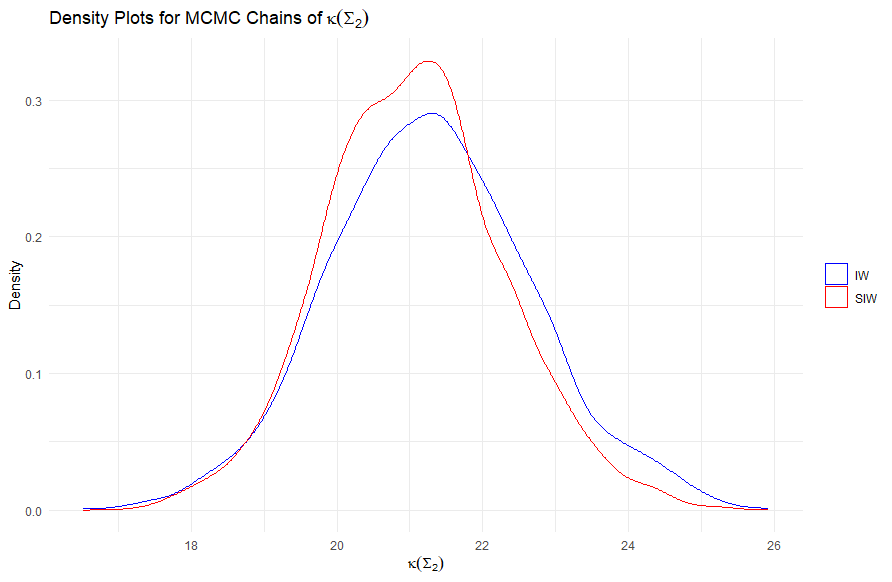} 
    \end{minipage}
           \centering
        \begin{minipage}{0.45\textwidth}
        \centering
        \includegraphics[width=\textwidth]{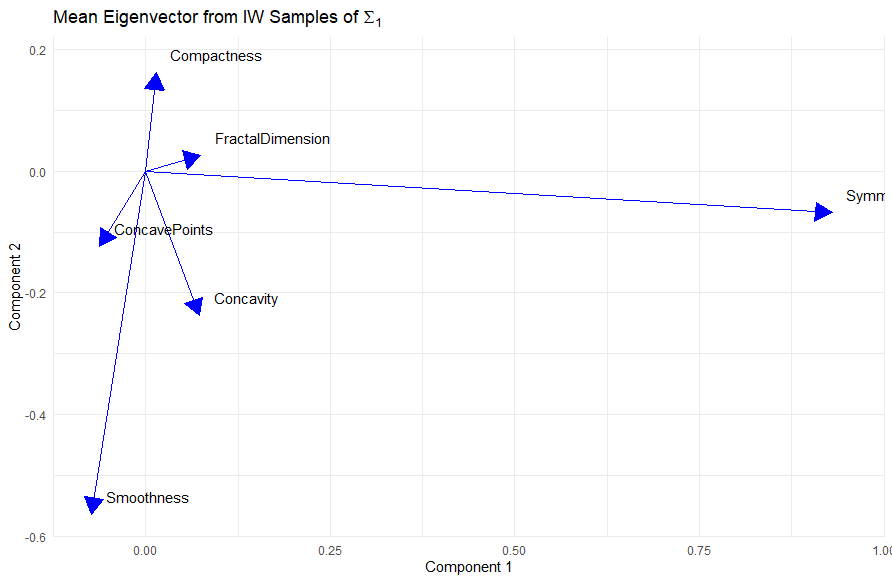} 
    \end{minipage}\hfill
    \begin{minipage}{0.45\textwidth}
        \centering
        \includegraphics[width=\textwidth]{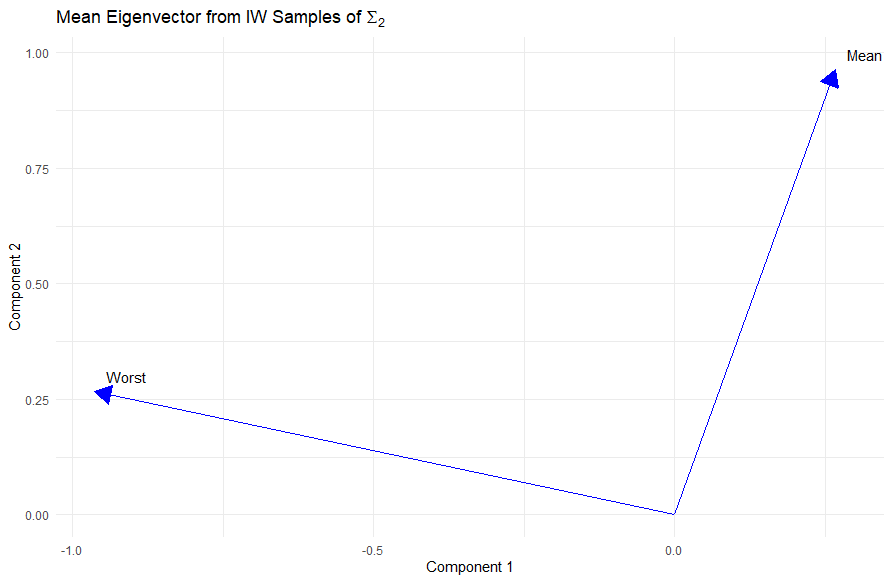} 
    \end{minipage}
    \centering
        \includegraphics[width=\textwidth]{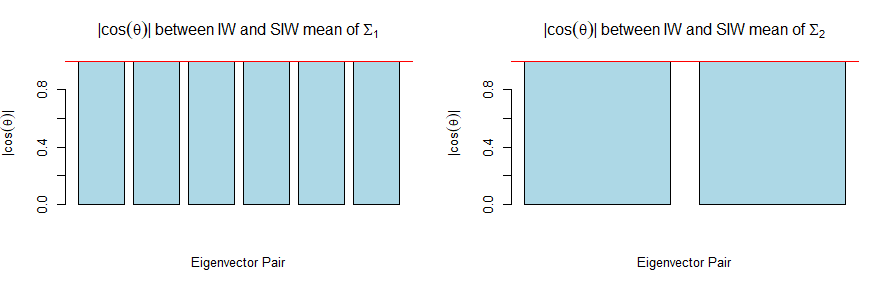} 
    \caption{Comparison of the posterior density of the condition numbers for $\Sigma_{1}$, $\Sigma_{2}$, the corresponding mean eigenvectors for the SIW posteriors under the regularized metric ( $\alpha = .95$), and mean eigenvector comparisons between the IW and SIW posteriors. Note the slightly heavier right tail of the SIW posterior for $\kappa(\Sigma_{1})$ in comparison to the IW posterior, potentially highlighting a sharper linear decay in $\Sigma_{1}$ than what the Inverse-Wishart can detect. }
    \label{fig: breast cancer analysis}
\end{figure}

In this section, we provide an analysis of a subset of the Wisconsin breast cancer dataset \cite{misc_breast_cancer_wisconsin_(diagnostic)_17}. From this dataset, we extract the features corresponding to $\mathcal{V} = \{$\texttt{smoothness, compactness, concavity, concave points, symmetry, fractal dimension}$\}$ for the factors $\mathcal{F} = \{$\texttt{mean,\\ worst}$\}$. Hence, each observation using these statistical factors can be represented in matrix form as $Y_{i} \in \mathcal{F} \times \mathcal{V} \in \mathbb{R}^{2 \times 6}$, or in vector form $y_{i} \in \mathbb{R}^{12}$. In this context, we are examining how the statistical factors co-vary alongside the covariance of the features themselves. We give a comparative analysis of this data with the IW, and SIW priors in Figure \ref{fig: breast cancer analysis}. Note the slightly heavier right tail of the SIW prior for $\kappa(\Sigma_{1})$ in comparison to the IW prior, potentially highlighting a sharp linear decay in $\Sigma_{1}$.

\section{Discussion}
In this paper, we present a geometric methodology for sampling separable covariance matrices by considering the induced geometry of the Kronecker structured covariance on the product manifold of the components. A choice of regularization was introduced which proved to be empirically effective, although computationally taxing in it's current implementation. A scheme was introduced to orthogonalize the metric to combat this, which, moreover, improves the stability of sampling.

The choice of the affine invariant metric, while conventional, may not be necessary in the study of covariance estimation. Other compelling alternatives to the affine-invariant metric have recently been introduced in the geometry literature, including the Bures-Wasserstein \cite{han2021riemannian}, and log-Cholesky \cite{lin2019riemannian} metrics.

In future work, we will consider extensions from the matrix case to the tensor case, where
\[
\Sigma = \otimes_{i = 1}^{d} \Sigma_{i} \quad d > 2.
\]
Moreover, it would be beneficial to consider more computationally convenient alternatives than naively extending the P-VL decomposition to the multiway case, as gradient complexity scales quadratically with the number of expansions introduced by a multiway analogue of the P-VL decomposition, making it prohibitively expensive.
\newpage
\bibliographystyle{plain}
\bibliography{sample-base}

\newpage
\include{supplement}

%




\end{document}

%% file: supplement.tex

\section{Supplementary Material} \label{sec: supplement}





\section*{Properties of Kronecker Products}
\begin{itemize}
\item{\it Transpose}:
\[
(A \otimes B)^{T} = A^{T} \otimes B^{T}.
\]
\item{\it Inverse}:
\[
(A \otimes B)^{-1} = A^{-1} \otimes B^{-1}.
\]
\item{ \it Mixed Product}: Let $A,B,C,D$ such that the matrix products $AC$ and $BD$ are valid. Then
\begin{equation} \label{eq: mixed product}
    (A \otimes B)(C \otimes D) = AC \otimes BD.
\end{equation}
\item{\it Mixed Trace}:
\begin{equation} \label{eq: mixed trace}
tr(A \otimes B) = tr(A) tr(B).
\end{equation}
\item{\it Mixed Determinant}: Let $A \in \mathbb{R}^{d_{A} \times d_{A}}$, $B \in \mathbb{R}^{d_{B} \times d_{B}}$
\[
\vert A \otimes B \vert = \vert A \vert^{d_{B}} \vert B \vert^{d_{A}}.
\]
\item{\it Rank}:
\[
rank(A \otimes B) = rank(A) rank(B).
\]
\item{\it Inner product representation}: Let $y \in \mathbb{R}^{n}$, then
\begin{equation} \label{eq:inner product representation}
y^{T} (A \otimes B) y = tr(A M^{T} B M)
\end{equation}
where $M$ is the matrix from reshaping $y$ into a $d_{A} \times d_{B}$ matrix.
\end{itemize}
\section*{The Affine-Invariant Pullback Geometry of $\mathcal{P}^{\otimes}(d_{1}, d_{2})$}
As discussed in Proposition 2.3 and Lemma 2.1, the pullback of the affine-invariant metric under the map $\phi: (\Sigma_{1}, \Sigma_{2}) \rightarrow \Sigma_{1} \otimes \Sigma_{2}$ is degenerate. In this section we will formally express this degeneracy.
\begin{definition}
    Let \( A \) and \( B \) be smooth manifolds. A smooth map \( \phi: A \rightarrow B \) is called an \emph{immersion} if the differential (or pushforward) of \(\phi\) at each point \( p \in A \),
\[
d\phi_p : T_p A \rightarrow T_{\phi(p)} B,
\]
is injective. In other words, the Jacobian matrix of \(\phi\) has full column rank (i.e., rank equal to \(\dim A\)) at every point \( p \in A \).
\end{definition}

First note that the map $\phi: (\Sigma_{1}, \Sigma_{2}) \rightarrow \Sigma_{1} \otimes \Sigma_{2}$ is a smooth map, as each entry of $\Sigma_{1} \otimes \Sigma_{2}$ is polynomial in each of its elements. We can straightforwardly derive the pushforward map using the bilinearity of the Kronecker product as:
\begin{align}
d \phi_{\Sigma_{1} \otimes \Sigma_{2}}(U,V) &= \lim_{t \rightarrow 0} \frac{\phi(\Sigma_{1} + tU, \Sigma_{2} + tV) - \phi(\Sigma_{1}, \Sigma_{2})}{t} \\
\phi(\Sigma_{1} + tU, \Sigma_{2} + tV) &= (\Sigma_{1} + tU)\otimes ( \Sigma_{2} + tV) = \Sigma_{1} \otimes \Sigma_{2} + tU \otimes \Sigma_{2} + \Sigma_{1} \otimes tV + t^{2} V \otimes U \\
\implies d \phi_{\Sigma_{1} \otimes \Sigma_{2}}(U,V) &= \lim_{t \rightarrow 0} \frac{\Sigma_{1} \otimes \Sigma_{2} + tU \otimes \Sigma_{2} + \Sigma_{1} \otimes tV + t^{2} V \otimes U - \Sigma_{1} \otimes \Sigma_{2}}{t} \\
&=  U \otimes \Sigma_{2} + \Sigma_{1} \otimes V.
\end{align}

\begin{definition}
    Let \( (A, g_A) \) and \( (B, g_B) \) be Riemannian manifolds, and let \( \phi: A \rightarrow B \) be a smooth map. Define the pullback metric on \( A \) by
    \[
    (\phi^*g_B)_p(v, w) = g_B(\phi(p))(d\phi_p(v), d\phi_p(w)),
    \]
    for all \( p \in A \) and \( v, w \in T_p A \).
\end{definition}

\begin{theorem} \label{theorem: pullback immersion}
    Let $\mathcal{M}$, $\mathcal{N}$ be two smooth manifolds, where $\mathcal{N}$ has the additional structure of a Riemannian metric, $g$. Let $\phi:\mathcal{M} \rightarrow \mathcal{N}$ be a smooth map, then the pullback metric \( \phi^*g \) is a non-degenerate Riemannian metric on \( \mathcal{M} \) if and only if $\phi$ is an immersion.
\end{theorem}
See \cite{lee2012smooth} for the proof of Theorem \ref{theorem: pullback immersion}.

Then the Jacobian we are characterizing is for the map $\phi: (vec(\Sigma_{1}), vec(\Sigma_{2})) \rightarrow vec(\Sigma_{1} \otimes \Sigma_{2})$. We can view the Jacobian as the convolution of the maps $\phi_{1}: (vec(\Sigma_{1}), vec(\Sigma_{2})) \rightarrow vec(\Sigma_{1}) \otimes vec(\Sigma_{2})$ and $\phi_{2}: vec(\Sigma_{1}) \otimes vec(\Sigma_{2}) \rightarrow vec(\Sigma_{1} \otimes \Sigma_{2})$, wherein $\phi = \phi_{2} \circ \phi_{1}$.

\begin{proposition} \label{prop: Jacobian}
    Let $\Sigma_{1} \in \mathcal{P}(n_{1})$, $\Sigma_{2} \in \mathcal{P}(n_{2})$. The map $\phi_{1}: (vec(\Sigma_{1}), vec(\Sigma_{2})) \rightarrow vec(\Sigma_{1}) \otimes vec(\Sigma_{2})$ has corresponding Jacobian:
    \[
J_1[(i- 1) \cdot n_{2} + j, k] = 
\begin{cases} 
\text{vec}(\Sigma_{2})[j] & \text{if } k = i, \\
\text{vec}(\Sigma_{1})[i] & \text{if } k = n_{1}^2 + j, \\
0 & \text{otherwise},
\end{cases}
\]
where \(i\) ranges over \(1, \ldots, n_{1}^2\) (the elements of \(\text{vec}(\Sigma_{1})\)), \(j\) over \(1, \ldots, n_{2}^2\) (the elements of \(\text{vec}(\Sigma_{2})\)), and \(k\) over \(1, \ldots, n_{1}^2 + n_{2}^2\) (the concatenated vector elements).
The mapping $\phi_{2}: vec(\Sigma_{1}) \otimes vec(\Sigma_{2}) \rightarrow vec(\Sigma_{1} \otimes \Sigma_{2})$ has corresponding Jacobian:
\[
J_2 = (\mathbb{I}_{n_{1}} \otimes K \otimes \mathbb{I}_{n_{2}}),
\]
where \(K\) is the commutation matrix of size \(n_{1} \times n_{2}\). This Jacobian effectively rearranges the entries to align with the vectorization of the Kronecker product. The Jacobian $J = J_{2} \circ J_{1} \in \mathbb{R}^{n_{1}^{2} n_{2}^{2} \times [n_{1}^{2} + n_{2}^{2}]}$ of $\phi = \phi_{2} \circ \phi_{1}$ has column rank $[n_{1}^{2} + n_{2}^{2}] - 1$ .

\begin{proof}
    First note
    \begin{equation} \label{eq: kronecker vector elements}
    \big(vec(\Sigma_{1}) \otimes vec(\Sigma_{2}) \big)_{(i - 1)*n_{2} + j} = vec(\Sigma_{1})_{i} vec(\Sigma_{2})_{j}.
    \end{equation}
    Then it's clear $J_{1}$ takes the form as stated. 
    After some permutations, we can view $J_{1}$ as:
    \[
    J_{1} = \begin{pmatrix}
        D & 1_{n_{1}^{2}} & 0_{n_{1}^{2}} & \cdots 0_{n_{1}^{2}}  \cdots &0_{n_{1}^{2}}\\
        D & 0_{n_{1}^{2}} & 1_{n_{1}^{2}} & \cdots 0_{n_{1}^{2}} \cdots & 0_{n_{1}^{2}}\\
        \vdots & \vdots & \vdots & \ddots & \vdots\\
        D & 0_{n_{1}^{2}} & 0_{n_{1}^{2}} & \cdots 0_{n_{1}^{2}} \cdots & 1_{n_{1}^{2}} 
    \end{pmatrix} = \begin{pmatrix}
        1_{n_{2}^{2}} \otimes D \vert I_{n_{2}^{2}}\otimes 1_{n_{1}^{2}}
    \end{pmatrix}.
    \]
    where $D = diag(vec(\Sigma_{1}))$. First note that summing the columns of $I_{n_{2}^{2}}\otimes 1_{n_{1}^{2}}$ returns $1_{n_{1}^{2} n_{2}^{2}}$. Let $D^{*} = vec(\Sigma_{1}[1,1], 0_{n_{1}^{2} - 1})$ and $D^{\dagger} = diag(vec(1, vec(\Sigma_{1})[-1]))$, then
    \begin{align*}
        J_{1}[\cdot, 1] = 1_{n_{2}^{2}} \otimes D^{*} = \Sigma_{1}[1,1]\big[1_{n_{1}^{2} n_{2}^{2}} - \sum_{j = n_{1}^{2}}^{2}(\frac{1}{D_{jj}}1_{n_{2}^{2}} \otimes D^{\dagger})[,j]\big].
    \end{align*}
    
    Hence, $J_{1} \in \mathbb{R}^{n_{1}^{2} n_{2}^{2} \times [n_{1}^{2} + n_{2}^{2}]}$ has column rank $[n_{1}^{2} + n_{2}^{2}] - 1$.

    The proof of $J_{2}$ follows immediately from Lemma 3 of \cite{magnus1985matrix}. Moreover, \cite{magnus1979commutation} shows the commutation matrix $K$ is full rank. Given the rank of a Kronecker product is the product of the ranks of the components, $J_{2}$ is then full rank.

    As there is no linear dependence between $J_{1}$ and $J_{2}$, the column rank of $J = J_{2} \circ J_{1} \in \mathbb{R}^{n_{1}^{2} n_{2}^{2} \times [n_{1}^{2} + n_{2}^{2}]}$ has column rank $[n_{1}^{2} + n_{2}^{2}] - 1$.
    
\end{proof}
\end{proposition}

Hence, Proposition \ref{prop: Jacobian} shows us the map $\phi: (vec(\Sigma_{1}), vec(\Sigma_{2})) \rightarrow vec(\Sigma_{1} \otimes \Sigma_{2})$ is not injective and in fact has column rank $n_{1}^{2} + n_{2}^{2} - 1$. This highlights that any pullback metric will be degenerate when naively implemented, and moreover that if we want to achieve a full rank metric tensor by directly modifying the domain $\{\Sigma_{1}, \Sigma_{2}\}$, simply enforcing the constraint $\vert \Sigma_{2} \vert = 1$ will suffice to give the domain a rank of $n_{1}^{2} + n_{2}^{2} - 1$. Otherwise, we will need to resort to the regularization scheme to achieve a postive definite metric.

\section*{Proofs of Propositions and Lemmas}
    \supplementaryproof{2.1}{ Note that the time derivative follows a product rule for the kronecker product:
    \[
    V(\tau) = \Sigma'(\tau) = \Sigma_{1}'(\tau) \otimes \Sigma_{2}(\tau) + \Sigma_{1}(\tau) \otimes \Sigma_{2}'(\tau) = V_{1}(\tau) \otimes \Sigma_{2}(\tau) + \Sigma_{1} (\tau) \otimes V_{2}(\tau).
    \]
    The norm under the affine invariant metric is given by:
        \begin{align}
            \|V(\tau)\|_{\Sigma_{1} \otimes \Sigma_{2}} &= g_{\Sigma_{1} \otimes \Sigma_{2}}(V(\tau), V(\tau)) = -\frac{\partial^{2}}{\partial t \partial s} \psi( \Sigma_{1} \otimes \Sigma_{2} + t(V_{1} \otimes \Sigma_{2}) + s( V_{2} \otimes \Sigma_{2})) |_{s=t = 0} \nonumber \\
            &= -\frac{\partial^{2}}{\partial t \partial s}tr(log( \Sigma_{1} \otimes \Sigma_{2} + t(V_{1} \otimes \Sigma_{2}) + s( V_{2} \otimes \Sigma_{2})))|_{s=t = 0} \nonumber \\
            &= -\frac{\partial}{\partial t} tr([\Sigma_{1} \otimes \Sigma_{2} + s(V_{1}(t) \otimes \Sigma_{2}(t)) + t(\Sigma_{1}(t) \otimes V_{2}(t))]^{-1}(V_{1}(t) \otimes \Sigma_{2}(t)))\vert_{s = t = 0} \nonumber \\
            &= tr([\Sigma_{1} \otimes \Sigma_{2} + s(V_{1}(t) \otimes \Sigma_{2}(t)) + t(\Sigma_{1}(t) \otimes V_{2}(t))]^{-1}(\Sigma_{1}(t) \otimes V_{2}(t))[\Sigma_{1} \otimes \Sigma_{2} \nonumber \\
            &+ s(V_{1}(t) \otimes \Sigma_{2}(t)) + t(\Sigma_{1}(t) \otimes V_{2}(t))]^{-1} (\Sigma_{1}(t) \otimes V_{2}(t))) \vert_{s = t = 0} \nonumber \\
            &= tr(\Sigma_{1}^{-1}(t) V_{1}(t) \Sigma_{1}(t) V_{1}(t) \Sigma_{1}^{-1}(t) \otimes I_{2}) \nonumber \\
            &+ tr(I_{1} \otimes \Sigma_{2}^{-1}(t) V_{2}(t) \Sigma_{2}(t) V_{2}(t)) + tr(\Sigma_{1}^{-1} V_{1})tr(\Sigma_{2}^{-1} V_{2}) \nonumber \\
            &= \|V_{1}(t)\|_{\Sigma_{1}}^{AI} + \| V_{2}(t)\|_{\Sigma_{2}}^{AI} + 2*vec(V_{1})^{T} vec(\Sigma_{1}^{-1})vec(\Sigma_{2}^{-1})^{T} vec(V_{2}). \nonumber
        \end{align}}

\supplementaryproof{2.2}{
        The Schur decomposition tells us for a block matrix:
        \[
        P = \begin{pmatrix}
            A & B \\
            C & D
        \end{pmatrix}
        \]
        that $\vert P \vert = \vert D \vert \vert A - BD^{-1}C\vert$. Hence,
        \begin{equation} \label{eq: Block Determinant}
        \vert G^{\otimes}(\Sigma_{1}, \Sigma_{2}) \vert = \vert d_{1} \Sigma_{2}^{-1} \otimes \Sigma_{2}^{-1} \vert \vert d_2 \Sigma_{1} \otimes \Sigma_{1} - vec(\Sigma_{1}^{-1}) vec(\Sigma_{2}^{-1})^{T} [\frac{1}{d_{1}}\Sigma_{2} \otimes  \Sigma_{2}] vec(\Sigma_{2}^{-1}) vec(\Sigma_{1}^{-1})^{T} \vert.
        \end{equation}
        Note that
        \[
        vec(\Sigma_{2}^{-1})^{T} \Sigma_{2}^{-1} \otimes \Sigma_{2}^{-1} vec(\Sigma_{2}^{-1}) = tr(\Sigma_{2}^{-1} \Sigma_{2} \Sigma_{2}^{-1} \Sigma_{2}) = d_{2}.
        \]
        hence, 
        \begin{equation} 
        vec(\Sigma_{1}^{-1}) vec(\Sigma_{2}^{-1})^{T} [\frac{1}{d_{1}}\Sigma_{2} \otimes  \Sigma_{2}] vec(\Sigma_{2}^{-1}) vec(\Sigma_{1}^{-1})^{T} = \frac{d_2}{d_1} vec(\Sigma_{1}^{-1}) vec(\Sigma_{1}^{-1})^{T}.
        \end{equation}
        And therefore (\ref{eq: Block Determinant}) reduces to:
        \begin{equation} \label{eq: Rank 1 purturbation}
        \vert G(\Sigma_{1}, \Sigma_{2}) \vert = \vert d_{1} \Sigma_{2}^{-1} \otimes \Sigma_{2}^{-1} \vert \vert d_{2} \Sigma_{1}^{-1} \otimes \Sigma_{1}^{-1}  -  \frac{d_2}{d_1} vec(\Sigma_{1}^{-1}) vec(\Sigma_{1}^{-1})^{T}\vert ,
        \end{equation}
        the second determinant in equation (\ref{eq: Rank 1 purturbation}) is a rank 1 perturbation. and hence has a determinant given by \cite{meyer2023matrix}:
        \begin{align}
        &\vert  d_{2} \Sigma_{1}^{-1} \otimes \Sigma_{1}^{-1}  - \frac{d_2}{d_1} vec(\Sigma_{1}^{-1}) vec(\Sigma_{1}^{-1})^{T}\vert \nonumber \\
        &= \vert d_{2} \Sigma_{1}^{-1} \otimes \Sigma_{1}^{-1} \vert [1 - \frac{1}{d_1} vec(\Sigma_{1}^{-1})^{T} \Sigma_{1} \otimes \Sigma_{1} vec(\Sigma_{1}^{-1}) ] = 0, \label{eq: Zero Determinant}
        \end{align}
        where the last line follows from 
        \begin{equation}
        vec(\Sigma_{1}^{-1})^{T} \Sigma_{1} \otimes \Sigma_{1} vec(\Sigma_{1}^{-1}) = tr(\Sigma_{1} \Sigma_{1}^{-1} \Sigma_{1} \Sigma_{1}^{-1}) = d_1 \label{eq: det 2}.
        \end{equation}
}
\supplementarylemma{2.1}{
First note $d_{2} \Sigma_{1}^{-1} \otimes \Sigma_{1}^{-1}$ is positive definite. Then by \cite{zhang2006schur}, it suffices to prove positive definiteness of:
\begin{align*}
    &d_{1} \Sigma_{2}^{-1}\otimes\Sigma_{2}^{-1} -  \frac{\alpha^{2}}{d_{2}} vec(\Sigma_{2}^{-1})vec(\Sigma_{1}^{-1})^{T} [\Sigma_{1}\otimes\Sigma_{1}] vec(\Sigma_{1}^{-1})vec(\Sigma_{2}^{-1})^{T}\\
    &= d_{1}[ \Sigma_{2}^{-1}\otimes\Sigma_{2}^{-1} - \frac{\alpha^{2}}{d_{2}} vec(\Sigma_{2}^{-1}) vec(\Sigma_{2}^{-1})^{T}].
\end{align*}
By the Sherman Morrison formula, $[ \Sigma_{2}^{-1}\otimes\Sigma_{2}^{-1} - \lambda vec(\Sigma_{2}^{-1}) vec(\Sigma_{2}^{-1})^{T}]$ is positive definite if and only if:
\begin{align*}
&1 + \lambda \big( vec(\Sigma_{2}^{-1})^{T} \Sigma_{2}\otimes\Sigma_{2} vec(\Sigma_{2}^{-1}) \big) > 0 \text{ for } \lambda \geq 0 \\
&\implies \lambda \geq \frac{-1}{d_{2}} .
\end{align*}
As $\lambda = -\frac{\alpha^{2}}{d_{2}}$ with $0 \leq \alpha < 1$, the proof is then complete.
}

\supplementaryproof{2.3}{
        Recall the metric choices are:
        \begin{align}
            \label{eq: regularized metric} G_{\alpha}^{\otimes}(\Sigma_{1}, \Sigma_{2}) &= \begin{pmatrix}
         d_{2} \Sigma_{1}^{-1}(\tau) \otimes \Sigma_{1}^{-1}(\tau)  &  \alpha*  vec(\Sigma_{1}^{-1}(\tau))vec(\Sigma_{2}^{-1}(\tau))^{T} \\
         \alpha * vec(\Sigma_{2}^{-1}(\tau))vec(\Sigma_{1}^{-1}(\tau))^{T} &  d_{1} \Sigma_{2}^{-1}(\tau) \otimes \Sigma_{2}^{-1}(\tau) \end{pmatrix} \\
         \label{eq: Orthogonal metric} G^{O}(\Sigma_{1}, \Sigma_{2}) &= \begin{pmatrix}
            d_{1} \Sigma_{1}^{-1} \otimes \Sigma_{1}^{-1} & 0 \\
            0 & d_{2} \Sigma_{2}^{-1} \otimes \Sigma_{2}^{-1} \end{pmatrix}\\
            \label{eq: weighted metric} G^{W}(\Sigma_{1}, \Sigma_{2} \vert \omega) &= \begin{pmatrix}
    (\omega d_{2} + (1-\omega)) \Sigma_{1}^{-1} \otimes \Sigma_{1}^{-1} & 0 \\
    0 & (\omega d_{1} + (1-\omega)) \Sigma_{2}^{-1} \otimes \Sigma_{2}^{-1}
    \end{pmatrix}\\
    \label{eq: product metric} G^{\times}(\Sigma_{1}, \Sigma_{2}) &= \begin{pmatrix}
    \Sigma_{1}^{-1} \otimes \Sigma_{1}^{-1} & 0 \\
    0 & \Sigma_{2}^{-1} \otimes \Sigma_{2}^{-1}
\end{pmatrix}.
        \end{align}
        Let $\Sigma = \Sigma_{1} \otimes \Sigma_{2}$ and define
        \begin{equation} \label{eq: geodesic functional}
        \mathcal{L}(\Sigma) = \frac{1}{2} \int_{a}^{b} g_{\Sigma}(\dot{\Sigma}(t), \dot{\Sigma}(t)) dt
        \end{equation}
        per \cite{jost2008riemannian}, the extremal points of $\mathcal{L}$ are geodesics. Given the independence (diagonal) structure of metrics (\ref{eq: Orthogonal metric}), (\ref{eq: weighted metric}), and (\ref{eq: product metric}), these all have obvious solutions in the Cartesian product as $\{\Sigma_{1}(t), \Sigma_{2}(t)\}$ and hence $\Sigma(t) = \Sigma_{1}(t) \otimes \Sigma_{2}(t)$. To show the same for (\ref{eq: regularized metric}), 
        First note if $\Sigma = \Sigma_{1} \otimes \Sigma_{2}$, and $\Sigma_{1} \in \mathcal{P}(n_{1})$ and $\Sigma_{2} \in \mathcal{P}(n_{2})$ are parameterized by a coordinate basis $E_{1}$, $E_{2}$, respectively, then with respect to some coordinate $p_{1}^{\alpha}$, $p_{2}^{\beta}$, $\frac{\partial(P_{1} \otimes P_{2})}{\partial p_{1}^{\alpha}} = E_{\alpha} \otimes \Sigma_{2}$. Observe $L = \frac{1}{2} [d_{2}tr(\Sigma_{1}^{-1} \dot{\Sigma}_{1} \Sigma_{1}^{-1} \dot{\Sigma}_{1}) + d_{1}tr(\Sigma_{2}^{-1} \dot{\Sigma}_{2} \Sigma_{2}^{-1} \dot{\Sigma}_{2}) + 2\alpha tr(\Sigma_{1}^{-1} \dot{\Sigma}_{1}) tr(\Sigma_{2}^{-1} \dot{\Sigma}_{2})]$ is the Lagrangian of equation (\ref{eq: geodesic functional}) , then following the method of \cite{moakher2011riemannian}, the critical points of the Kronecker affine geodesic equation corresponds to the simultaneous solutions of the coupled differential equations:
    \begin{align}
        \frac{\partial L}{\partial p_{1}^{\alpha}} - \frac{d}{dt} \frac{\partial L}{\partial \dot{p}_{1}^{\alpha}} &= 0 \text{ for } \alpha \in \{1, \ldots, n_{1}(n_{1} + 1)/2\} \\
        \frac{\partial L}{\partial p_{2}^{\beta}} - \frac{d}{dt} \frac{\partial L}{\partial \dot{p}_{2}^{\beta}} &= 0  \text{ for } \beta \in \{1, \ldots, n_{2}(n_{2} + 1)/2\}. 
    \end{align}    

    It then follows 
\begin{align*}
\frac{\partial L}{\partial p_{1}^{\alpha}} &= -d_{2} tr(\Sigma_{1}^{-1} E_{\alpha}^{1} \Sigma_{1}^{-1} \dot{\Sigma}_{1} \Sigma_{1}^{-1} \dot{\Sigma}_{1}) - \alpha tr(\Sigma_{1}^{-1} E^{1}_{\alpha} \Sigma_{1}^{-1} \dot{\Sigma}_{1})tr(\Sigma_{2}^{-1} \dot{\Sigma}_{2}) \\
\frac{\partial L}{\partial p_{2}^{\beta}} &= -d_{1} tr(\Sigma_{2}^{-1} E_{\beta}^{2} \Sigma_{2}^{-1} \dot{\Sigma}_{2} \Sigma_{2}^{-1} \dot{\Sigma}_{2}) - \alpha tr(\Sigma_{1}^{-1} \dot{\Sigma}_{1})tr(\Sigma_{2}^{-1} E^{2}_{\beta} \Sigma_{2}^{-1} \dot{\Sigma}_{2})
\end{align*}
\begin{align*}
    \frac{\partial}{\partial t} \frac{\partial L}{\partial \dot{p}^{1}_{\alpha}} &= -2d_{2} tr(\Sigma_{1}^{-1} \dot{\Sigma}_{1} \Sigma_{1}^{-1} E_{\alpha}^{1} \Sigma_{1}^{-1} \dot{\Sigma}_{1}) + d_{2} tr(\Sigma_{1}^{-1} E_{\alpha}^{1} \Sigma_{1}^{-1} \ddot{\Sigma}_{1}) + \\
    & \alpha[-tr(\Sigma_{1}^{-1} \dot{\Sigma}_{1} \Sigma_{1}^{-1} E_{\alpha}^{1}) tr(\Sigma_{2}^{-1} \dot{\Sigma}_{2}) -  tr(\Sigma_{1}^{-1} E_{\alpha}^{1})tr(\Sigma_{2}^{-1} \dot{\Sigma}_{2} \Sigma_{2}^{-1} \dot{\Sigma}_{2}) + \\
    &tr(\Sigma_{1}^{-1} E_{\alpha}^{1})tr(\Sigma_{2}^{-1} \ddot{\Sigma}_{2})]
\end{align*}
\begin{align*}
    \frac{\partial}{\partial t} \frac{\partial L}{\partial \dot{p}^{2}_{\beta}} &= -2d_{1}tr(\Sigma_{2}^{-1} \dot{\Sigma}_{2} \Sigma_{2}^{-1} E_{\beta}^{2} \Sigma_{2}^{-1} \dot{\Sigma}_{2}) + d_{1}tr(\Sigma_{2}^{-1} E_{\beta}^{2} \Sigma_{2}^{-1} \ddot{\Sigma}_{2}) + \\
    & \alpha[-tr(\Sigma_{2}^{-1} \dot{\Sigma}_{2} \Sigma_{2}^{-1} E_{\beta}^{2}) tr(\Sigma_{1}^{-1} \dot{\Sigma}_{1}) -  tr(\Sigma_{2}^{-1} E_{\beta}^{2})tr(\Sigma_{1}^{-1} \dot{\Sigma}_{1} \Sigma_{1}^{-1} \dot{\Sigma}_{1}) + \\
    &tr(\Sigma_{2}^{-1} E_{\beta}^{2})tr(\Sigma_{1}^{-1} \ddot{\Sigma}_{1})].
\end{align*}
And so the geodesics will belong to the simultaneous solutions of :
\begin{align*}
    \frac{\partial L}{\partial p_{1}^{\alpha}} - \frac{d}{dt} \frac{\partial L}{\partial \dot{p}_{1}^{\alpha}} &= d_{2} tr(\Sigma_{1}^{-1} E_{\alpha}^{1} \Sigma_{1}^{-1} \dot{\Sigma}_{1} P_{1}^{-1} \dot{\Sigma}_{1}) -  d_{2} tr(\Sigma_{1}^{-1} E_{\alpha}^{1} P_{1}^{-1} \ddot{\Sigma}_{1}) +  \\
    &\alpha tr(\Sigma_{1}^{-1} E_{\alpha}^{1})tr(\Sigma_{2}^{-1} \dot{\Sigma}_{2} P_{2}^{-1} \dot{\Sigma}_{2}) - \alpha tr(\Sigma_{1}^{-1} E_{\alpha}^{1})tr(\Sigma_{2}^{-1} \ddot{\Sigma}_{2}) = 0
\end{align*}

\begin{align*}
    \frac{\partial L}{\partial p_{2}^{\beta}} - \frac{d}{dt} \frac{\partial L}{\partial \dot{p}_{2}^{\beta}} &= d_{1} tr(\Sigma_{2}^{-1} \dot{\Sigma}_{2} \Sigma_{2}^{-1} E_{\beta}^{2} \Sigma_{2}^{-1} \dot{\Sigma}_{2}) - d_{1}tr(\Sigma_{2}^{-1} E_{\beta}^{2} \Sigma_{2}^{-1} \ddot{\Sigma}_{2}) +  \\
    &\alpha tr(\Sigma_{2}^{-1} E_{\beta}^{2})tr(\Sigma_{1}^{-1} \dot{\Sigma}_{1} P_{1}^{-1} \dot{\Sigma}_{1}) - \alpha tr(\Sigma_{1}^{-1} E_{\alpha}^{1})tr(\Sigma_{2}^{-1} \ddot{\Sigma}_{2}) = 0.
\end{align*}

We can rewrite these equations as:
\begin{align}
    \frac{\partial L}{\partial p_{1}^{\alpha}} - \frac{d}{dt} \frac{\partial L}{\partial \dot{p}_{1}^{\alpha}} &= tr([d_{2} \Sigma_{1}^{-1} \dot{\Sigma}_{1}\Sigma_{1}^{-1} \dot{\Sigma}_{1} P_{1}^{-1} -  d_{2}   \Sigma_{1}^{-1} \ddot{\Sigma}_{1} \Sigma_{1}^{-1} + \label{eq: trace 1,1} \\
    &\alpha tr(\Sigma_{2}^{-1} \dot{\Sigma}_{2} \Sigma_{2}^{-1} \dot{\Sigma}_{2})\Sigma_{1}^{-1} - \alpha tr(\Sigma_{2}^{-1} \ddot{\Sigma}_{2}) \Sigma_{1}^{-1}] E_{\alpha}^{1}) = 0 \label{eq: trace 1,2}
\end{align}

\begin{align}
    \frac{\partial L}{\partial p_{2}^{\beta}} - \frac{d}{dt} \frac{\partial L}{\partial \dot{p}_{2}^{\beta}} &= tr([d_{1} \Sigma_{2}^{-1} \dot{\Sigma}_{2} \Sigma_{2}^{-1} \dot{\Sigma}_{2} \Sigma_{2}^{-1} - d_{1} \Sigma_{2}^{-1} \ddot{\Sigma}_{2} \Sigma_{2}^{-1} +  \label{eq: trace 2,1}\\
    &\alpha tr(\Sigma_{1}^{-1} \dot{\Sigma}_{1} \Sigma_{1}^{-1} \dot{\Sigma}_{1})\Sigma_{2}^{-1} - \alpha tr(\Sigma_{1}^{-1} \ddot{\Sigma}_{1}) \Sigma_{2}^{-1}] E_{\beta}^{2}) = 0. \label{eq: trace 2,2}
\end{align}
Note however that on the product manifold, the solutions to the geodesic equations would be given when
\begin{align*}
\ddot{P}_{1} - \dot{P}_{1} P_{1}^{-1} \dot{P}_{1} &= 0 \\
\ddot{P}_{2} - \dot{P}_{2} P_{2}^{-1} \dot{P}_{2} &= 0.
\end{align*}
In which case $\ddot{P}_{1} = \dot{P}_{1} P_{1}^{-1} \dot{P}_{1}$, $\ddot{P}_{2} = \dot{P}_{2} P_{2}^{-1} \dot{P}_{2}$, note however these are also simultaneous solutions to equations (\ref{eq: trace 1,1}-\ref{eq: trace 1,2}) and (\ref{eq: trace 2,1} - \ref{eq: trace 2,2}). This implies solutions to the geodesic equations under metric (\ref{eq: regularized metric}) can be computed as solutions under the product manifold.

Hence, for any of the metric choices, paths on $\Sigma$ may be described as $\Sigma(\tau) = \Sigma_{1}(\tau) \otimes \Sigma_{2}(\tau)$ with corresponding geodesic updates according to a product manifold with components independently endowed with the affine-invariant metric.
}
    
\supplementarylemma{2.3}{
        This immediately follows from the fact that $\vert G_{\alpha}^{\otimes} \vert$ is proportional to the determinants of the diagonal blocks. From \cite{moakher2011riemannian}, the introduction of the duplication matrices makes the determinants of the corresponding diagonal blocks proportional to $\vert \Sigma_{i} \vert^{d_{i} + 1}$.
}

For the following two propositions, we will make use of the notation $A:B = tr(A^{T} B)$ for matrices of appropriate dimension $A,B$.
\supplementaryproof{4.1}{
    Note that the prior density up to proportionality is defined as:
    \[
    P(\Sigma_{i}\vert d_{i} + 2, \frac{\gamma}{d_{i}} I_{d_{i}}) \propto \vert \Sigma_{i} \vert^{-\frac{(2d_{i} + 3)}{2}} e^{-\frac{1}{2} tr(\frac{\gamma}{d_{i}}I_{d_{i}} \Sigma_{i}^{-1})}.
    \]
    The likelihood is given by
    \begin{equation} \label{eq:separable likelihood}
    \mathcal{L}(Y|\Sigma_{1}, \Sigma_{2}) \propto \vert \Sigma_{i} \vert^{-n*d_{-i}/2} e^{-\frac{1}{2} tr(\Sigma_{1}^{-1} \otimes \Sigma_{2}^{-1} YY^{T})}.
    \end{equation}
    Using the Pitsianis decomposition:
    \[
    YY^{T} = \sum_{i = 1}^{d_{1}d_{2}} A_{i} \otimes B_{i}.
    \]
    Then equation (\ref{eq:separable likelihood}) can be expressed as:
    \begin{align}
        \mathcal{L}(Y|\Sigma_{1}, \Sigma_{2}) &\propto \vert \Sigma_{1} \vert^{-nd_{2}/2} \vert \Sigma_{2} \vert^{-nd_{1}/2} e^{-\frac{1}{2} tr(\Sigma_{1}^{-1} \otimes \Sigma_{2}^{-1} \sum_{i = 1}^{d_{1} d_{2}} A_{i} \otimes B_{i})}  \\
        &= \vert \Sigma_{1} \vert^{-nd_{2}/2} \vert \Sigma_{2} \vert^{-nd_{1}/2} e^{-\frac{1}{2} tr(\sum_{i = 1}^{d_{1} d_{2}} (\Sigma_{1}^{-1} A_{i}) \otimes (\Sigma_{2}^{-1} B_{i}))}.
    \end{align}
    Note that the term in the trace is expressible as:
    \begin{align}
        &tr(\sum_{i = 1}^{d_{1} d_{2}} (\Sigma_{1}^{-1} A_{i}) \otimes (\Sigma_{2}^{-1} B_{i})) \nonumber \\
        &= \sum_{i = 1}^{d_{1} d_{2}} tr(\Sigma_{1}^{-1} A_{i}) tr(\Sigma_{2}^{-1}B_{i}) \\
        &= tr( \sum_{i = 1}^{d_{1} d_{2}} 
 \Sigma_{1}^{-1}A_{i}tr(\Sigma_{2}^{-1} B_{i}))
    \end{align}
    or alternatively
    \begin{equation}
        =tr( \sum_{i = 1}^{d_{1} d_{2}} 
 \Sigma_{2}^{-1}B_{i}tr(\Sigma_{1}^{-1}A_{i})).
    \end{equation}

    Hence, the full conditional distributions for $\Sigma_{1}$ and $\Sigma_{2}$ can be expressed by:
    \begin{align}
        p(\Sigma_{1} | Y, \Sigma_{2}) &\propto \vert \Sigma_{1} \vert^{-nd_{2} +(2d_{1} + 3)/2} e^{-\frac{1}{2}tr(\Sigma_{1}^{-1} (\frac{\gamma}{d_{1}}I_{d_{1}} + \sum_{i = 1}^{d_{1} d_{2}} A_{i} tr(\Sigma_{2}^{-1}B_{i})))} \\
        p(\Sigma_{2} | Y, \Sigma_{1}) &\propto \vert \Sigma_{2} \vert^{-nd_{1} +(2d_{2} + 3)/2}e^{-\frac{1}{2}tr(\Sigma_{2}^{-1} (\frac{\gamma}{d_{2}}I_{d_{2}} + \sum_{i = 1}^{d_{1} d_{2}} B_{i} tr(\Sigma_{1}^{-1}A_{i})))}.
    \end{align}
}

\supplementaryproof{4.2}{
First note that for a differentiable function $f$ it follows from the chain rule that
\[
df = (\frac{\partial f}{\partial A} : d A) + (\frac{\partial f}{\partial B}:dB)
\]
and observe $df$ is expressed by:
\begin{align}
f &= (\Sigma_{1}^{-1} \otimes \Sigma_{2}^{-1}) YY^{T} = tr(\sum_{i = 1}^{r} (\Sigma_{1}^{-1} \otimes \Sigma_{2}^{-1}) (V_{k} \otimes Z_{k})) \nonumber \\
& (\Sigma_{1}^{-1} \otimes \Sigma_{2}^{-1})(V_{k} \otimes Z_{k}) = (\Sigma_{1}^{-1} V_{k}) \otimes (\Sigma_{2}^{-1} Z_{k}) \nonumber \\
&\rightarrow f = tr(\sum_{k = 1}^{r} (\Sigma_{1}^{-1} V_{k}) \otimes (\Sigma_{2}^{-1} Z_{k})) = \sum_{k = 1}^{r} tr((\Sigma_{1}^{-1} V_{k}) \otimes (\Sigma_{2}^{-1}Z_{k})) \nonumber \\
&= \sum_{k = 1}^{r} tr(\Sigma_{1}^{-1} V_{k})tr(\Sigma_{2}^{-1} Z_{k}). \nonumber 
\end{align}
Note that if follows
\begin{align}
&d [tr(\Sigma_{1}^{-1}V_{k})tr(\Sigma_{2}^{-1}Z_{k})] \nonumber \\
&= tr(d \Sigma_{1}^{-1} V_{k}) tr(\Sigma_{2}^{-1}Z_{k}) + tr(\Sigma_{1}^{-1}V_{k})tr(d \Sigma_{2}^{-1}Z_{k}) \nonumber
\end{align}
and $d \Sigma_{i}^{-1} = - \Sigma_{i}^{-1} d\Sigma_{i} \Sigma_{i}^{-1}$, therefore we may express $df$ as:
\begin{align}
\rightarrow df &= -\sum_{k = 1}^{r} tr(\Sigma_{1}^{-1} d \Sigma_{1} \Sigma_{1}^{-1} V_{k}) tr(\Sigma_{2}^{-1} Z_{k}) + tr(\Sigma_{1}^{-1} V_{k} ) tr(\Sigma_{2}^{-1} d\Sigma_{2} \Sigma_{2}^{-1} Z_{k} ) \nonumber \\
&= -\sum_{k = 1}^{r} tr( \Sigma_{2}^{-1}Z_{k}) tr(\Sigma_{1}^{-1} V_{k} \Sigma_{1}^{-1} d\Sigma_{1})  + tr(\Sigma_{1}^{-1} V_{k} ) tr(\Sigma_{2}^{-1} Z_{k} \Sigma_{2}^{-1} d \Sigma_{2}) \nonumber \\ 
&= - \sum_{k = 1}^{r} tr(tr(\Sigma_{2}^{-1} Z_{k}) \Sigma_{1}^{-1} V_{k} \Sigma_{1}^{-1} d\Sigma_{1}) + tr(tr(\Sigma_{1}^{-1} V_{k}) \Sigma_{2}^{-1} Z_{k} \Sigma_{2}^{-1} d\Sigma_{2}) \nonumber \\
&= tr(-\sum_{k = 1}^{r} tr(\Sigma_{2}^{-1} Z_{k}) \Sigma_{1}^{-1} V_{k} \Sigma_{1}^{-1} d\Sigma_{1}) \nonumber \\
&+ tr(-\sum_{k = 1}^{r}tr(\Sigma_{1}^{-1} V_{k}) \Sigma_{2}^{-1}Z_{k}\Sigma_{2}^{-1} d\Sigma_{2}). \nonumber
\end{align}
Hence it then follows that
\begin{align}
df &= tr(\frac{\partial f}{\partial \Sigma_{1}} d\Sigma_{1}) + tr(\frac{\partial f}{\partial \Sigma_{2}} d\Sigma_{2}). \nonumber
\end{align}
This then implies
\begin{align}
\frac{\partial f}{\partial \Sigma_{1}} &= -\sum_{k = 1}^{r} tr(\Sigma_{2}^{-1} Z_{k}) \Sigma_{1}^{-1} V_{k} \Sigma_{1}^{-1} \nonumber \\
\frac{\partial f}{\partial \Sigma_{2}} &=  -\sum_{k = 1}^{r}tr(\Sigma_{1}^{-1} V_{k}) \Sigma_{2}^{-1}Z_{k}\Sigma_{2}^{-1}. \nonumber
\end{align}
}

\section*{Full Empirical Comparisons} \label{sec: supplement empirical comparisons}

In this section, we demonstrate the validity of SGLMC on several generated datasets with varying levels of dimension between $\Sigma_{1}$ and $\Sigma_{2}$. In each setting, we generate $\Sigma_{1}$ and $\Sigma_{2}$ as:
\begin{align}
    \Sigma_{i} &\sim \mathcal{I}\mathcal{W}(\nu = d_{i} + 10, \frac{\sqrt{\gamma}}{d_{i}} I_{d_{i}}) \nonumber \\
    y_{1}, \ldots, y_{n} &\sim \mathcal{N}(0, \Sigma_{1} \otimes \Sigma_{2}) \nonumber
\end{align}
with $\gamma = 5$, $n = 300$. We give comparisons of global matrix summaries, such as the traces, determinants, and posterior densities, as well as summaries of the properties of the components of the Kronecker product such as their eigenvectors, and condition numbers. The dimensions of the Experiments were correspondingly $(d_{1},d_{2}) \in \{(3,6), (15,6), (15,15)$ in Figures \ref{fig: d1 = 3, d2 = 6}, \ref{fig: d1= 15, d2 = 6, alpha = .95}, and \ref{fig: d1 = 15, d2 = 15, alpha = .95}, respectively.

In Experiments \ref{fig: d1 = 3, d2 = 6} and \ref{fig: d1= 15, d2 = 6, alpha = .95} , $5000$ samples were drawn from the Gibbs sampler, and $5000$ samples drawn from stan. The Gibbs sampler was given a burn-in of $30000$,and stan's "warm=up" parameter was set to $1000$. The only tuning parameter adjusted for stan was {\it int time}, which was set to $.05$. Stan would not run for Experiment \ref{fig: d1 = 15, d2 = 15, alpha = .95}, which we omitted for that example, but the parameters for the Gibbs sampler was the same.

In Experiment \ref{fig: d1 = 3, d2 = 6}, SGLMC was given an adaptation of $n_{adapt} = 1000$, a burn-in of $500$, and $5000$ samples generated afterwards. In Experiments \ref{fig: d1= 15, d2 = 6, alpha = .95} and \ref{fig: d1 = 15, d2 = 15, alpha = .95}, we used $500$ adaptation and $500$ burn-in iterations, with $2000$ samples generated. We set the regularization parameter to $\alpha = .95$, with the acceptance probability for the dual averaging of $\epsilon$ set to $\alpha_{0} = .8$. 

The comparisons can be found in Figure \ref{fig: d1= 15, d2 = 6, alpha = .95}.

\begin{figure}[ht]
    \centering
    \begin{minipage}{0.45\textwidth}
        \centering
        \includegraphics[width=\textwidth]{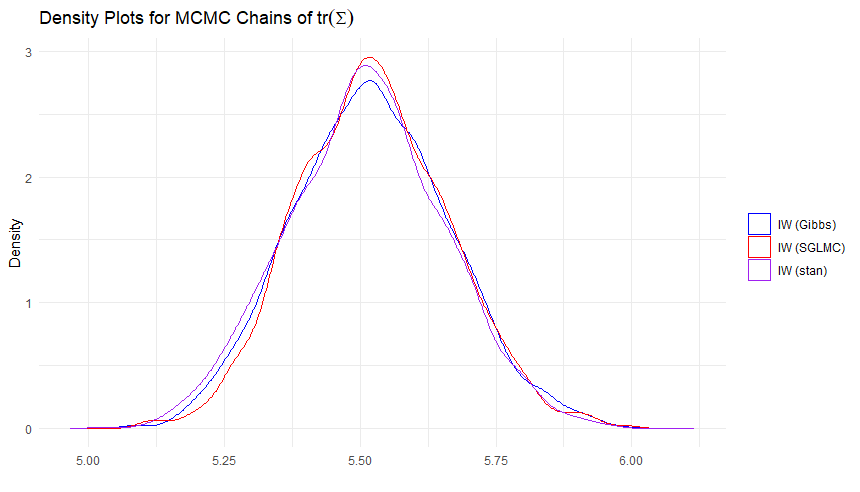} 
    \end{minipage}\hfill
    \begin{minipage}{0.45\textwidth}
        \centering
        \includegraphics[width=\textwidth]{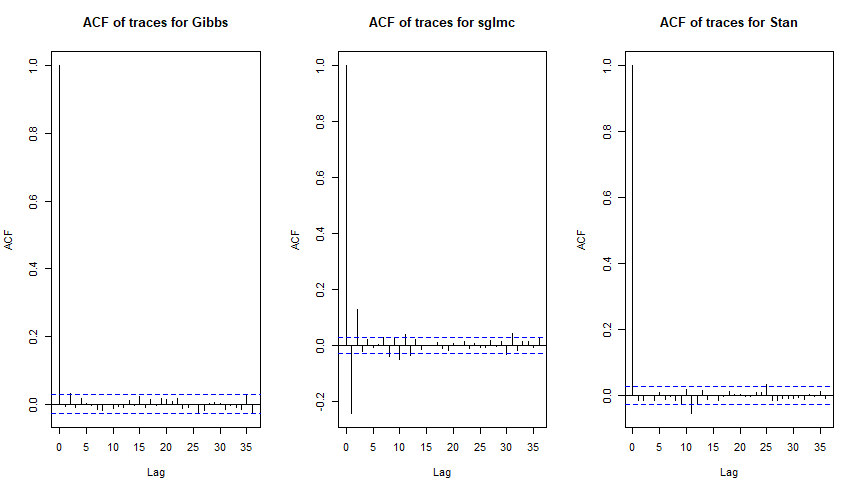} 
    \end{minipage}
           \centering
        \begin{minipage}{0.45\textwidth}
        \centering
        \includegraphics[width=\textwidth]{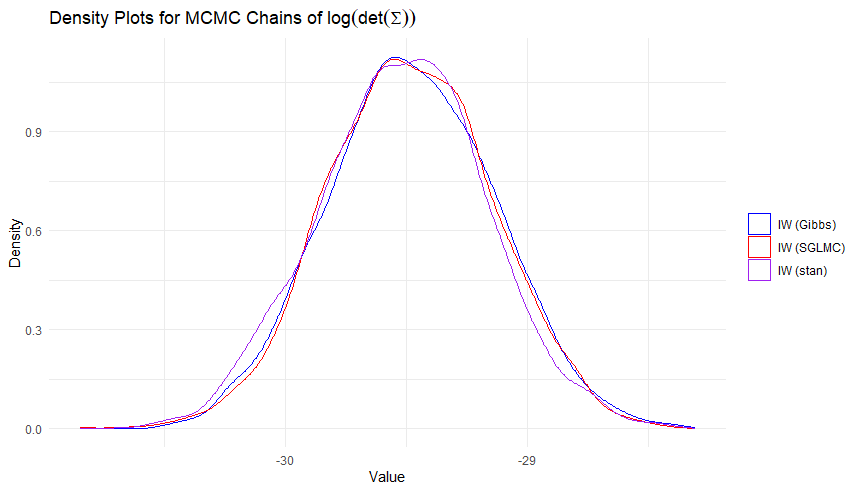} 
    \end{minipage}\hfill
    \begin{minipage}{0.45\textwidth}
        \centering
        \includegraphics[width=\textwidth]{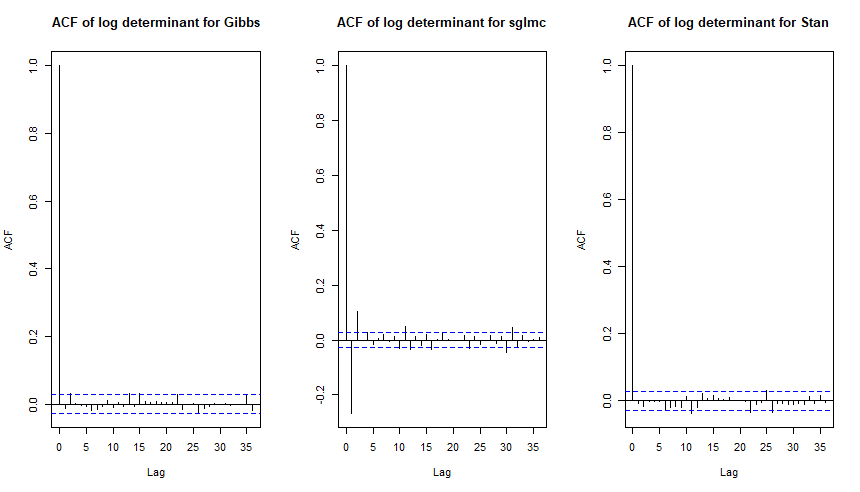} 
    \end{minipage}
    \centering
    \begin{minipage}{0.45\textwidth}
        \centering
        \includegraphics[width=\textwidth]{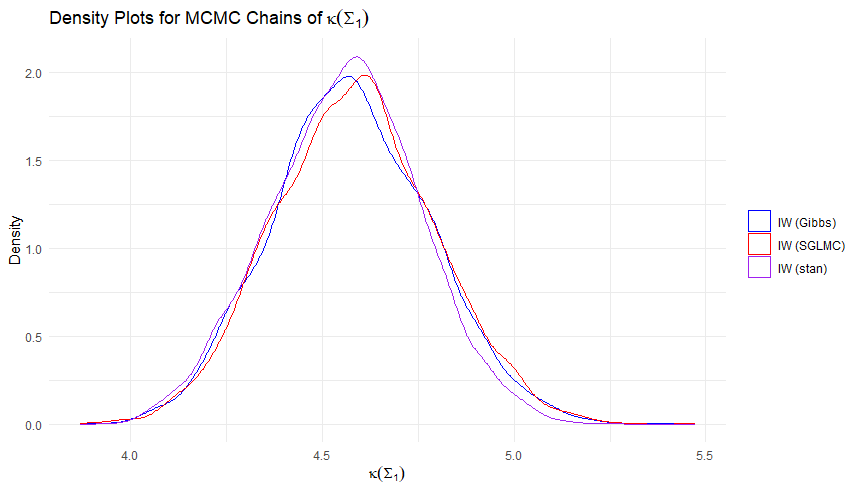} 
    \end{minipage}\hfill
    \begin{minipage}{0.45\textwidth}
        \centering
        \includegraphics[width=\textwidth]{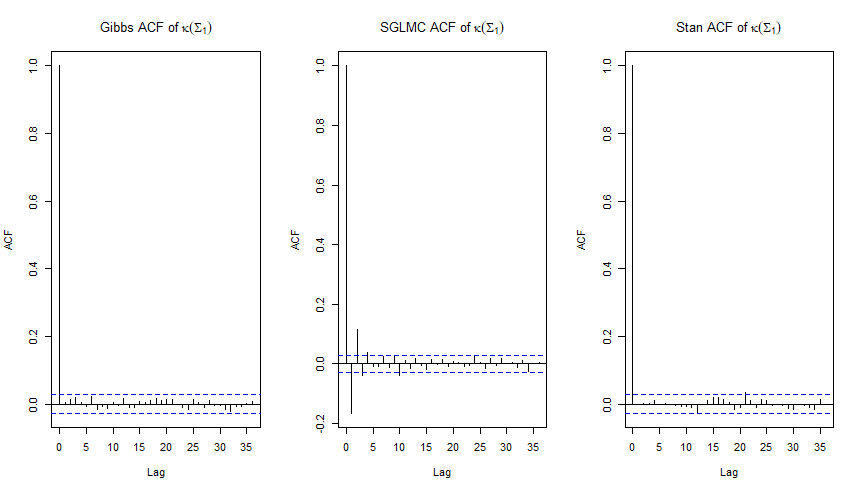} 
    \end{minipage}
           \centering
        \begin{minipage}{0.45\textwidth}
        \centering
        \includegraphics[width=\textwidth]{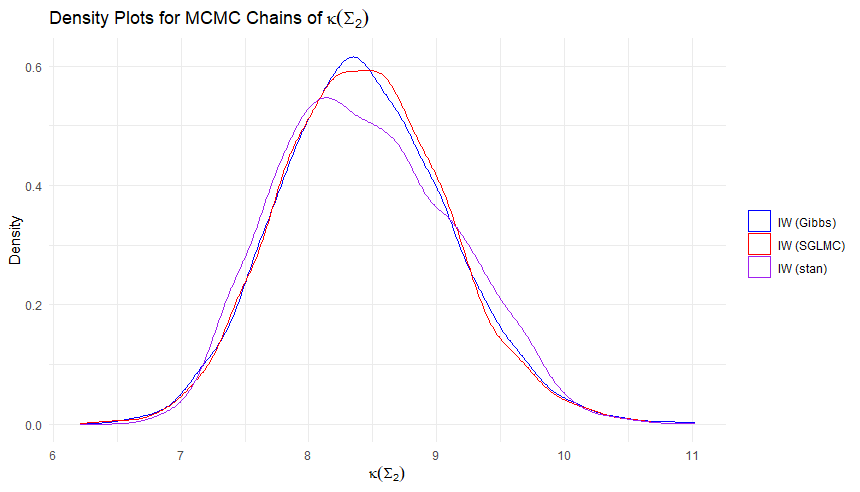} 
    \end{minipage}\hfill
    \begin{minipage}{0.45\textwidth}
        \centering
        \includegraphics[width=\textwidth]{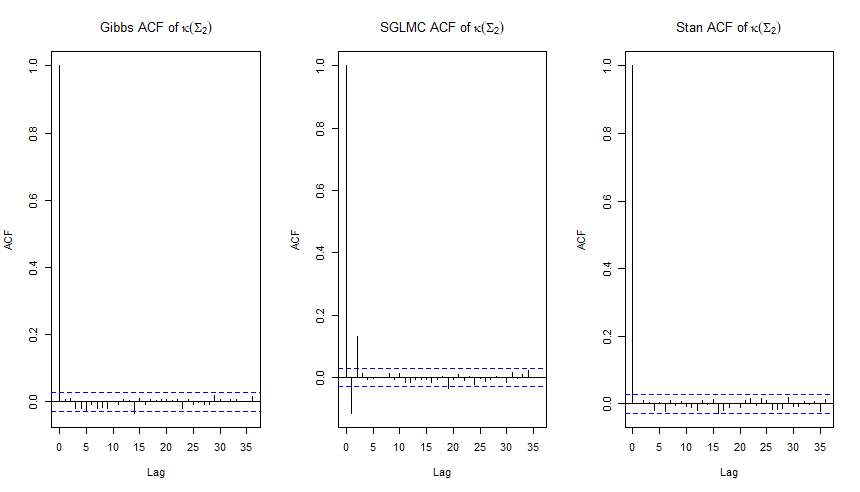} 
    \end{minipage}
               \centering
        \begin{minipage}{0.45\textwidth}
        \centering
        \includegraphics[width=\textwidth]{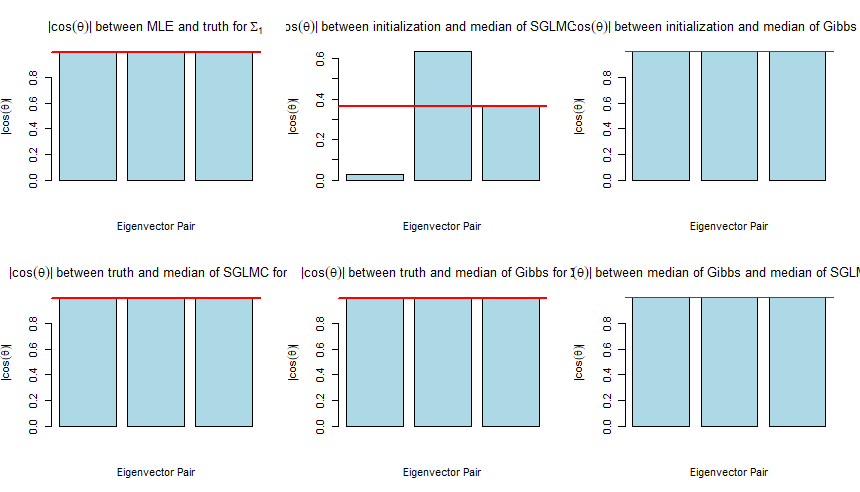} 
    \end{minipage}\hfill
    \begin{minipage}{0.45\textwidth}
        \centering
        \includegraphics[width=\textwidth]{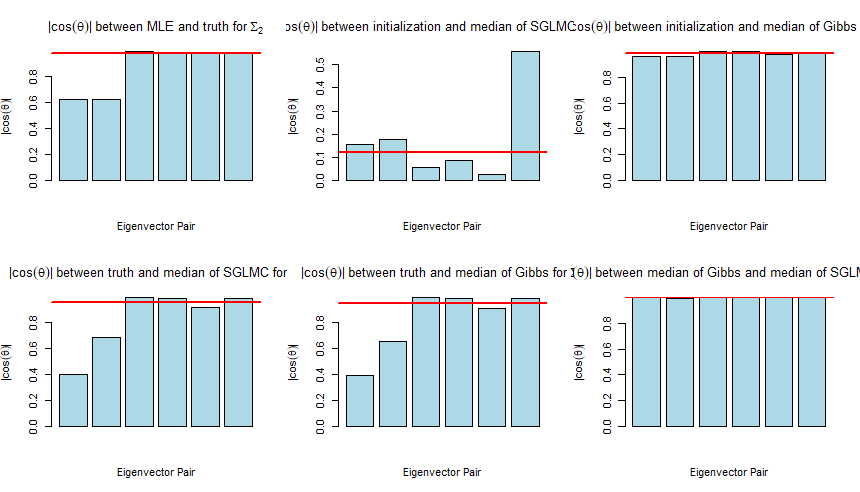} 
    \end{minipage}
    \caption{Density plots of the traces, determinants for $\Sigma$, their corresponding ACFs of the global $\Sigma$, and the condition numbers and eigenvectors comparisons for $\Sigma_{1}, \Sigma_{2}$ for SGLMC, Gibbs, and stan ($d_{1} = 3$, $d_{2} = 6$, $\alpha = .95$).}
    \label{fig: d1 = 3, d2 = 6}
\end{figure}

\begin{figure}[ht]
    \centering
    \begin{minipage}{0.45\textwidth}
        \centering
        \includegraphics[width=\textwidth]{images/15x6/densitytraceplots.png} 
    \end{minipage}\hfill
    \begin{minipage}{0.45\textwidth}
        \centering
        \includegraphics[width=\textwidth]{images/15x6/acftrace.png} 
    \end{minipage}
           \centering
        \begin{minipage}{0.45\textwidth}
        \centering
        \includegraphics[width=\textwidth]{images/15x6/densitydetplots.png} 
    \end{minipage}\hfill
    \begin{minipage}{0.45\textwidth}
        \centering
        \includegraphics[width=\textwidth]{images/15x6/acfdet.png} 
    \end{minipage}
    \centering
    \begin{minipage}{0.45\textwidth}
        \centering
        \includegraphics[width=\textwidth]{images/15x6/densitykappa1.png} 
    \end{minipage}\hfill
    \begin{minipage}{0.45\textwidth}
        \centering
        \includegraphics[width=\textwidth]{images/15x6/acfkappa1.png} 
    \end{minipage}
           \centering
        \begin{minipage}{0.45\textwidth}
        \centering
        \includegraphics[width=\textwidth]{images/15x6/densitykappa2.png} 
    \end{minipage}\hfill
    \begin{minipage}{0.45\textwidth}
        \centering
        \includegraphics[width=\textwidth]{images/15x6/acfkappa2.png} 
    \end{minipage}
               \centering
        \begin{minipage}{0.45\textwidth}
        \centering
        \includegraphics[width=\textwidth]{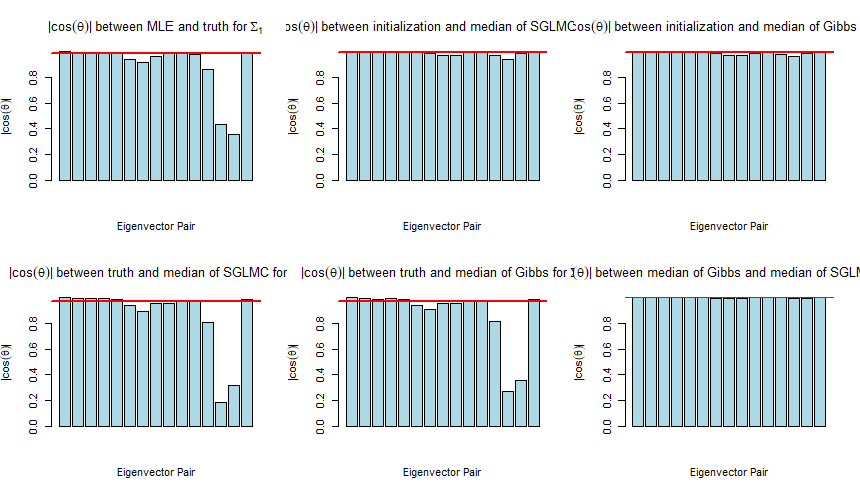} 
    \end{minipage}\hfill
    \begin{minipage}{0.45\textwidth}
        \centering
        \includegraphics[width=\textwidth]{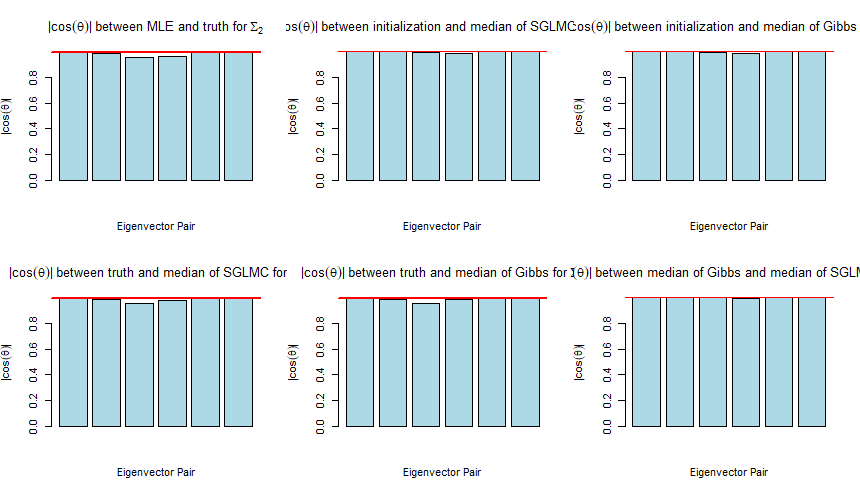} 
    \end{minipage}
    \caption{Density plots of the traces, determinants for $\Sigma$, their corresponding ACFs of the global $\Sigma$, and the condition numbers and eigenvectors comparisons for $\Sigma_{1}, \Sigma_{2}$ for SGLMC, Gibbs, and stan ($d_{1} = 15$, $d_{2} = 6$, $\alpha = .95$).}    
    \label{fig: d1= 15, d2 = 6, alpha = .95}
\end{figure}

\begin{figure}[ht]
    \centering
    \begin{minipage}{0.45\textwidth}
        \centering
        \includegraphics[width=\textwidth]{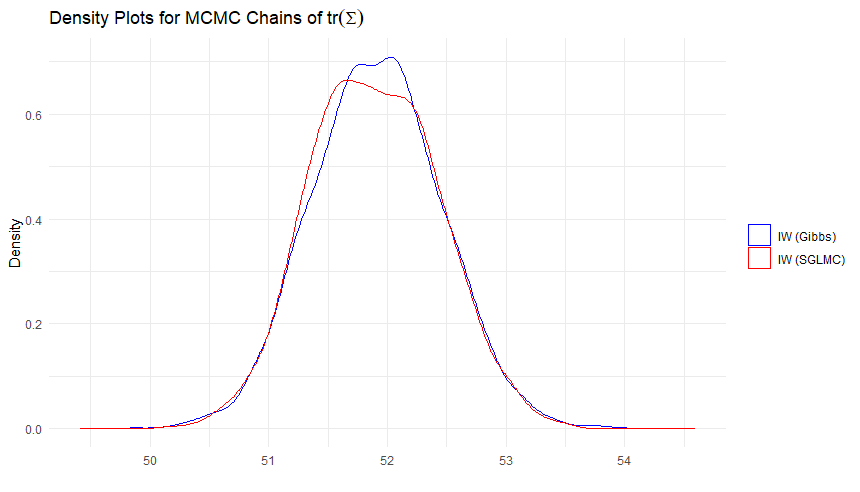} 
    \end{minipage}\hfill
    \begin{minipage}{0.45\textwidth}
        \centering
        \includegraphics[width=\textwidth]{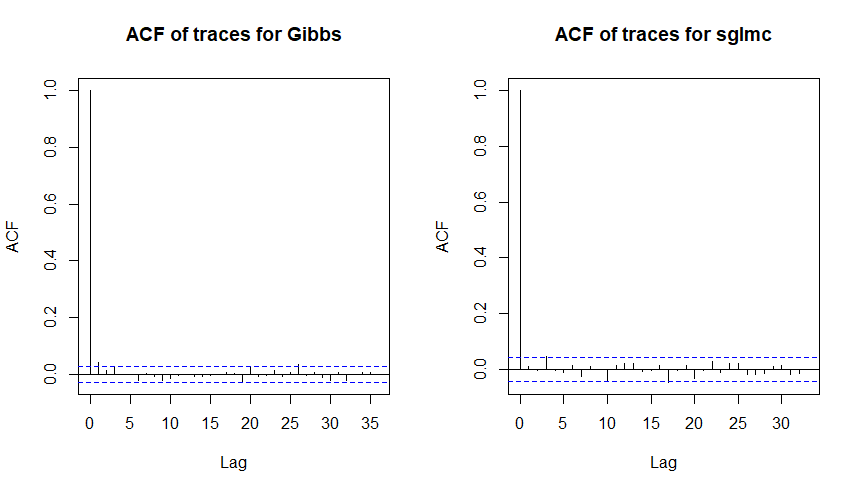} 
    \end{minipage}
           \centering
        \begin{minipage}{0.45\textwidth}
        \centering
        \includegraphics[width=\textwidth]{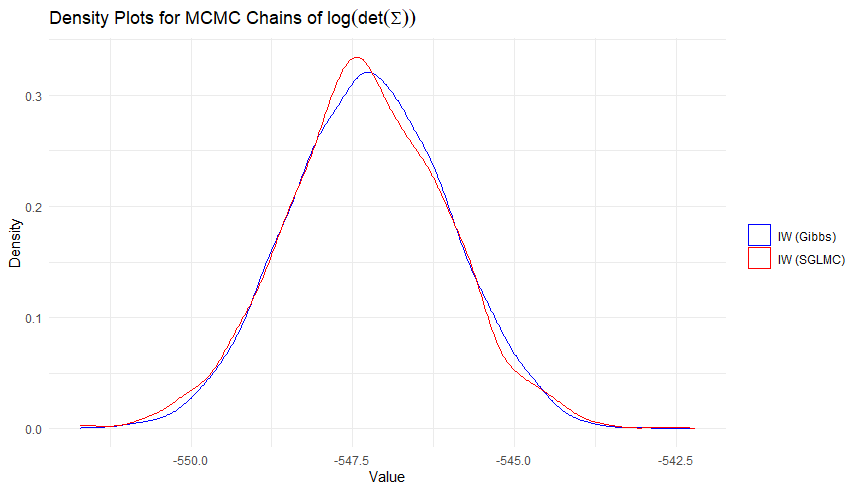} 
    \end{minipage}\hfill
    \begin{minipage}{0.45\textwidth}
        \centering
        \includegraphics[width=\textwidth]{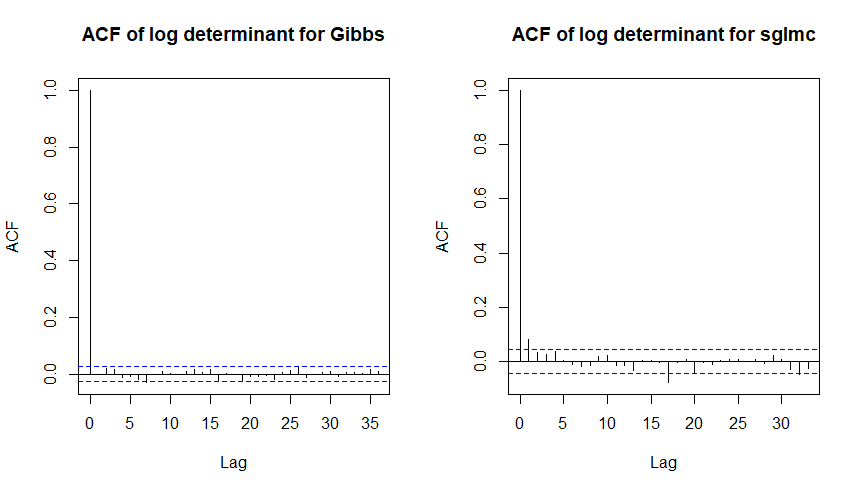} 
    \end{minipage}
    \centering
    \begin{minipage}{0.45\textwidth}
        \centering
        \includegraphics[width=\textwidth]{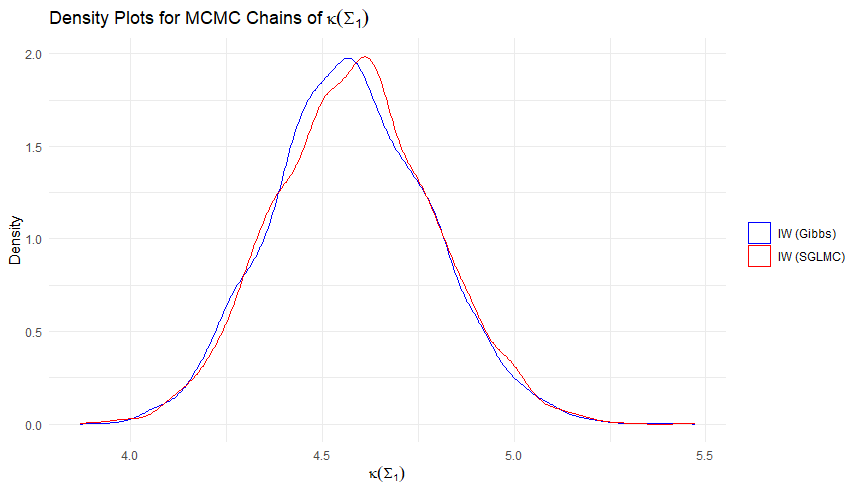} 
    \end{minipage}\hfill
    \begin{minipage}{0.45\textwidth}
        \centering
        \includegraphics[width=\textwidth]{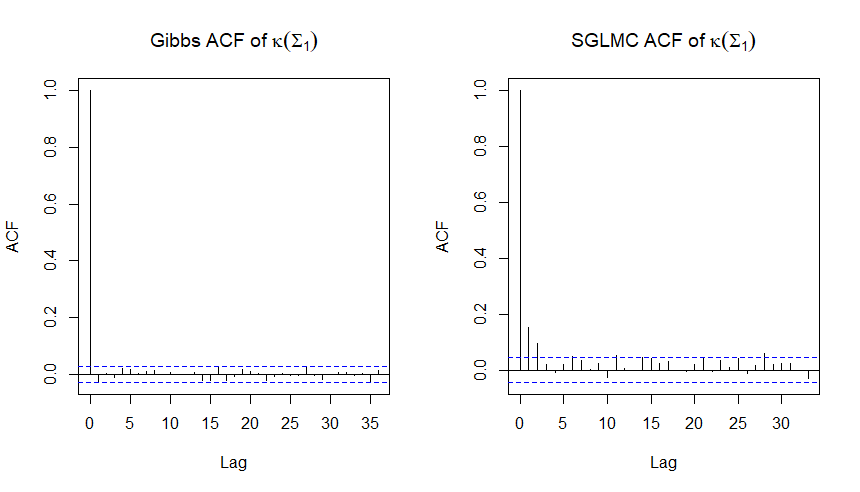} 
    \end{minipage}
           \centering
        \begin{minipage}{0.45\textwidth}
        \centering
        \includegraphics[width=\textwidth]{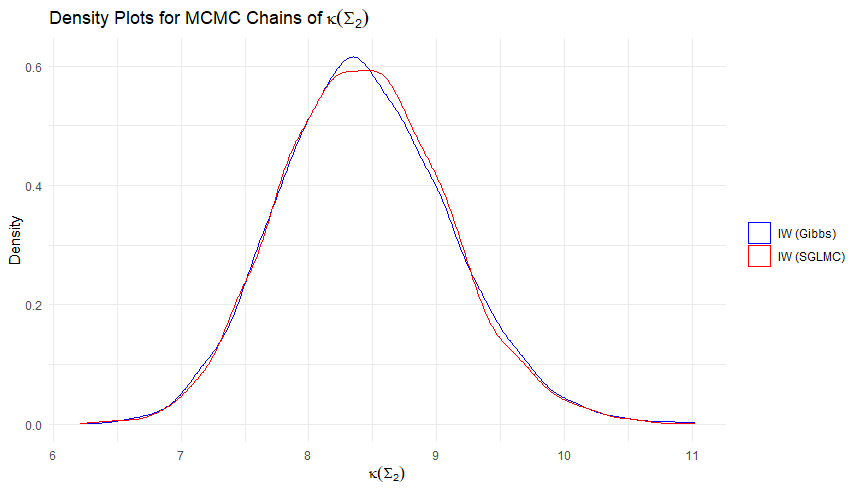} 
    \end{minipage}\hfill
    \begin{minipage}{0.45\textwidth}
        \centering
        \includegraphics[width=\textwidth]{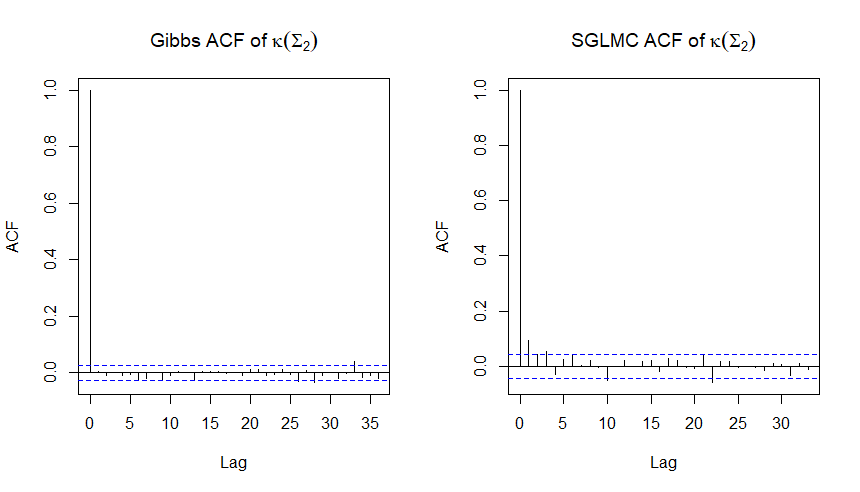} 
    \end{minipage}
               \centering
        \begin{minipage}{0.45\textwidth}
        \centering
        \includegraphics[width=\textwidth]{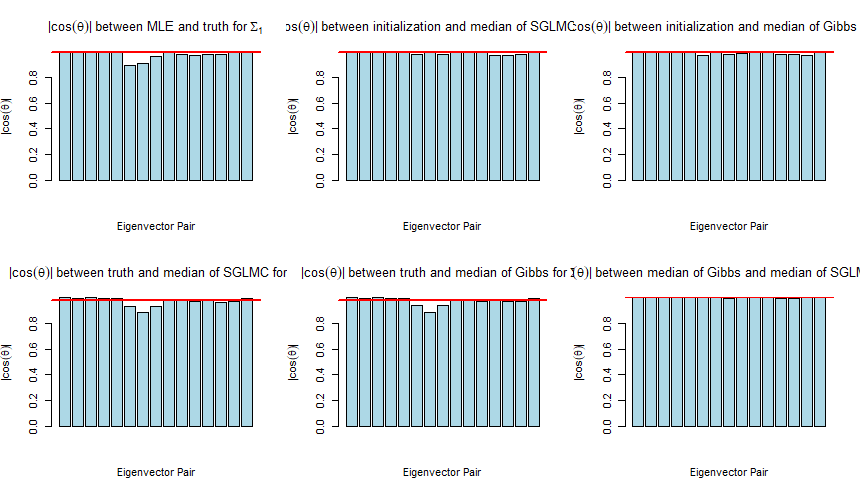} 
    \end{minipage}\hfill
    \begin{minipage}{0.45\textwidth}
        \centering
        \includegraphics[width=\textwidth]{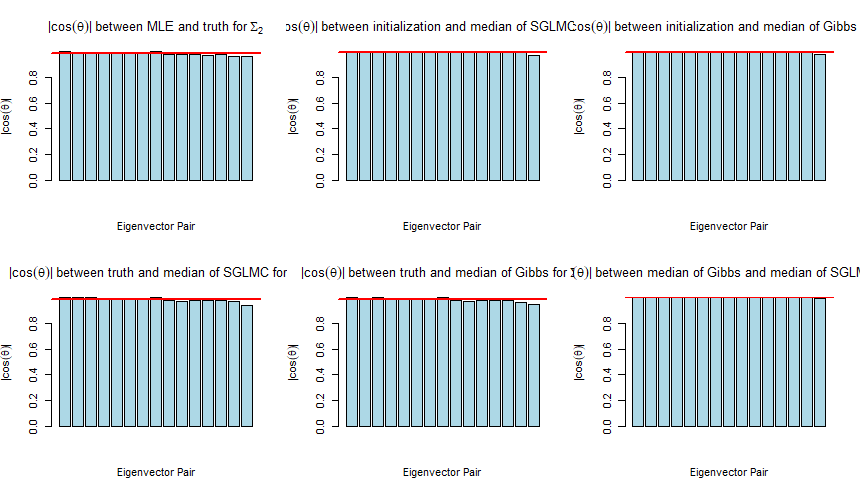} 
    \end{minipage}\hfill
    \caption{Density plots of the traces, determinants for $\Sigma$, their corresponding ACFs of the global $\Sigma$, and the condition numbers and eigenvectors comparisons for $\Sigma_{1}, \Sigma_{2}$ for SGLMC, Gibbs, and stan ($d_{1} = 15$, $d_{2} = 15$, $\alpha = .95$).}
    \label{fig: d1 = 15, d2 = 15, alpha = .95}
\end{figure}

\subsection{Comparisons of Regularization parameter}
\begin{figure}[ht]  
    \centering
    \begin{minipage}{0.45\textwidth}
        \centering
        \includegraphics[width=\textwidth]{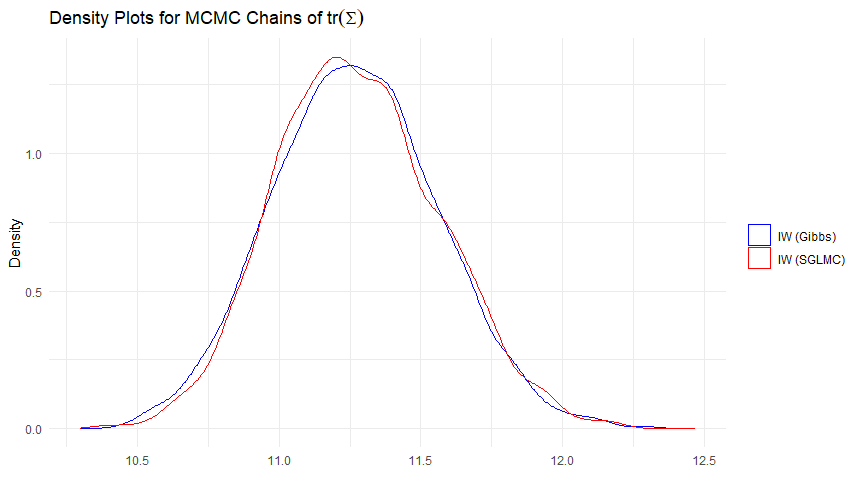} 
    \end{minipage}\hfill
    \begin{minipage}{0.45\textwidth}
        \centering
        \includegraphics[width=\textwidth]{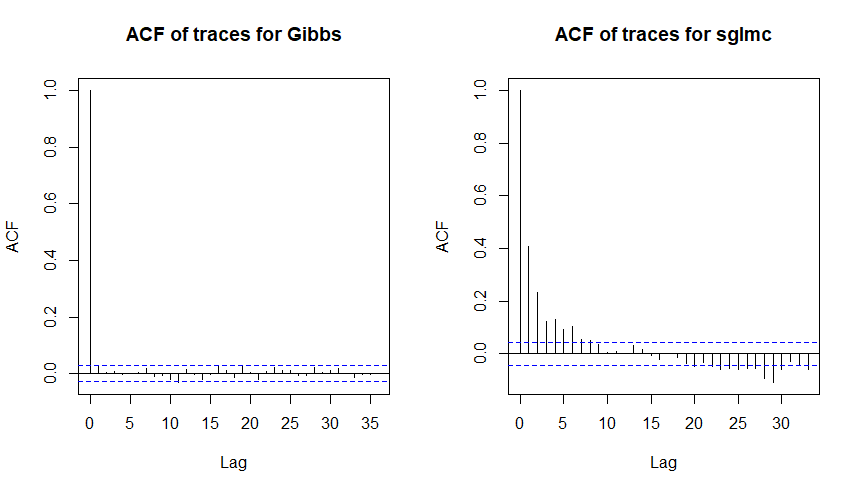} 
    \end{minipage}
           \centering
        \begin{minipage}{0.45\textwidth}
        \centering
        \includegraphics[width=\textwidth]{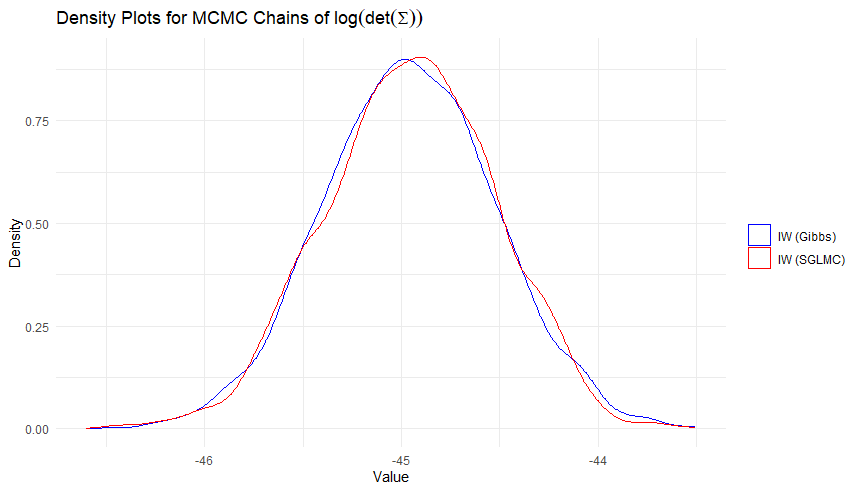} 
    \end{minipage}\hfill
    \begin{minipage}{0.45\textwidth}
        \centering
        \includegraphics[width=\textwidth]{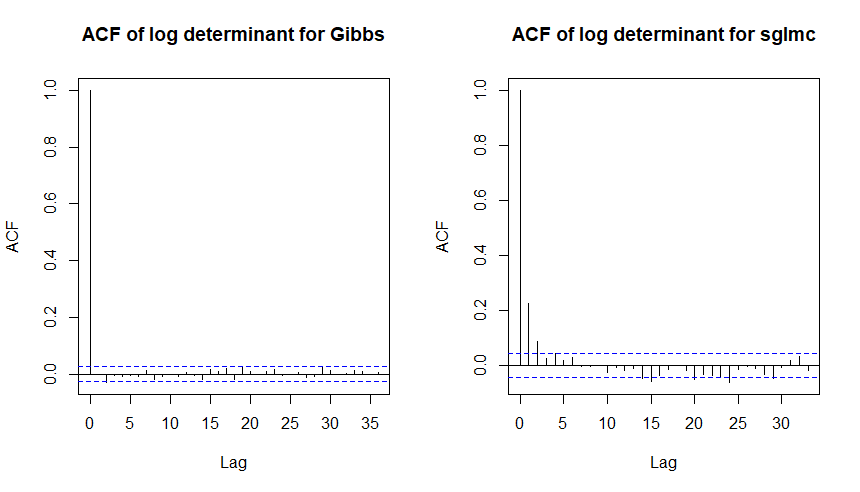} 
    \end{minipage}
        \centering
    \begin{minipage}{0.45\textwidth}
        \centering
        \includegraphics[width=\textwidth]{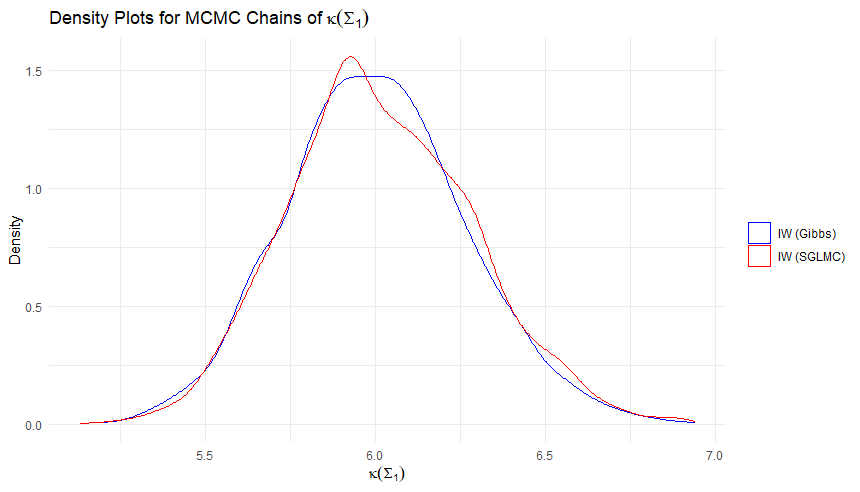} 
    \end{minipage}\hfill
    \begin{minipage}{0.45\textwidth}
        \centering
        \includegraphics[width=\textwidth]{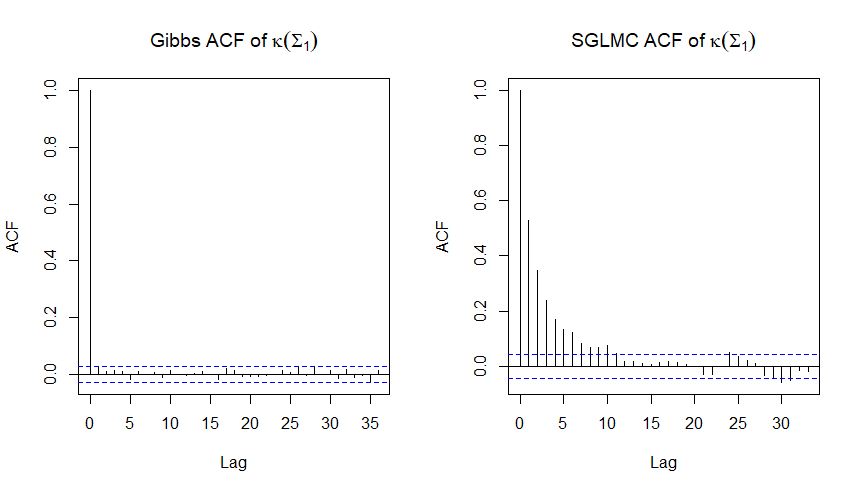} 
    \end{minipage}
        \centering
    \begin{minipage}{0.45\textwidth}
        \centering
        \includegraphics[width=\textwidth]{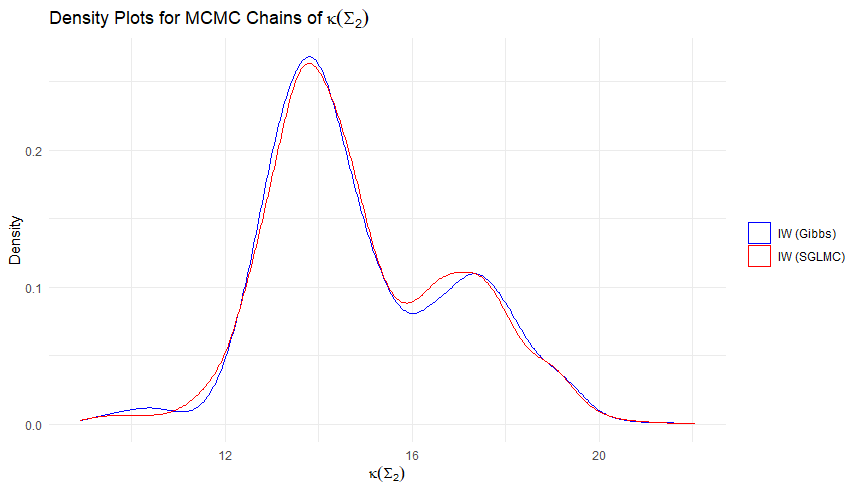} 
    \end{minipage}\hfill
    \begin{minipage}{0.45\textwidth}
        \centering
        \includegraphics[width=\textwidth]{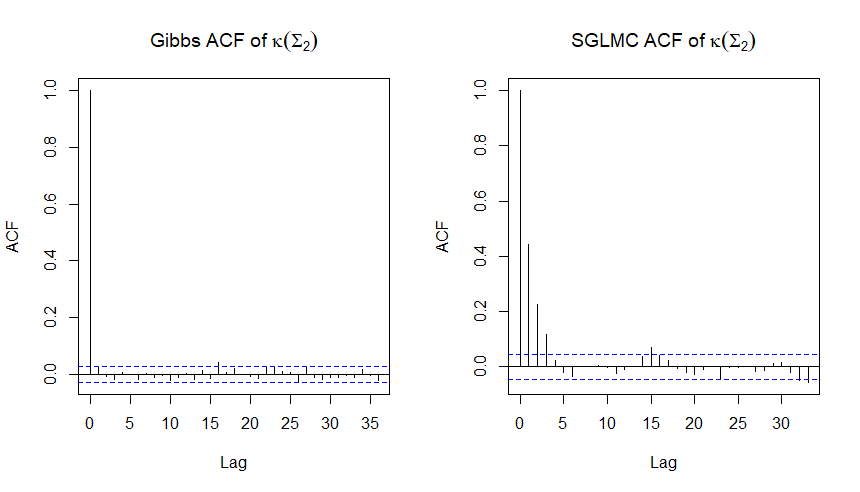} 
    \end{minipage}
        \begin{minipage}{0.45\textwidth}
        \centering
        \includegraphics[width=\textwidth]{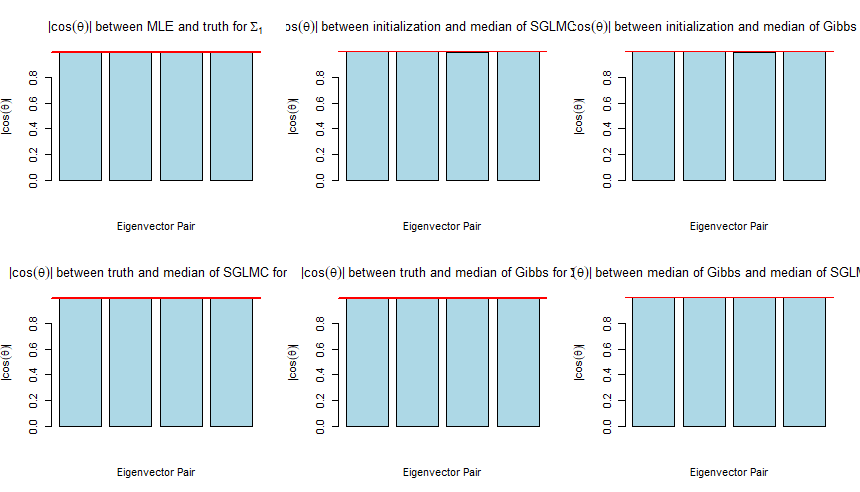} 
    \end{minipage}\hfill
    \begin{minipage}{0.45\textwidth}
        \centering
        \includegraphics[width=\textwidth]{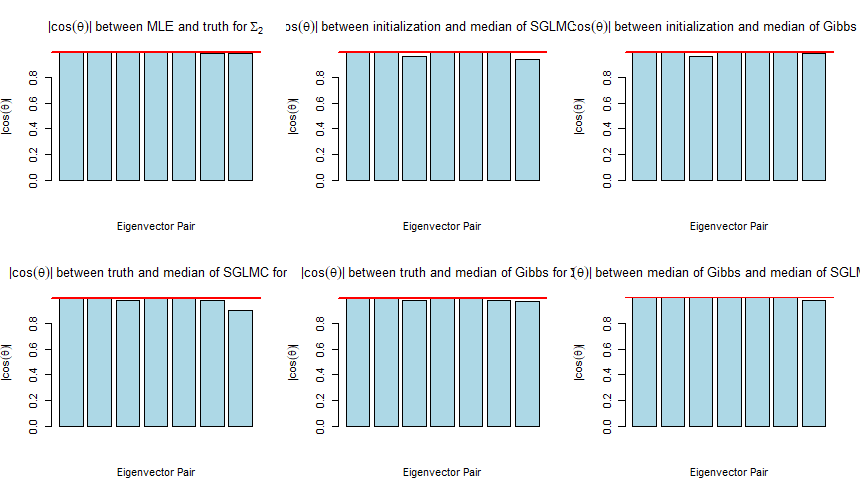} 
    \end{minipage}
    
    \caption{Density plots of the traces, determinants for $\Sigma$, their corresponding ACFs of the global $\Sigma$, and the condition number comparisons for $\Sigma_{1}, \Sigma_{2}$ for SGLMC, Gibbs, and stan ($d_{1} = 4$, $d_{2} = 7$, $\alpha = 0$).}
    \label{fig: alpha = 0}
\end{figure}

\begin{figure}[ht]  
    \centering
    \begin{minipage}{0.45\textwidth}
        \centering
        \includegraphics[width=\textwidth]{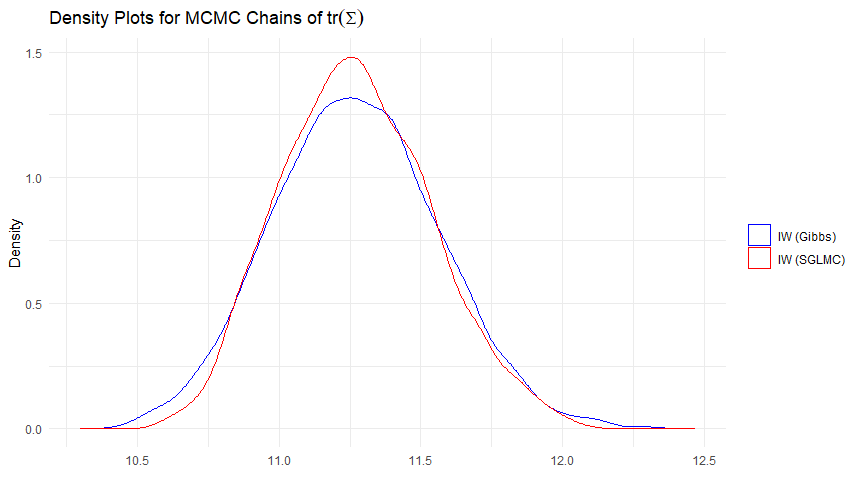} 
    \end{minipage}\hfill
    \begin{minipage}{0.45\textwidth}
        \centering
        \includegraphics[width=\textwidth]{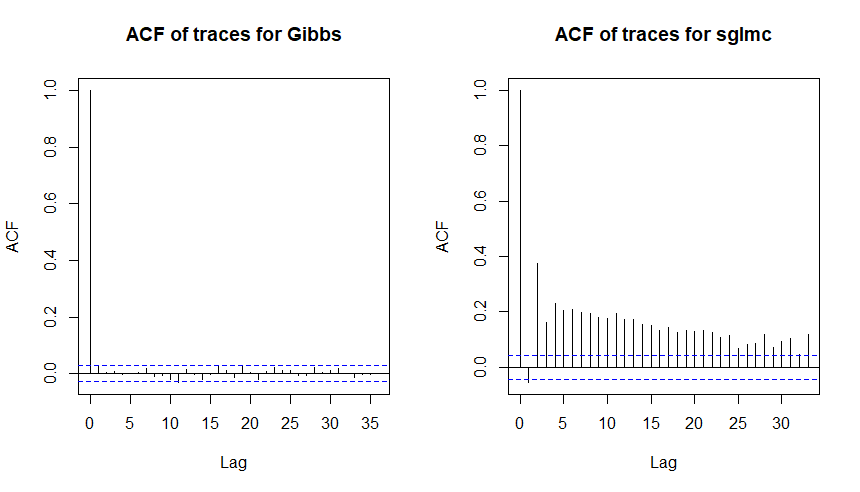} 
    \end{minipage}
           \centering
        \begin{minipage}{0.45\textwidth}
        \centering
        \includegraphics[width=\textwidth]{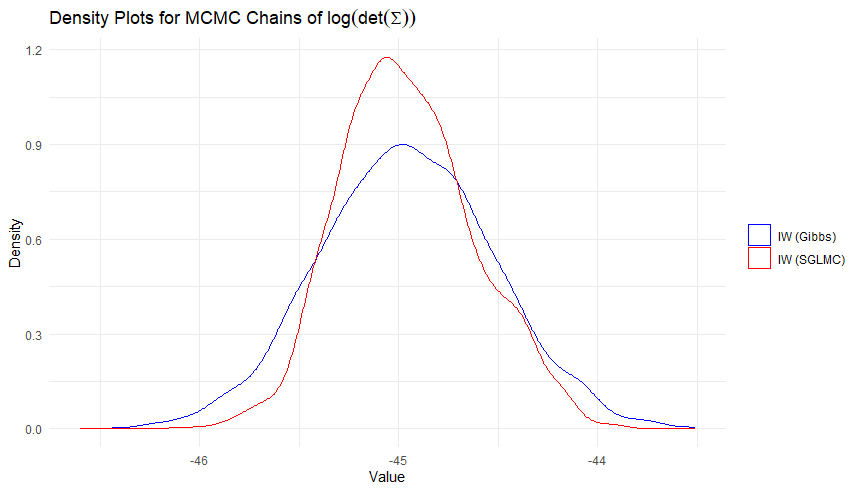} 
    \end{minipage}\hfill
    \begin{minipage}{0.45\textwidth}
        \centering
        \includegraphics[width=\textwidth]{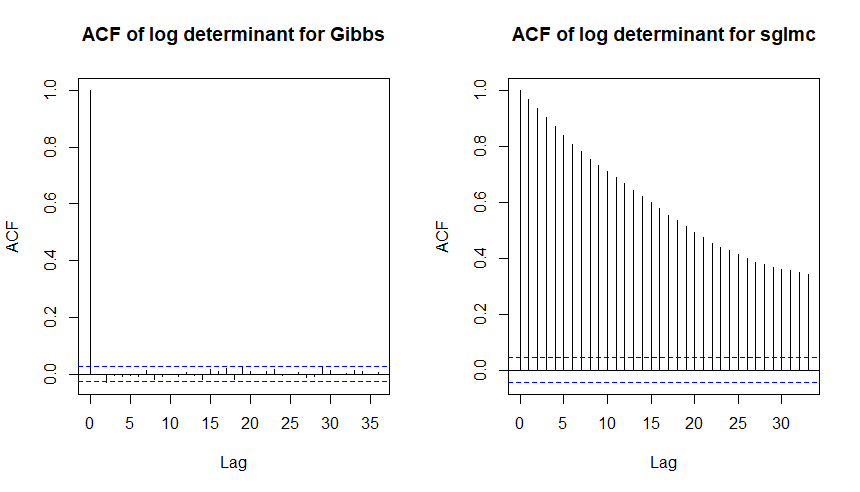} 
    \end{minipage}
        \centering
    \begin{minipage}{0.45\textwidth}
        \centering
        \includegraphics[width=\textwidth]{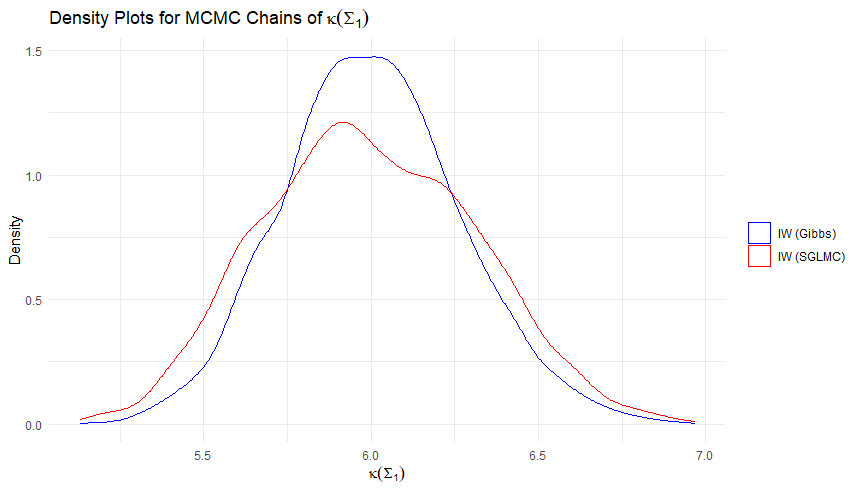} 
    \end{minipage}\hfill
    \begin{minipage}{0.45\textwidth}
        \centering
        \includegraphics[width=\textwidth]{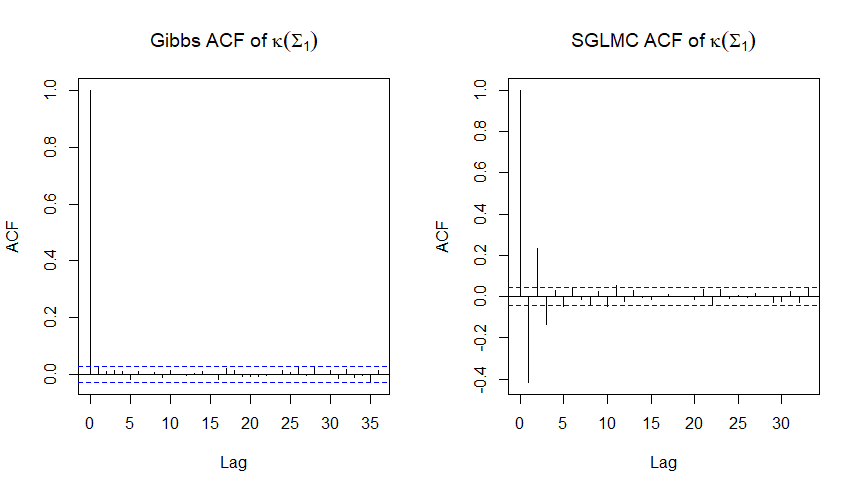} 
    \end{minipage}
        \centering
    \begin{minipage}{0.45\textwidth}
        \centering
        \includegraphics[width=\textwidth]{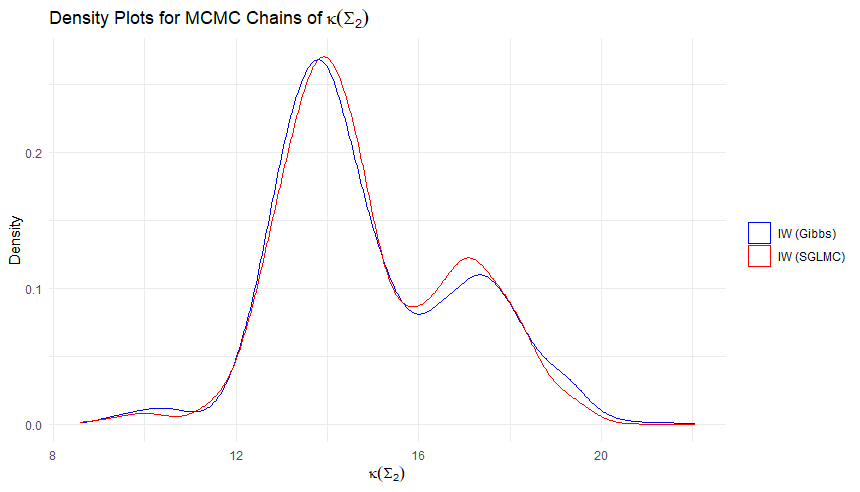} 
    \end{minipage}\hfill
    \begin{minipage}{0.45\textwidth}
        \centering
        \includegraphics[width=\textwidth]{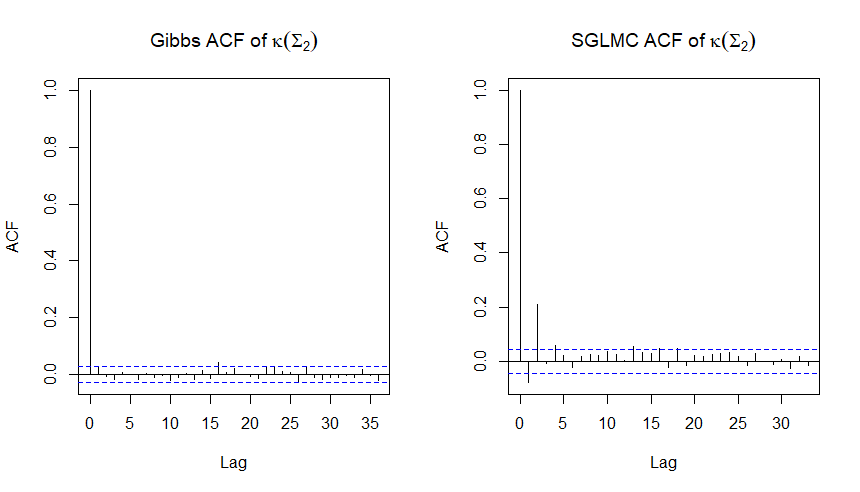} 
    \end{minipage}
        \begin{minipage}{0.45\textwidth}
        \centering
        \includegraphics[width=\textwidth]{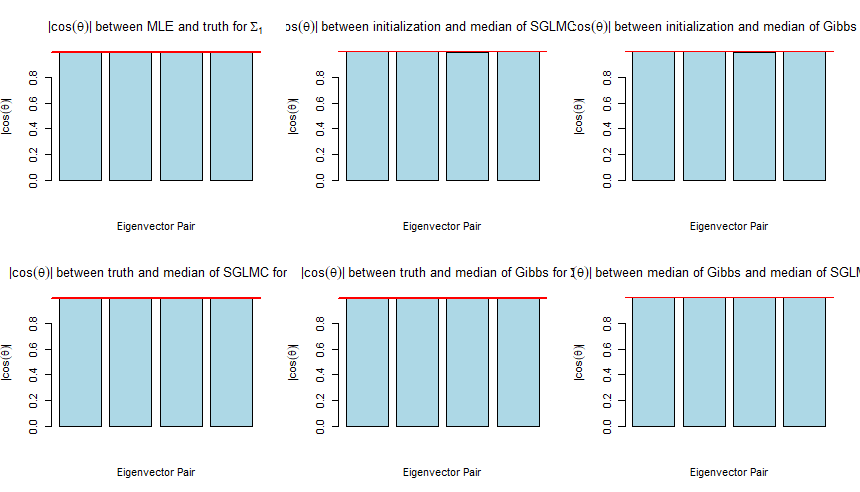} 
    \end{minipage}\hfill
    \begin{minipage}{0.45\textwidth}
        \centering
        \includegraphics[width=\textwidth]{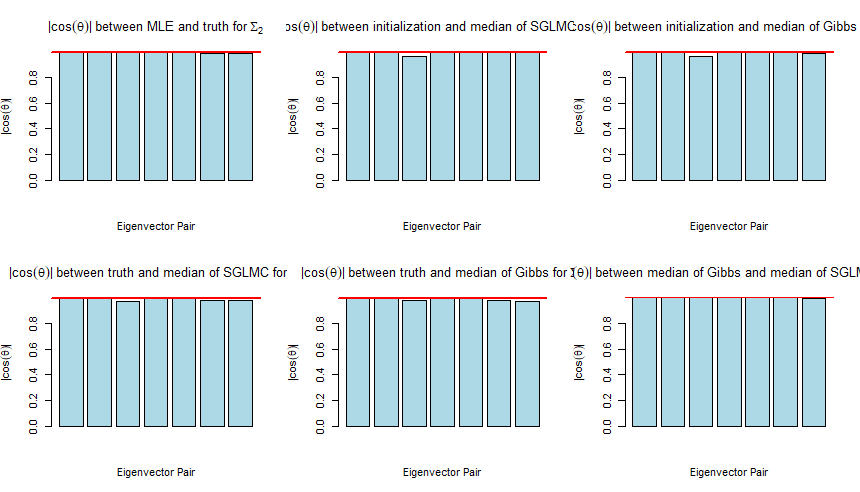} 
    \end{minipage}
    \caption{Density plots of the traces, determinants for $\Sigma$, their corresponding ACFs of the global $\Sigma$, and the condition number comparisons for $\Sigma_{1}, \Sigma_{2}$ for SGLMC, Gibbs, and stan ($d_{1} = 4$, $d_{2} = 7$, $\alpha = .1$).}
    \label{fig: alpha = .1}
\end{figure}

\begin{figure}[ht]
    \centering
    \begin{minipage}{0.45\textwidth}
        \centering
        \includegraphics[width=\textwidth]{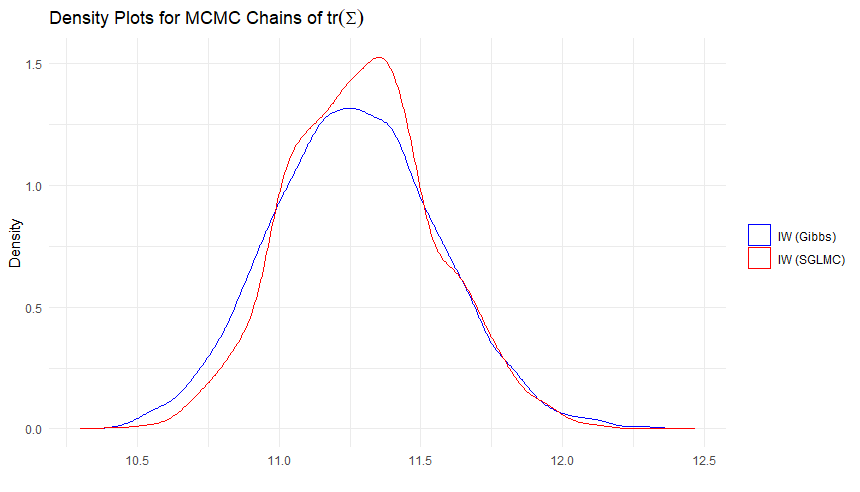} 
    \end{minipage}\hfill
    \begin{minipage}{0.45\textwidth}
        \centering
        \includegraphics[width=\textwidth]{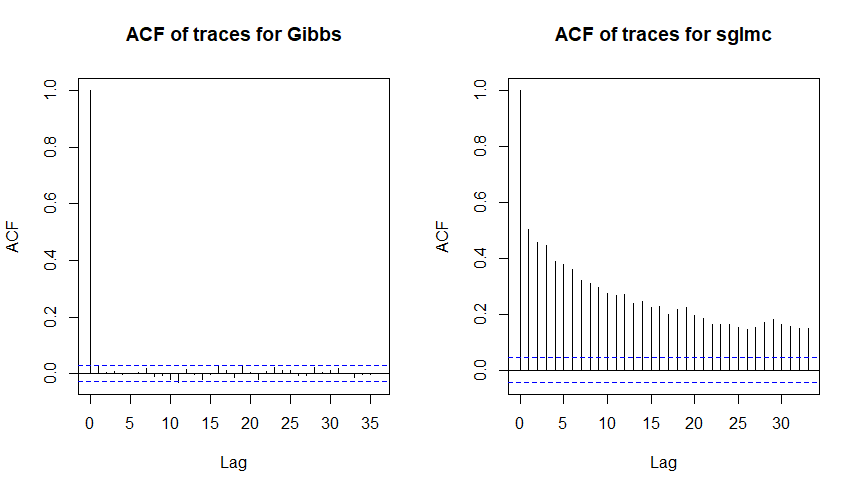} 
    \end{minipage}
           \centering
        \begin{minipage}{0.45\textwidth}
        \centering
        \includegraphics[width=\textwidth]{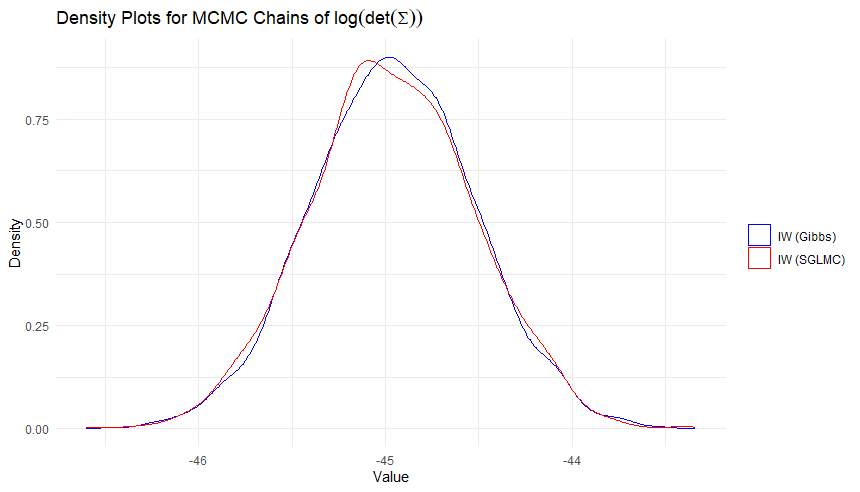} 
    \end{minipage}\hfill
    \begin{minipage}{0.45\textwidth}
        \centering
        \includegraphics[width=\textwidth]{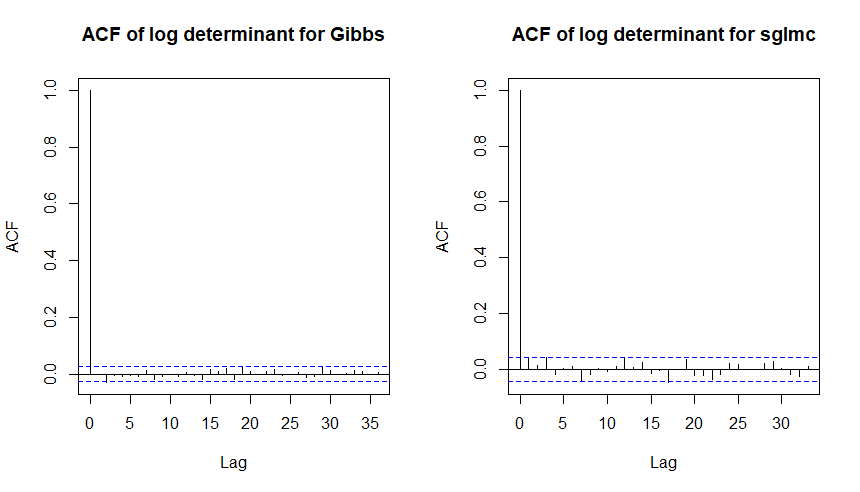} 
    \end{minipage}
        \centering
    \begin{minipage}{0.45\textwidth}
        \centering
        \includegraphics[width=\textwidth]{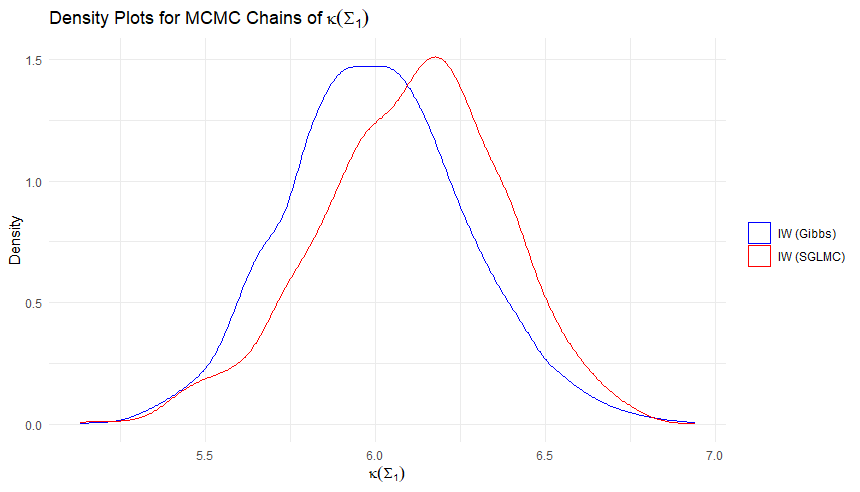} 
    \end{minipage}\hfill
    \begin{minipage}{0.45\textwidth}
        \centering
        \includegraphics[width=\textwidth]{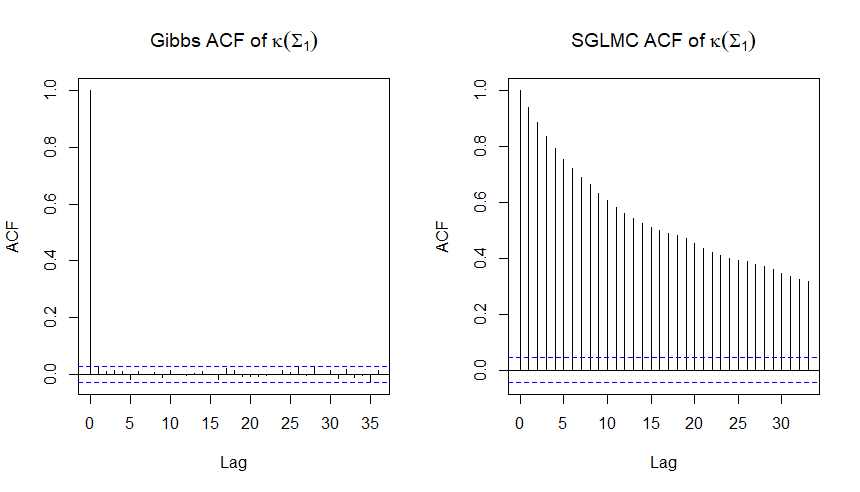} 
    \end{minipage}
        \centering
    \begin{minipage}{0.45\textwidth}
        \centering
        \includegraphics[width=\textwidth]{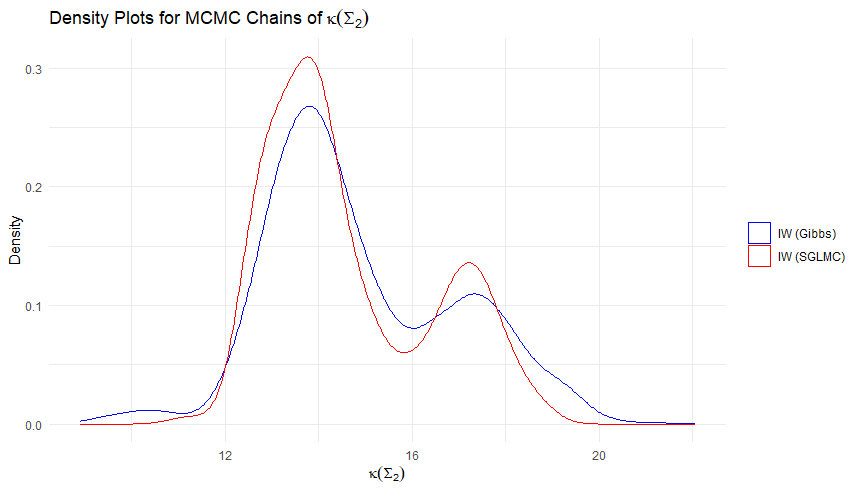} 
    \end{minipage}\hfill
    \begin{minipage}{0.45\textwidth}
        \centering
        \includegraphics[width=\textwidth]{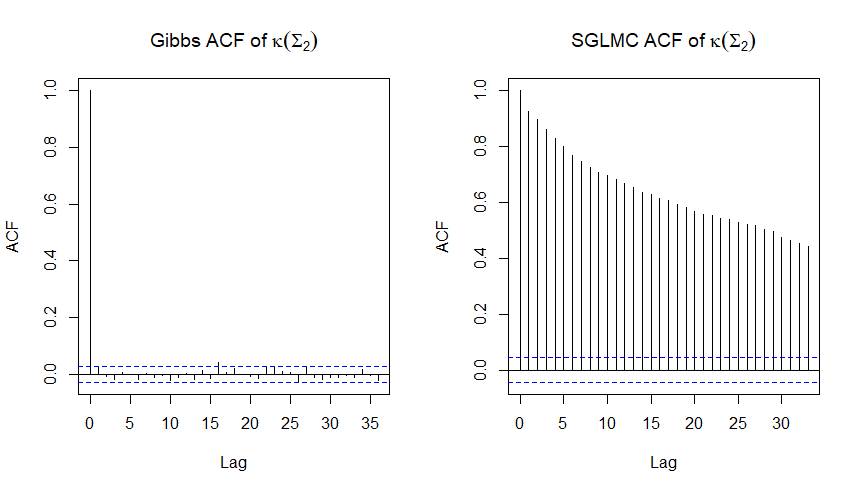} 
    \end{minipage}
        \begin{minipage}{0.45\textwidth}
        \centering
        \includegraphics[width=\textwidth]{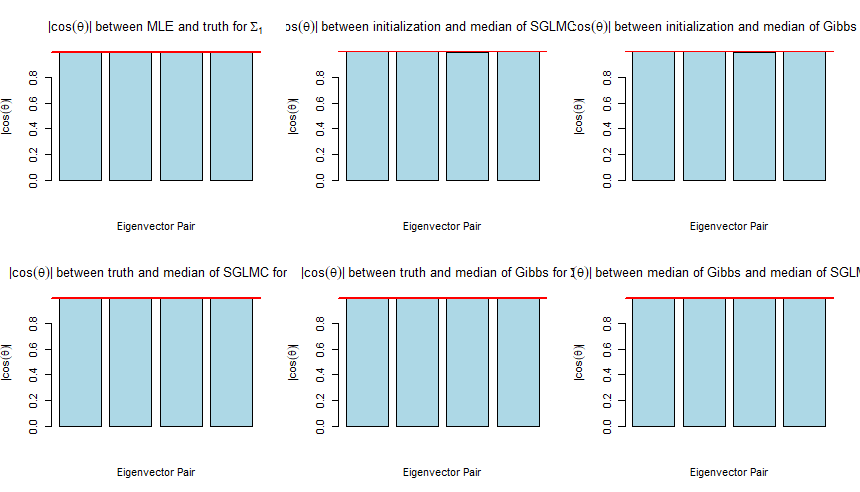} 
    \end{minipage}\hfill
    \begin{minipage}{0.45\textwidth}
        \centering
        \includegraphics[width=\textwidth]{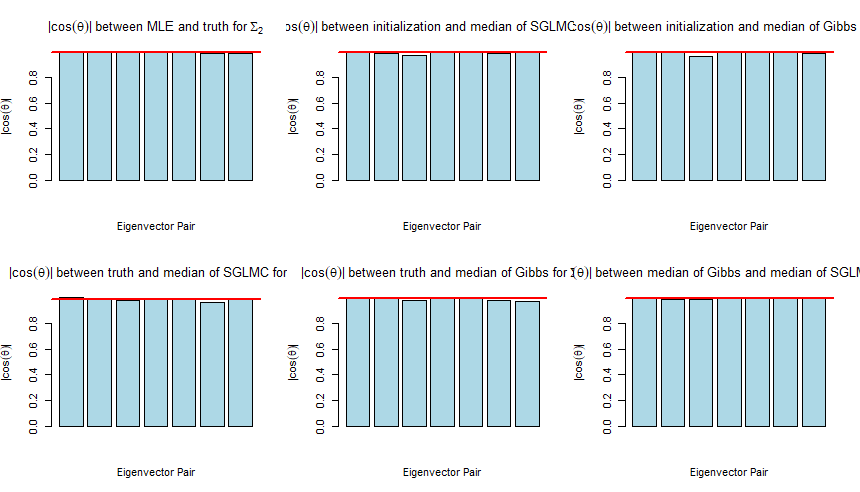} 
    \end{minipage}
    \caption{Density plots of the traces, determinants for $\Sigma$, their corresponding ACFs of the global $\Sigma$, and the condition number comparisons for $\Sigma_{1}, \Sigma_{2}$ for SGLMC, Gibbs, and stan ($d_{1} = 4$, $d_{2} = 7$, $\alpha = .25$).}
    \label{fig: alpha = .25}
\end{figure}

\begin{figure}[ht]
    \centering
    \begin{minipage}{0.45\textwidth}
        \centering
        \includegraphics[width=\textwidth]{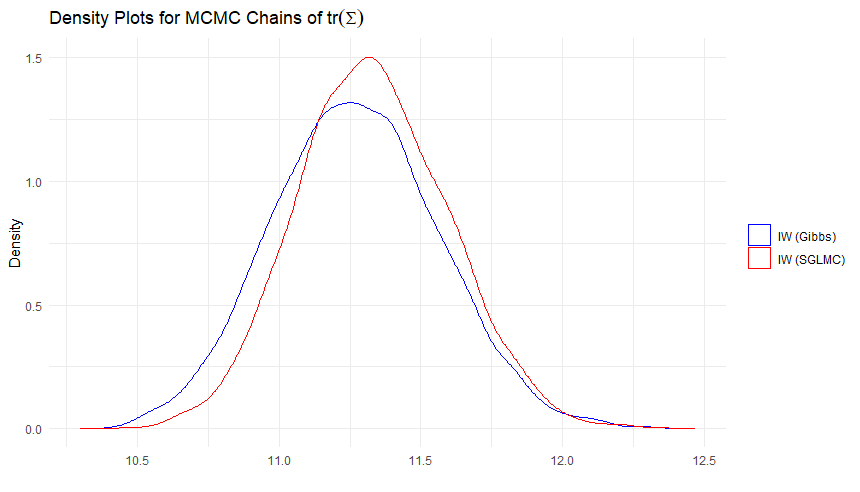} 
    \end{minipage}\hfill
    \begin{minipage}{0.45\textwidth}
        \centering
        \includegraphics[width=\textwidth]{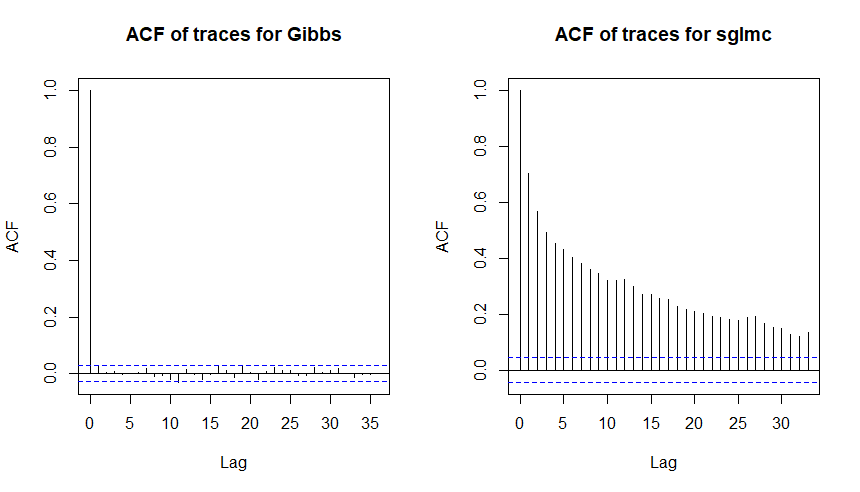} 
    \end{minipage}
           \centering
        \begin{minipage}{0.45\textwidth}
        \centering
        \includegraphics[width=\textwidth]{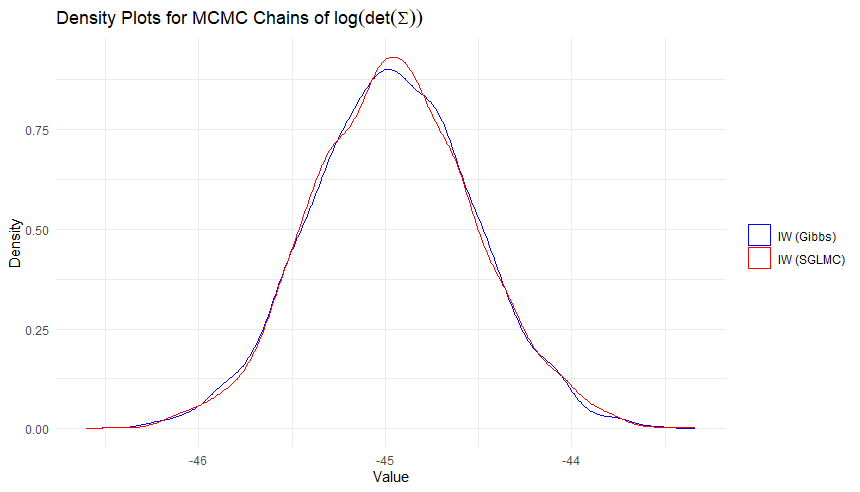} 
    \end{minipage}\hfill
    \begin{minipage}{0.45\textwidth}
        \centering
        \includegraphics[width=\textwidth]{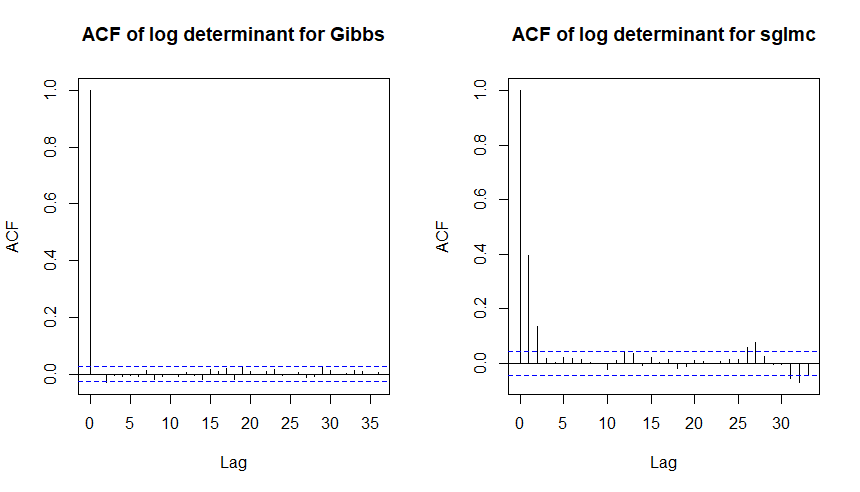} 
    \end{minipage}
        \centering
    \begin{minipage}{0.45\textwidth}
        \centering
        \includegraphics[width=\textwidth]{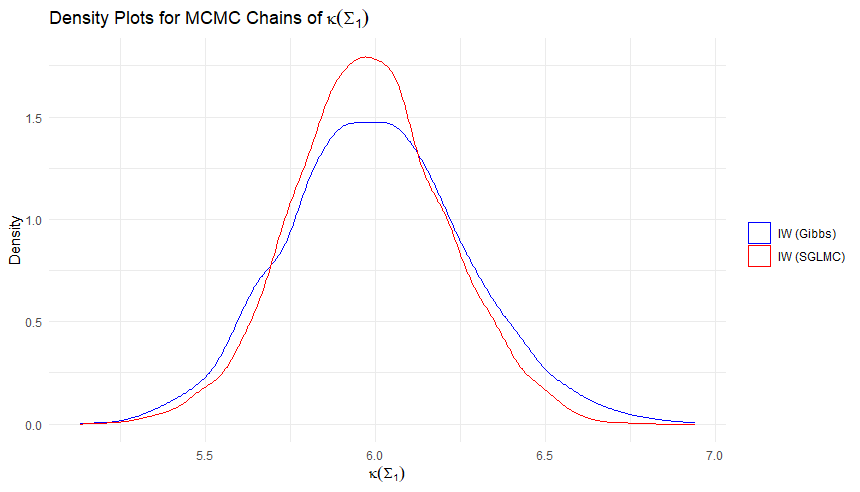} 
    \end{minipage}\hfill
    \begin{minipage}{0.45\textwidth}
        \centering
        \includegraphics[width=\textwidth]{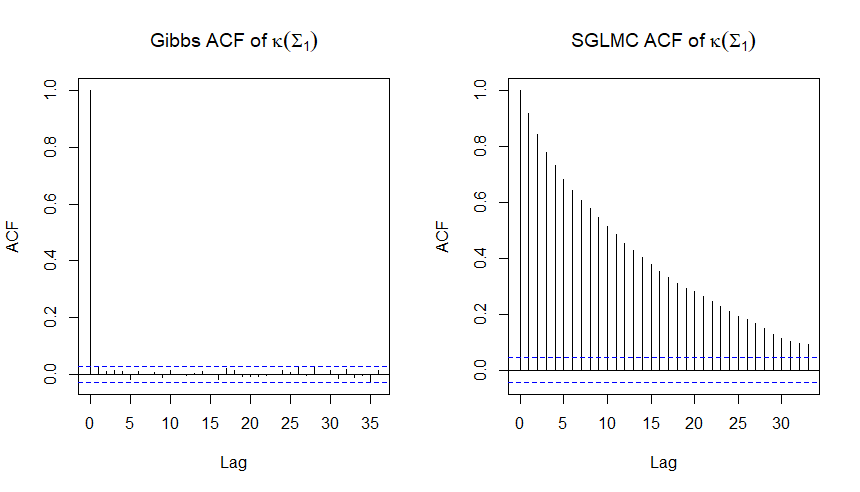} 
    \end{minipage}
        \centering
    \begin{minipage}{0.45\textwidth}
        \centering
        \includegraphics[width=\textwidth]{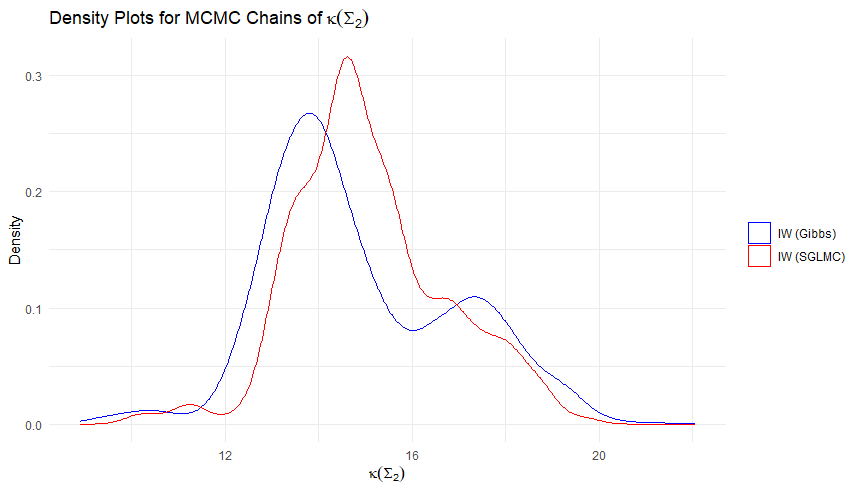} 
    \end{minipage}\hfill
    \begin{minipage}{0.45\textwidth}
        \centering
        \includegraphics[width=\textwidth]{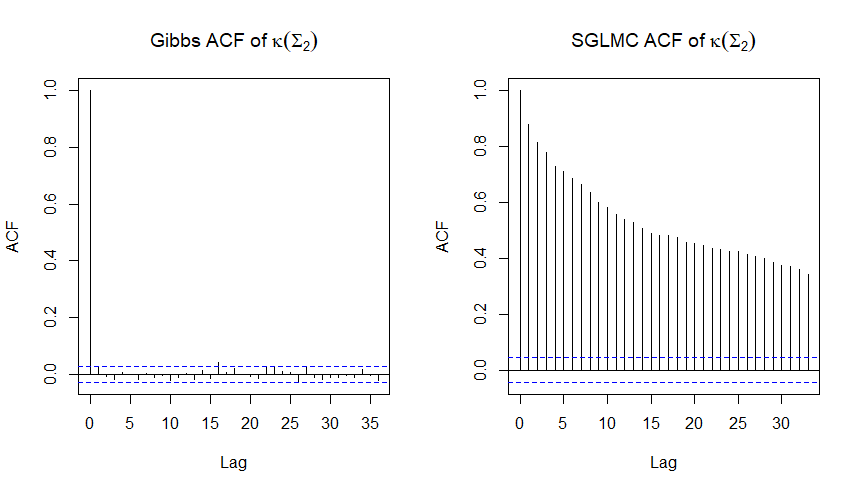} 
    \end{minipage}
        \begin{minipage}{0.45\textwidth}
        \centering
        \includegraphics[width=\textwidth]{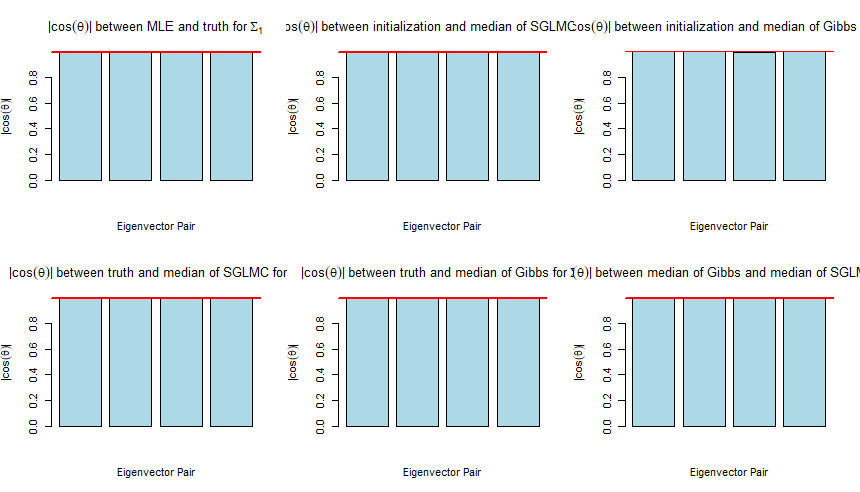} 
    \end{minipage}\hfill
    \begin{minipage}{0.45\textwidth}
        \centering
        \includegraphics[width=\textwidth]{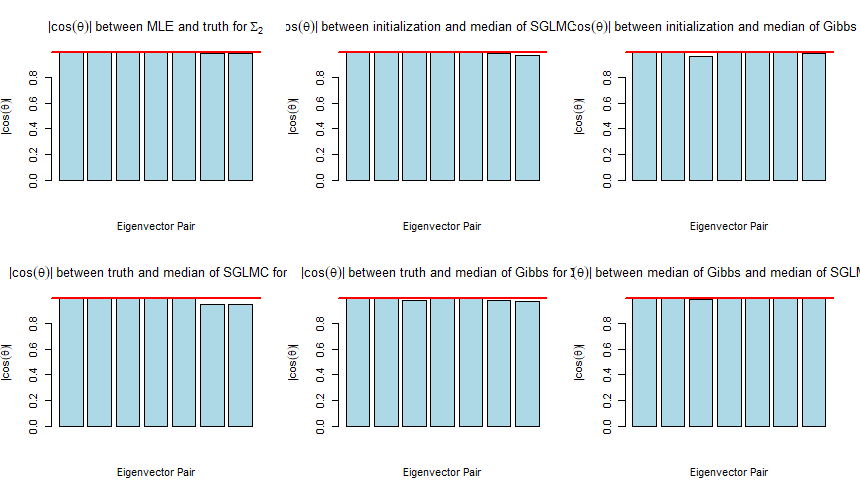} 
    \end{minipage}
    \caption{Density plots of the traces, determinants for $\Sigma$, their corresponding ACFs of the global $\Sigma$, and the condition number comparisons for $\Sigma_{1}, \Sigma_{2}$ for SGLMC, Gibbs, and stan ($d_{1} = 4$, $d_{2} = 7$, $\alpha = .5$).}
    \label{fig: alpha = .5}
\end{figure}

\begin{figure}[ht]
    \centering
    \begin{minipage}{0.45\textwidth}
        \centering
        \includegraphics[width=\textwidth]{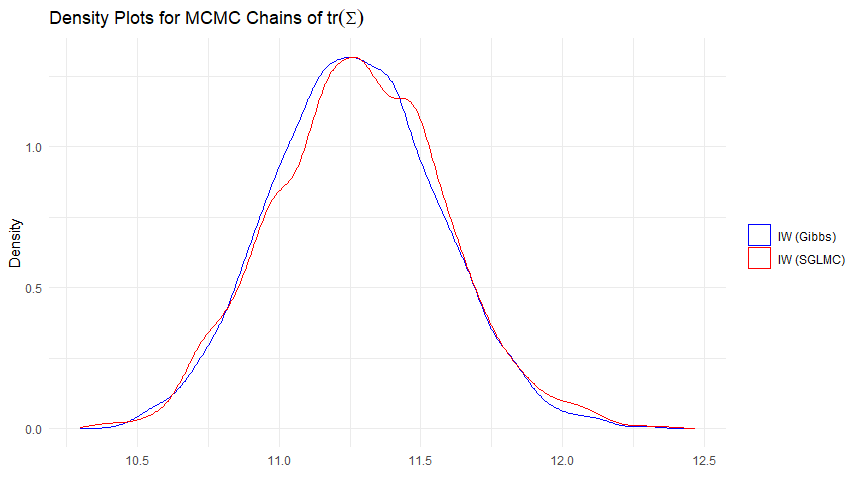} 
    \end{minipage}\hfill
    \begin{minipage}{0.45\textwidth}
        \centering
        \includegraphics[width=\textwidth]{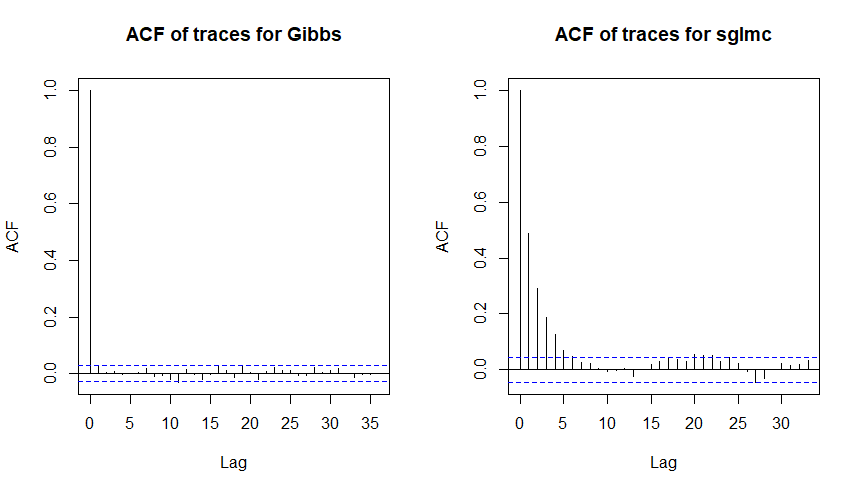} 
    \end{minipage}
           \centering
        \begin{minipage}{0.45\textwidth}
        \centering
        \includegraphics[width=\textwidth]{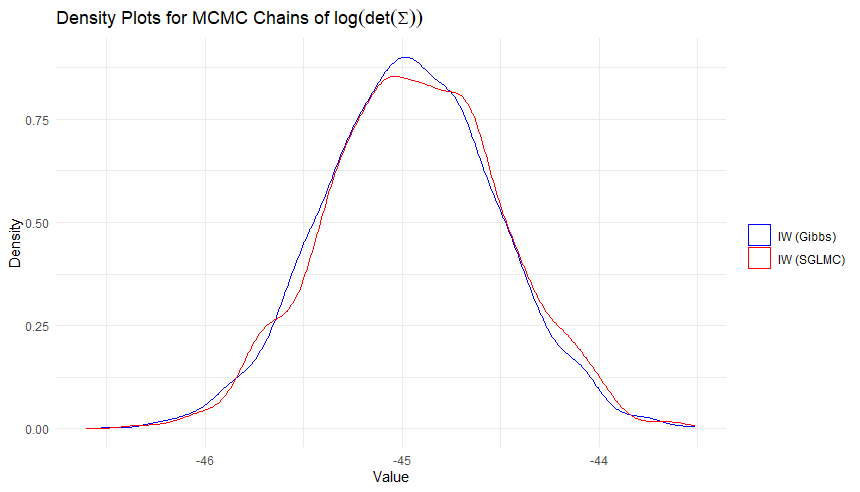} 
    \end{minipage}\hfill
    \begin{minipage}{0.45\textwidth}
        \centering
        \includegraphics[width=\textwidth]{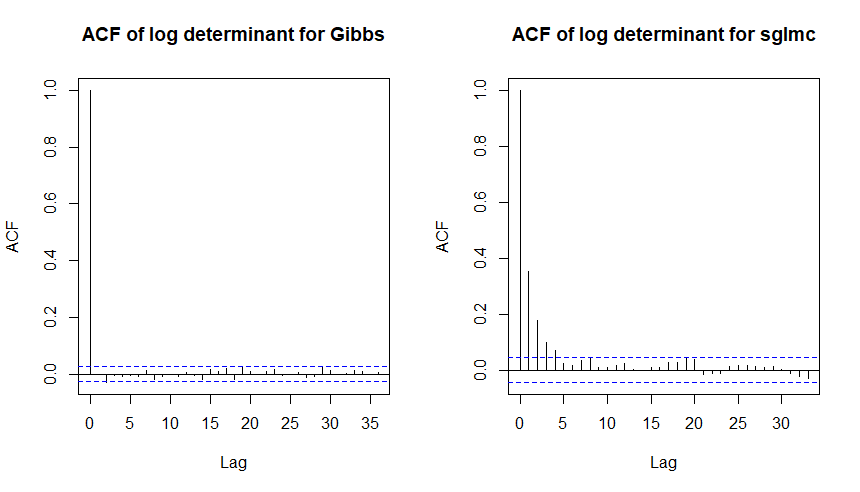} 
    \end{minipage}
        \centering
    \begin{minipage}{0.45\textwidth}
        \centering
        \includegraphics[width=\textwidth]{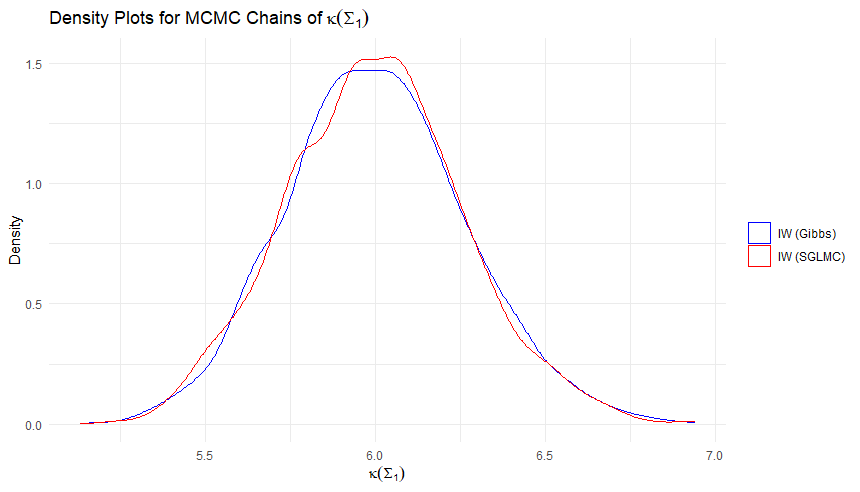} 
    \end{minipage}\hfill
    \begin{minipage}{0.45\textwidth}
        \centering
        \includegraphics[width=\textwidth]{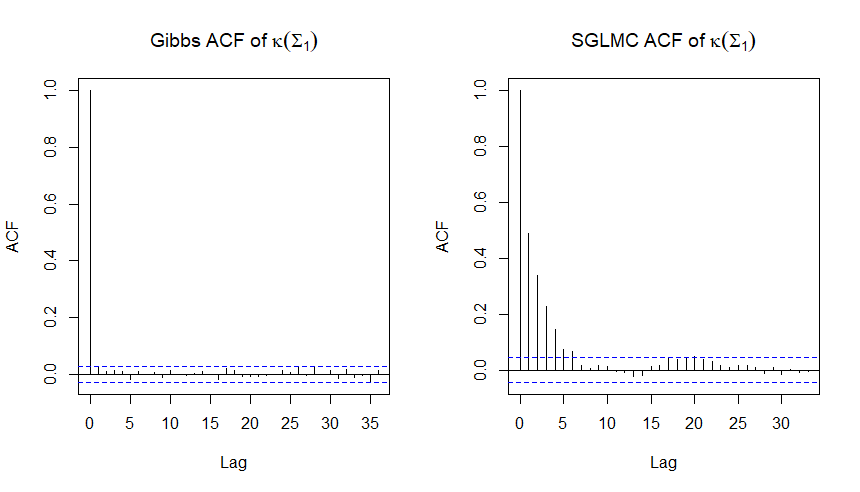} 
    \end{minipage}
        \centering
    \begin{minipage}{0.45\textwidth}
        \centering
        \includegraphics[width=\textwidth]{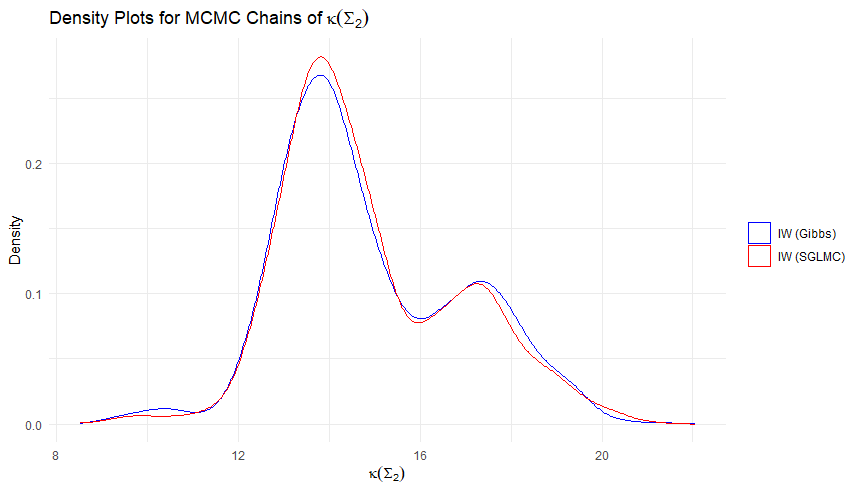} 
    \end{minipage}\hfill
    \begin{minipage}{0.45\textwidth}
        \centering
        \includegraphics[width=\textwidth]{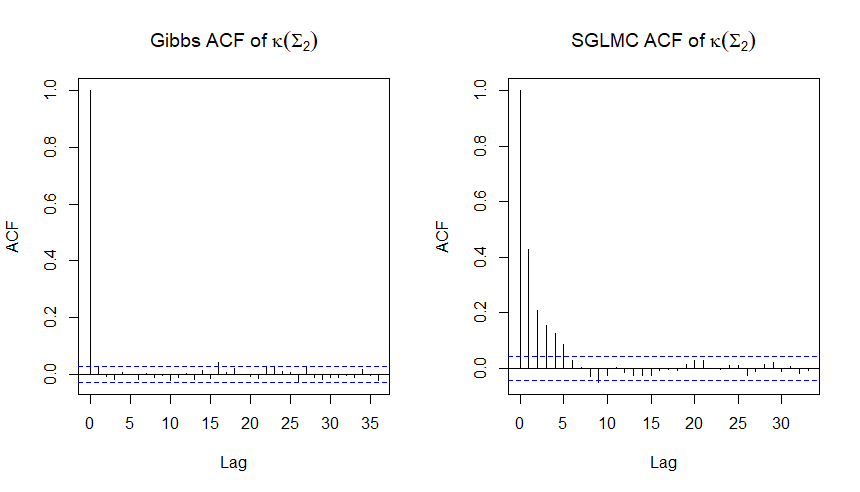} 
    \end{minipage}
        \begin{minipage}{0.45\textwidth}
        \centering
        \includegraphics[width=\textwidth]{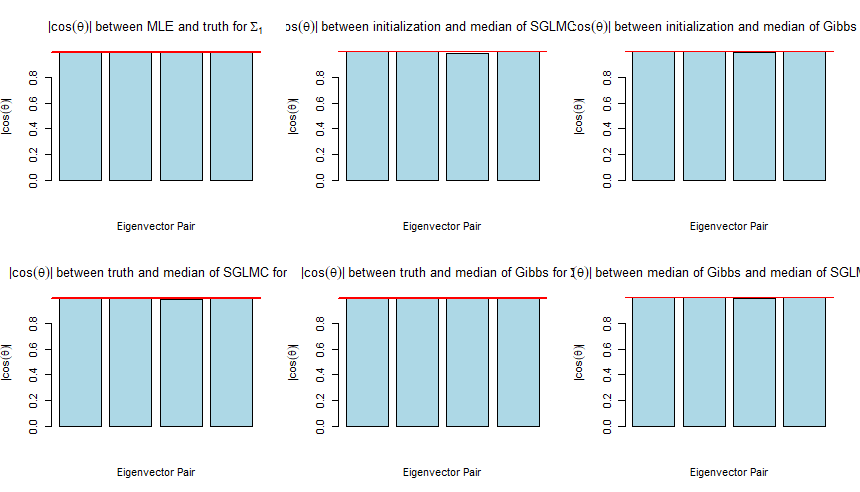} 
    \end{minipage}\hfill
    \begin{minipage}{0.45\textwidth}
        \centering
        \includegraphics[width=\textwidth]{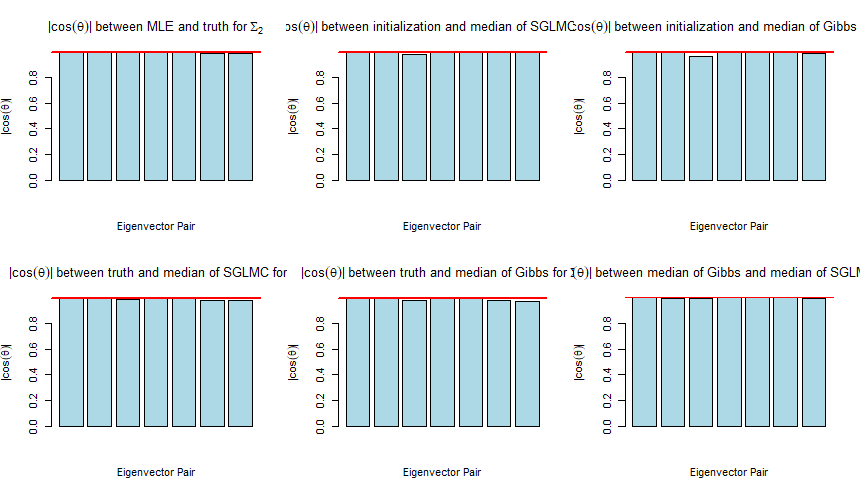} 
    \end{minipage}
    \caption{Density plots of the traces, determinants for $\Sigma$, their corresponding ACFs of the global $\Sigma$, and the condition number comparisons for $\Sigma_{1}, \Sigma_{2}$ for SGLMC, Gibbs, and stan ($d_{1} = 4$, $d_{2} = 7$, $\alpha = .9$).}
    \label{fig: alpha = .9}
\end{figure}

\begin{figure}[ht]
    \centering
    \begin{minipage}{0.45\textwidth}
        \centering
        \includegraphics[width=\textwidth]{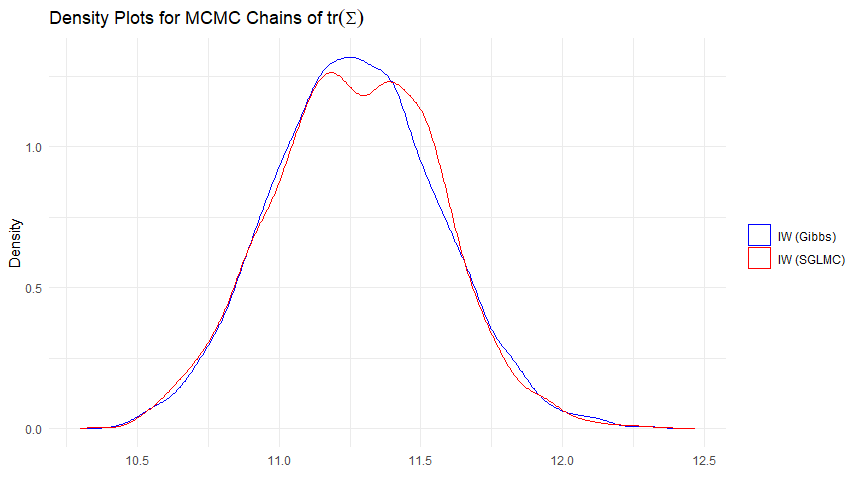} 
    \end{minipage}\hfill
    \begin{minipage}{0.45\textwidth}
        \centering
        \includegraphics[width=\textwidth]{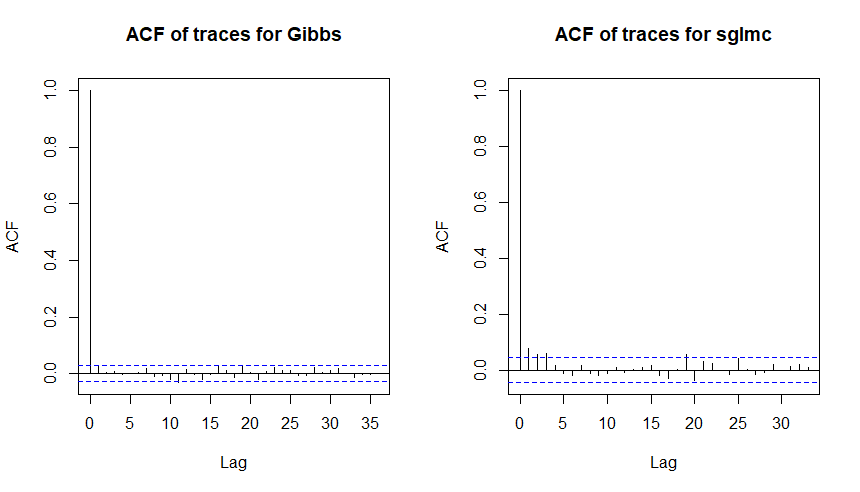} 
    \end{minipage}
           \centering
        \begin{minipage}{0.45\textwidth}
        \centering
        \includegraphics[width=\textwidth]{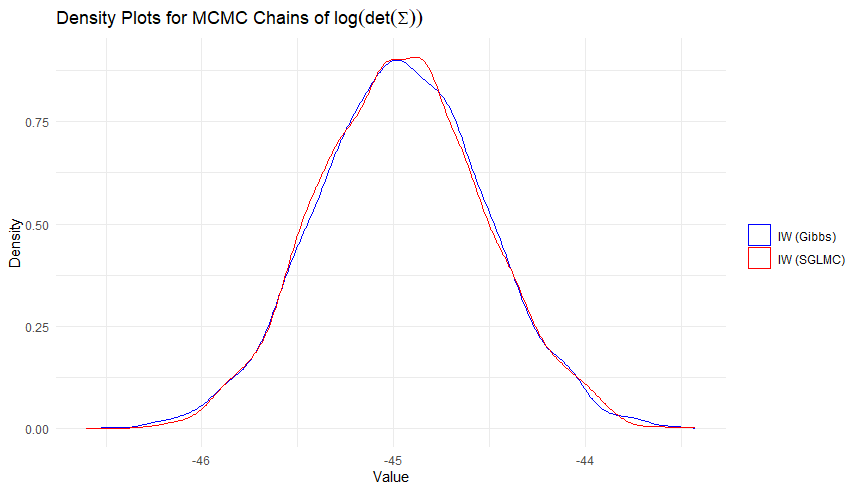} 
    \end{minipage}\hfill
    \begin{minipage}{0.45\textwidth}
        \centering
        \includegraphics[width=\textwidth]{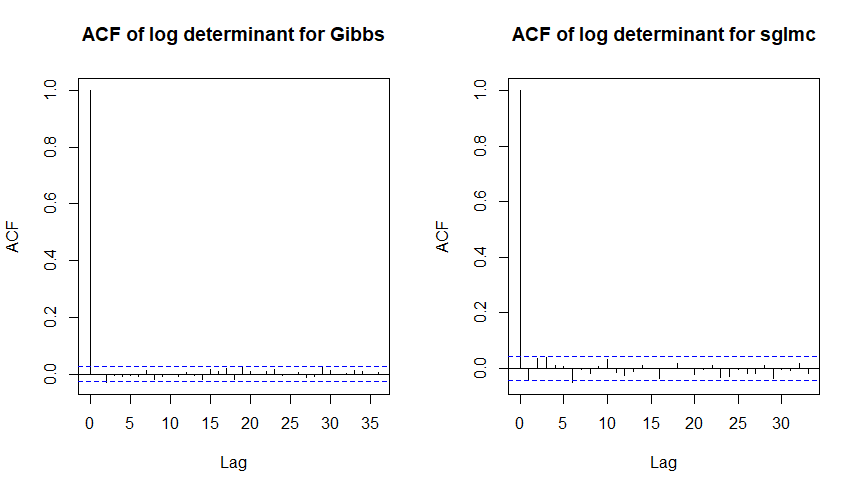} 
    \end{minipage}
        \centering
    \begin{minipage}{0.45\textwidth}
        \centering
        \includegraphics[width=\textwidth]{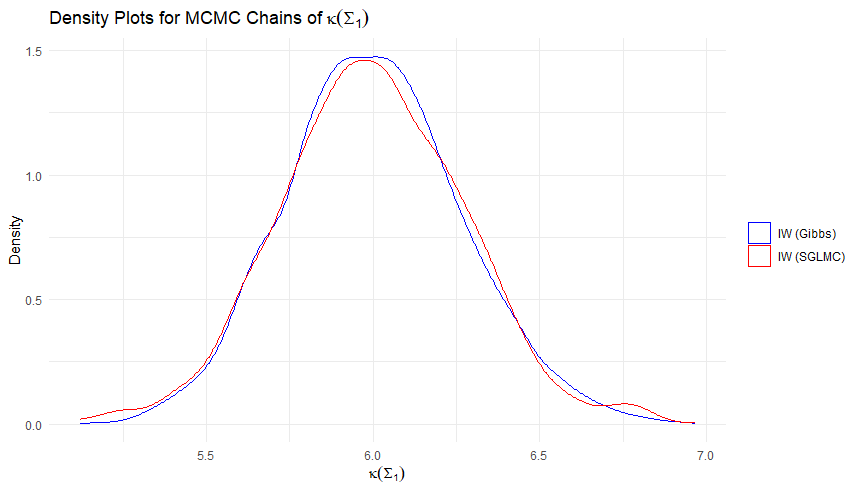} 
    \end{minipage}\hfill
    \begin{minipage}{0.45\textwidth}
        \centering
        \includegraphics[width=\textwidth]{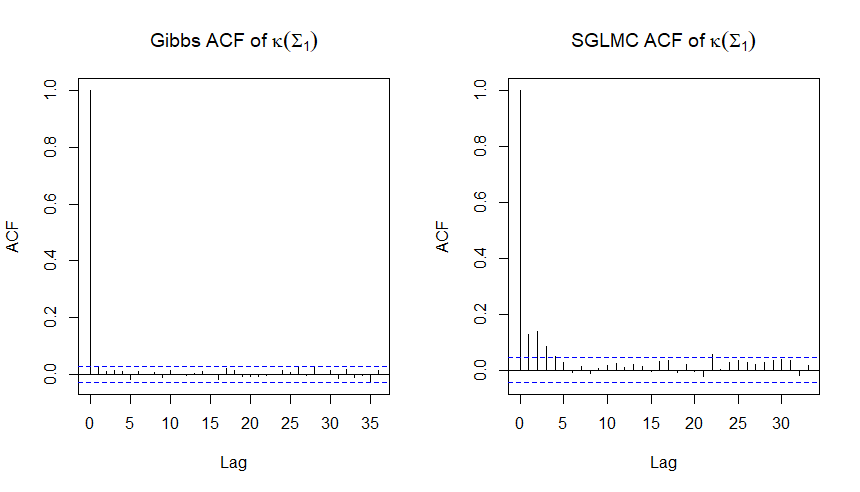} 
    \end{minipage}
        \centering
    \begin{minipage}{0.45\textwidth}
        \centering
        \includegraphics[width=\textwidth]{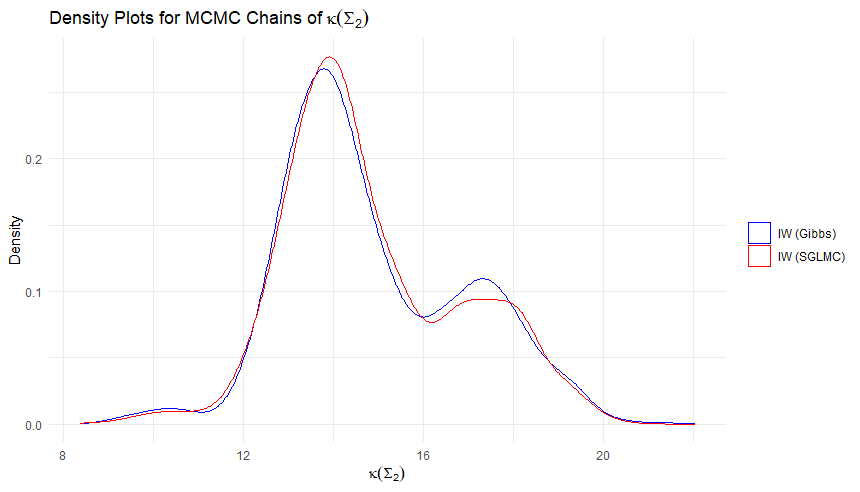} 
    \end{minipage}\hfill
    \begin{minipage}{0.45\textwidth}
        \centering
        \includegraphics[width=\textwidth]{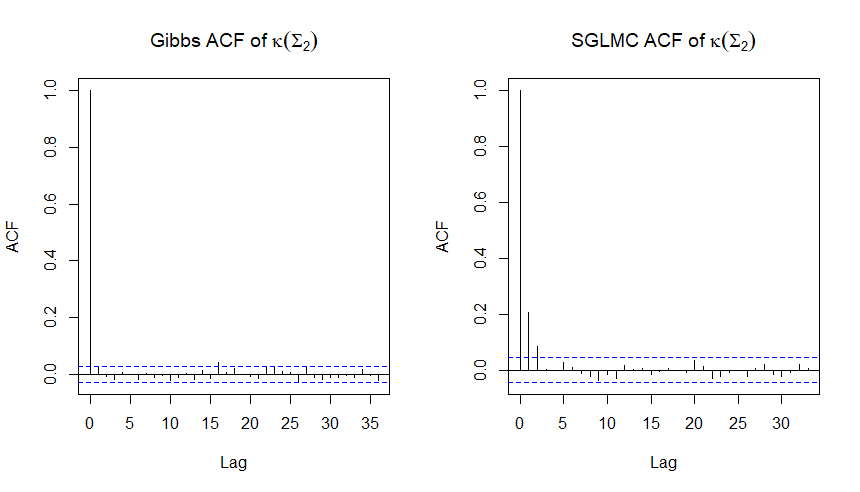} 
    \end{minipage}
        \begin{minipage}{0.45\textwidth}
        \centering
        \includegraphics[width=\textwidth]{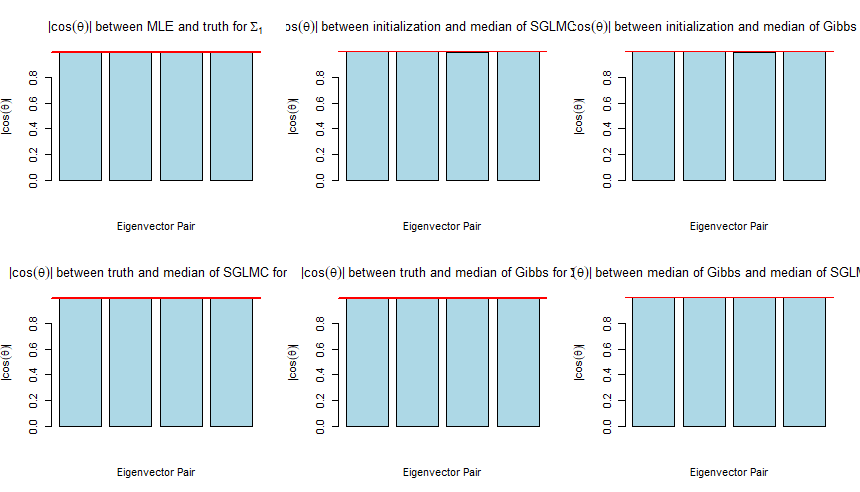} 
    \end{minipage}\hfill
    \begin{minipage}{0.45\textwidth}
        \centering
        \includegraphics[width=\textwidth]{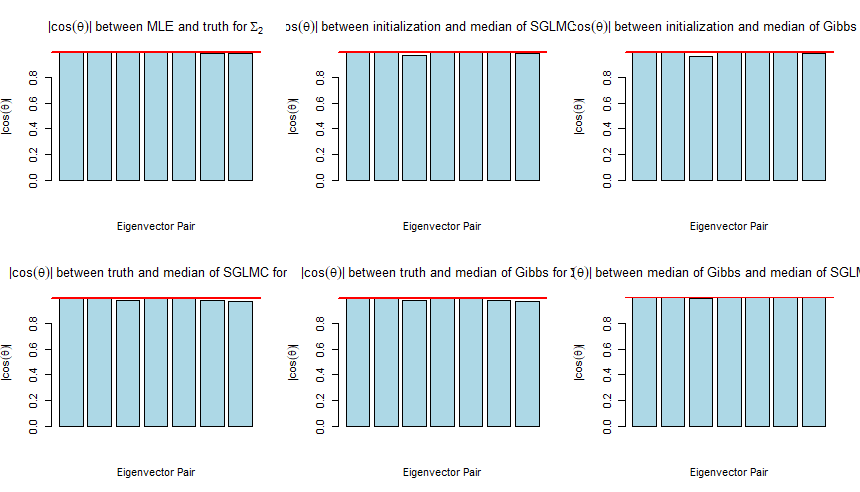} 
    \end{minipage}
    
    \caption{Density plots of the traces, determinants for $\Sigma$, their corresponding ACFs of the global $\Sigma$, and the condition number comparisons for $\Sigma_{1}, \Sigma_{2}$ for SGLMC, Gibbs, and stan ($d_{1} = 4$, $d_{2} = 7$, $\alpha = .95$).}
    \label{fig: alpha = .95}
\end{figure}

For $d_1 = 4, d_2 = 7$, we compare the regularization parameter $\alpha$ under the settings $\alpha \in \mathcal{A} = \{0, .1, .25, .5, .75, .9, .95\}$. For each setting we ran for a burn-in of $500$, an adaptation of $500$, and $1500$ samples generated after. Each setting is compared with the results from the Gibbs sampler. We use $\alpha_{0} = .6$ in each comparison. The comparisons for each level in $\mathcal{A}$ can be found in Figures \ref{fig: alpha = 0}, \ref{fig: alpha = .1}, \ref{fig: alpha = .25}, \ref{fig: alpha = .5}, \ref{fig: alpha = ,75}, \ref{fig: alpha = .9}, and \ref{fig: alpha = .95}, respectively. We found that while densities largely remained unchanged between these global and local matrix statistics,  the autocorrelations associated with those statistics may change fairly significantly between different settings of $\alpha$. We would however like to emphasize that in any setting with $\alpha \geq .5$, the autocorrelation effects may be significantly reduced under different settings of the acceptance rate for the dual averaging algorithm of $\epsilon$, as discussed in the next section.

\begin{figure}[ht]
    \centering
    \begin{minipage}{0.45\textwidth}
        \centering
        \includegraphics[width=\textwidth]{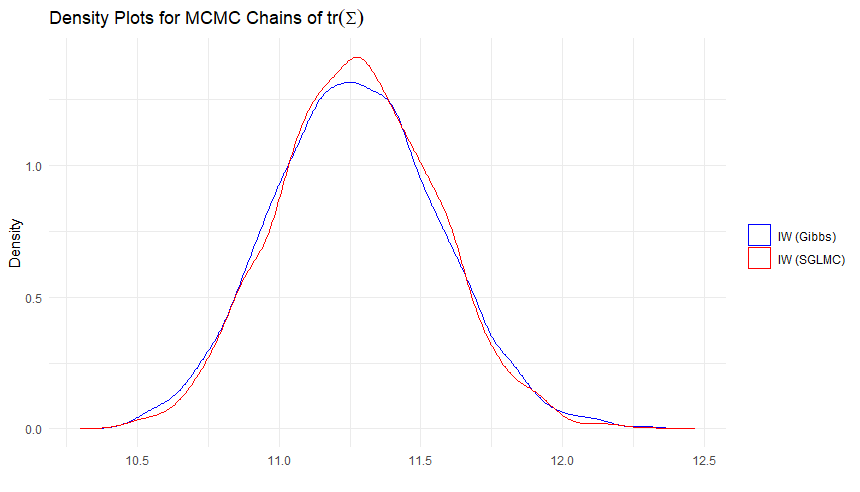} 
    \end{minipage}\hfill
    \begin{minipage}{0.45\textwidth}
        \centering
        \includegraphics[width=\textwidth]{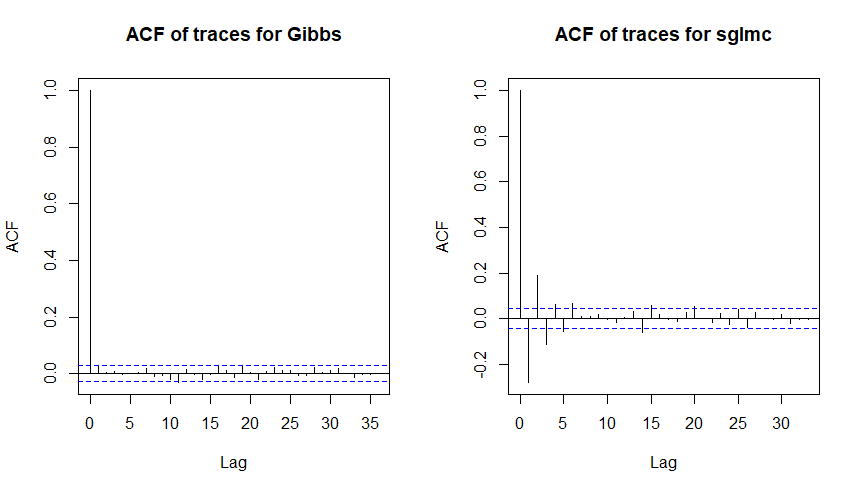} 
    \end{minipage}
           \centering
        \begin{minipage}{0.45\textwidth}
        \centering
        \includegraphics[width=\textwidth]{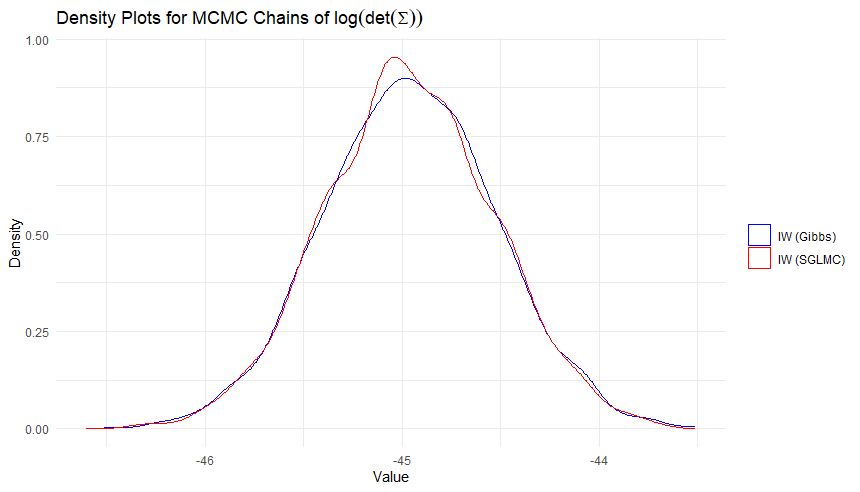} 
    \end{minipage}\hfill
    \begin{minipage}{0.45\textwidth}
        \centering
        \includegraphics[width=\textwidth]{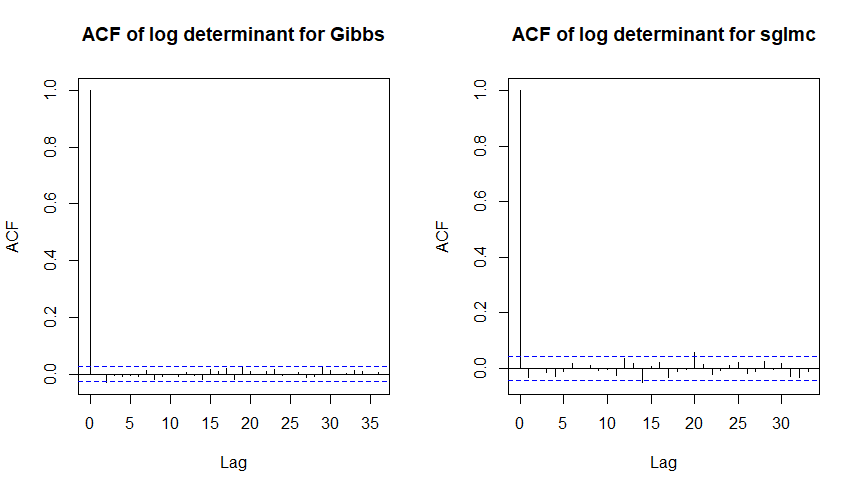} 
    \end{minipage}
        \centering
    \begin{minipage}{0.45\textwidth}
        \centering
        \includegraphics[width=\textwidth]{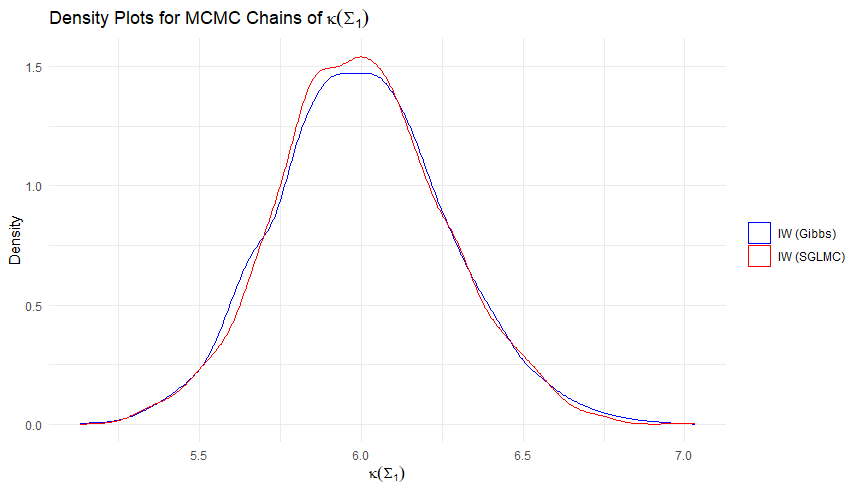} 
    \end{minipage}\hfill
    \begin{minipage}{0.45\textwidth}
        \centering
        \includegraphics[width=\textwidth]{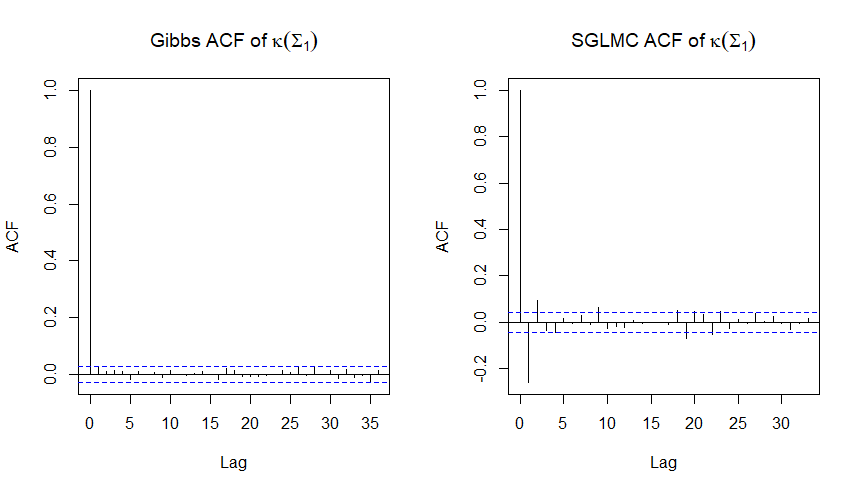} 
    \end{minipage}
        \centering
    \begin{minipage}{0.45\textwidth}
        \centering
        \includegraphics[width=\textwidth]{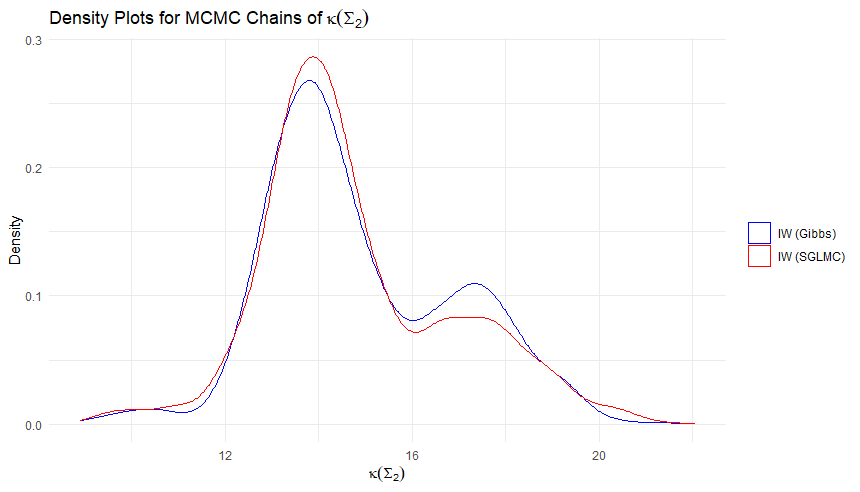} 
    \end{minipage}\hfill
    \begin{minipage}{0.45\textwidth}
        \centering
        \includegraphics[width=\textwidth]{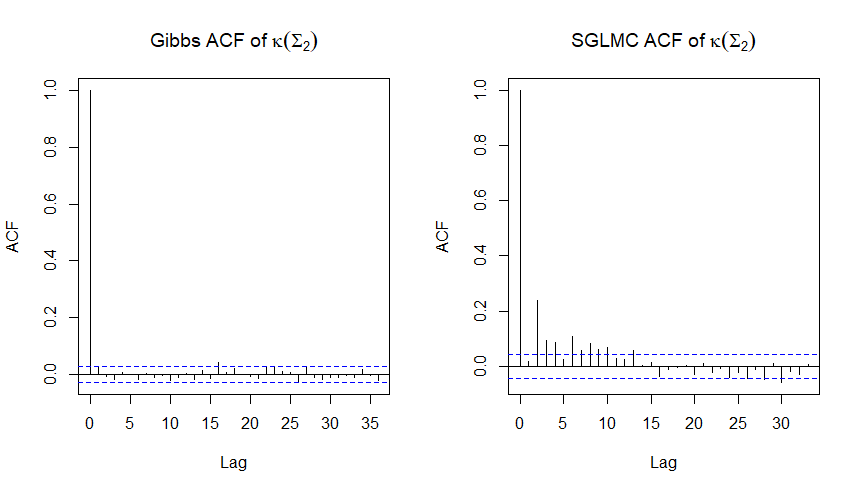} 
    \end{minipage}
        \begin{minipage}{0.45\textwidth}
        \centering
        \includegraphics[width=\textwidth]{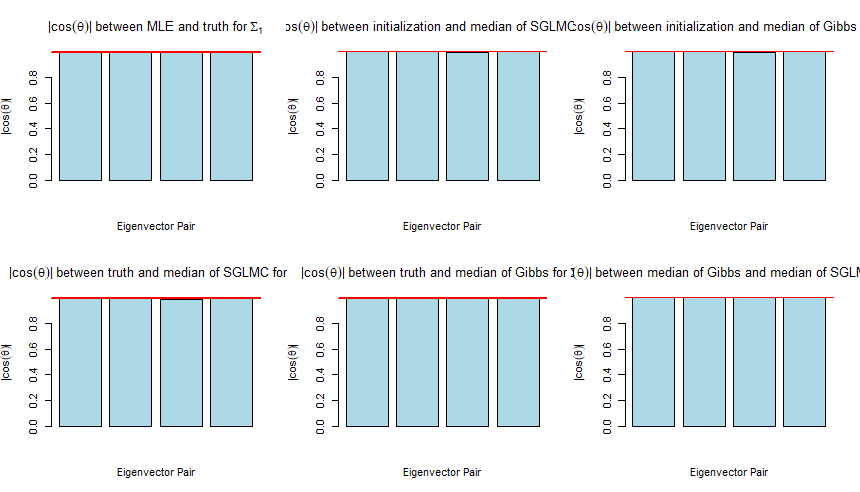} 
    \end{minipage}\hfill
    \begin{minipage}{0.45\textwidth}
        \centering
        \includegraphics[width=\textwidth]{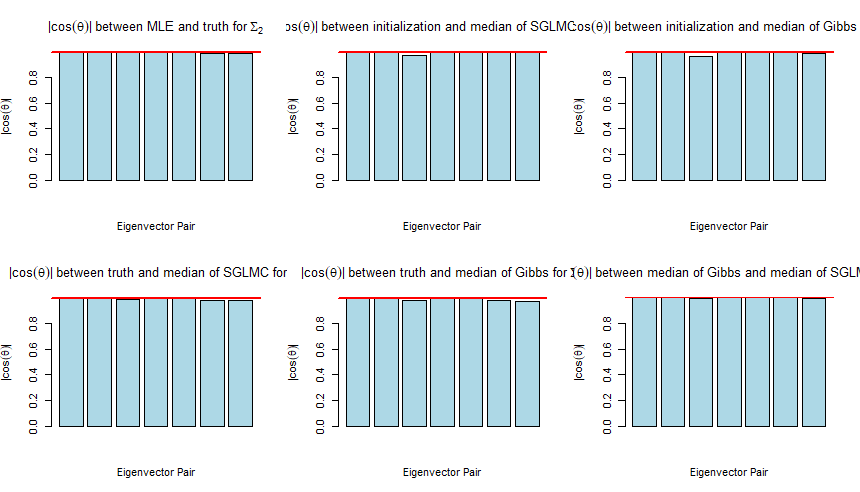} 
    \end{minipage}

    \caption{Density plots of the traces, determinants for $\Sigma$, their corresponding ACFs of the global $\Sigma$, and the condition number comparisons for $\Sigma_{1}, \Sigma_{2}$ for SGLMC, Gibbs, and stan ($d_{1} = 4$, $d_{2} = 7$, $\alpha = .75$).}
    \label{fig: alpha = ,75}
\end{figure}

\subsection{Robustifying samples from Regularization}
\begin{figure}[ht] 
    \centering
    \begin{minipage}{0.45\textwidth}
        \centering
        \includegraphics[width=\textwidth]{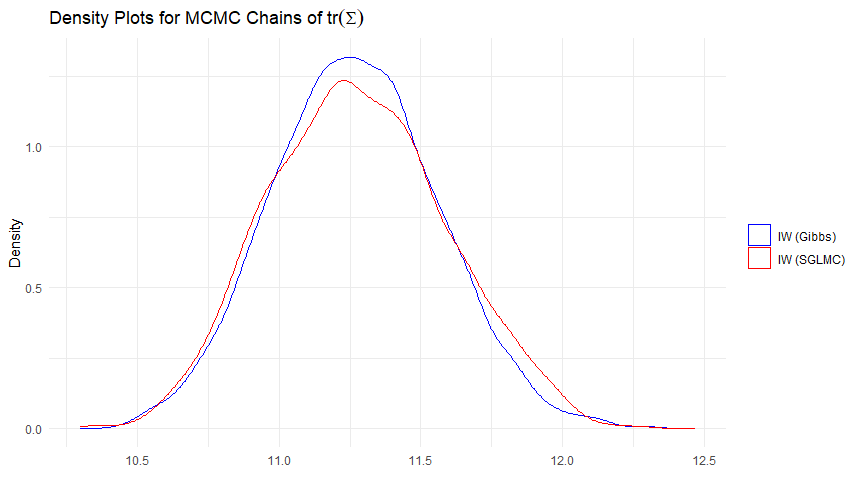} 
    \end{minipage}\hfill
    \begin{minipage}{0.45\textwidth}
        \centering
        \includegraphics[width=\textwidth]{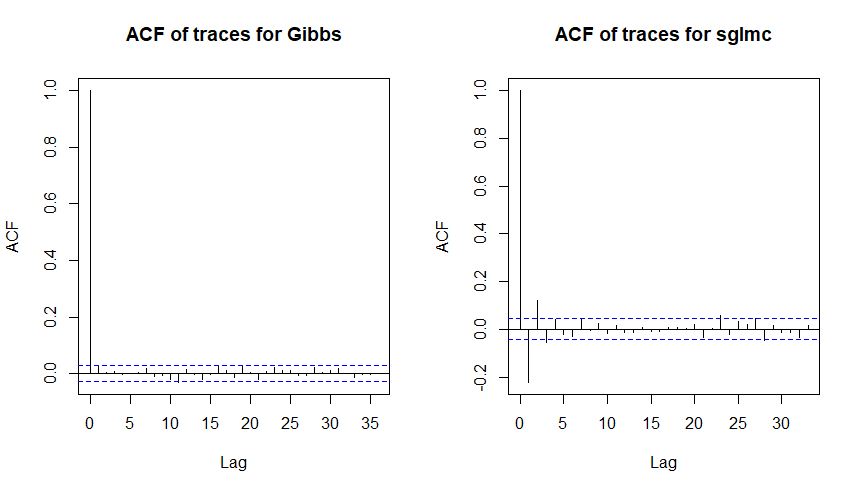} 
    \end{minipage}
           \centering
        \begin{minipage}{0.45\textwidth}
        \centering
        \includegraphics[width=\textwidth]{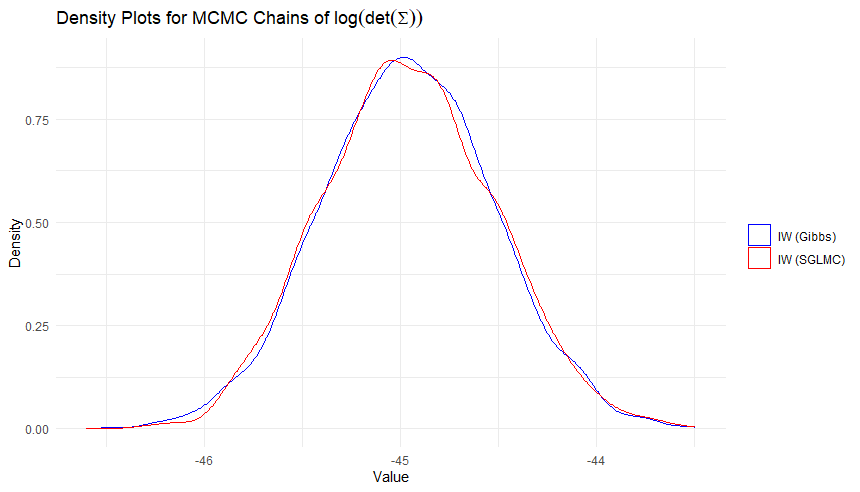} 
    \end{minipage}\hfill
    \begin{minipage}{0.45\textwidth}
        \centering
        \includegraphics[width=\textwidth]{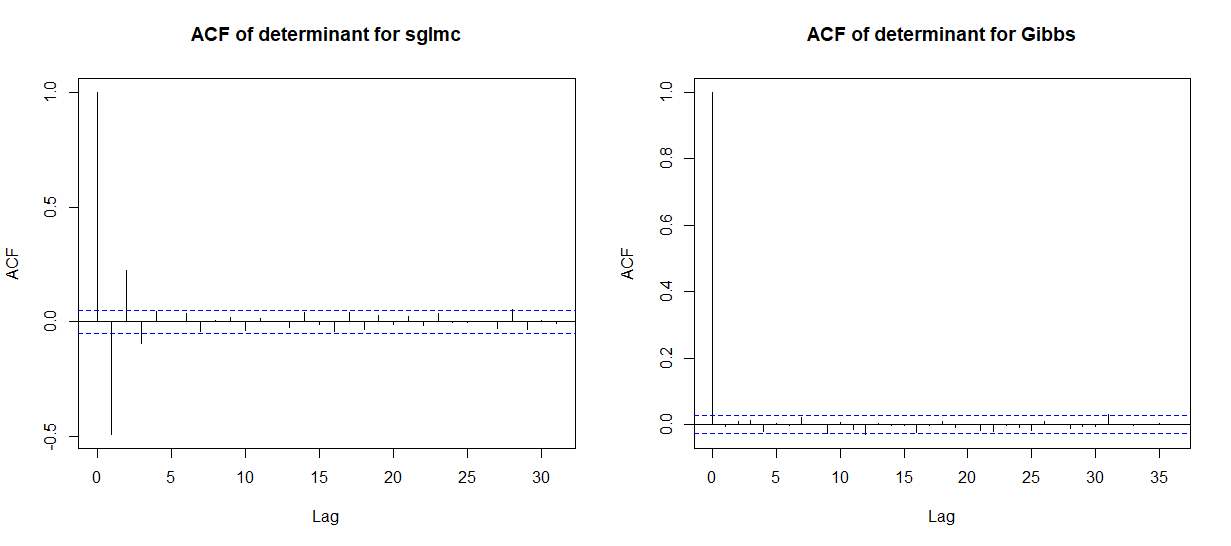} 
    \end{minipage}
        \centering
    \begin{minipage}{0.45\textwidth}
        \centering
        \includegraphics[width=\textwidth]{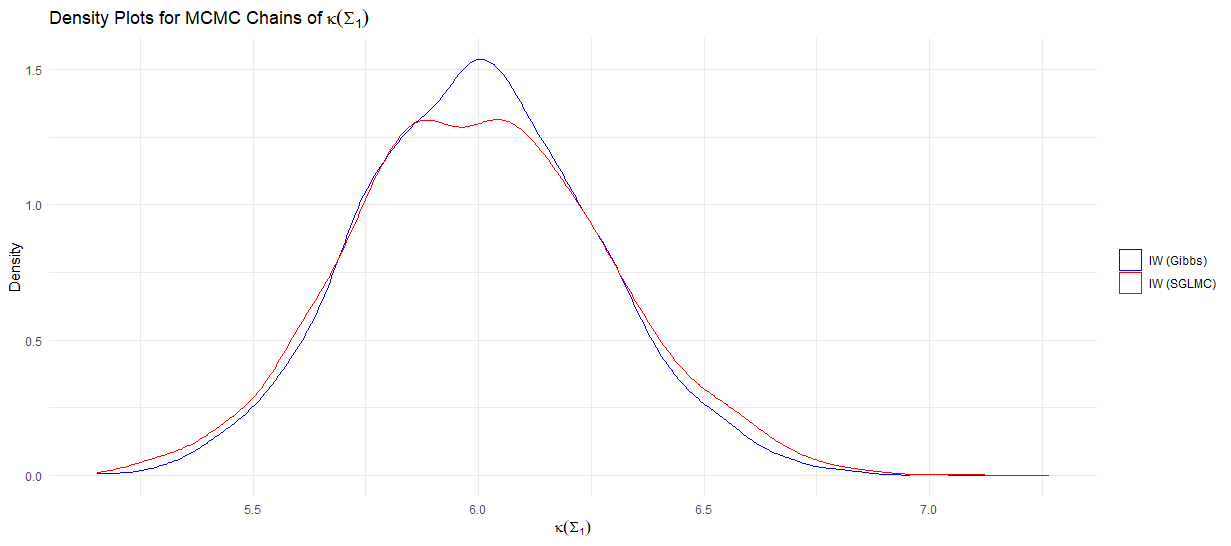} 
    \end{minipage}\hfill
    \begin{minipage}{0.45\textwidth}
        \centering
        \includegraphics[width=\textwidth]{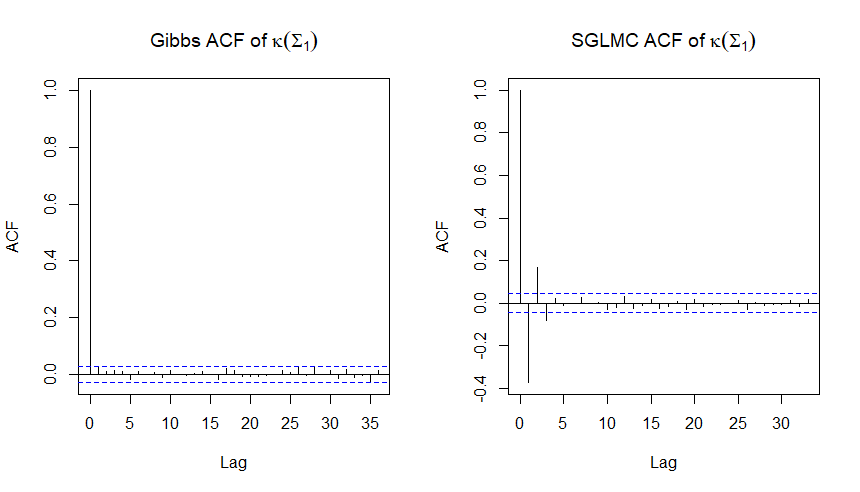} 
    \end{minipage}
        \centering
    \begin{minipage}{0.45\textwidth}
        \centering
        \includegraphics[width=\textwidth]{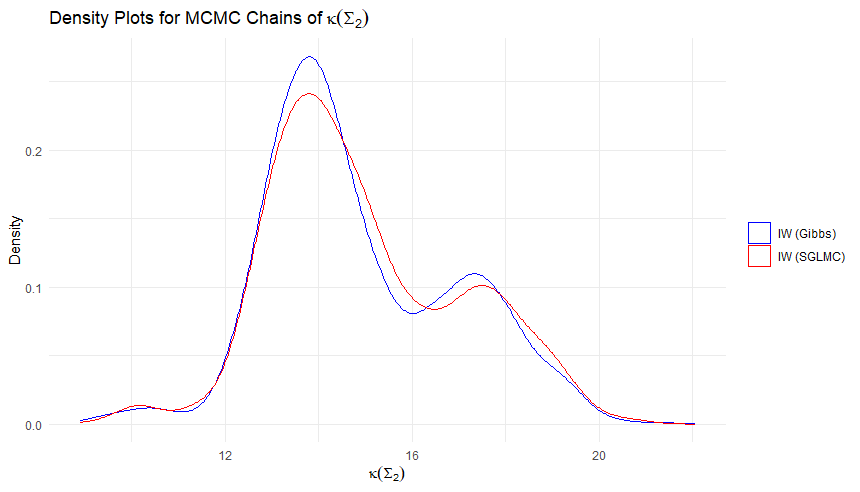} 
    \end{minipage}\hfill
    \begin{minipage}{0.45\textwidth}
        \centering
        \includegraphics[width=\textwidth]{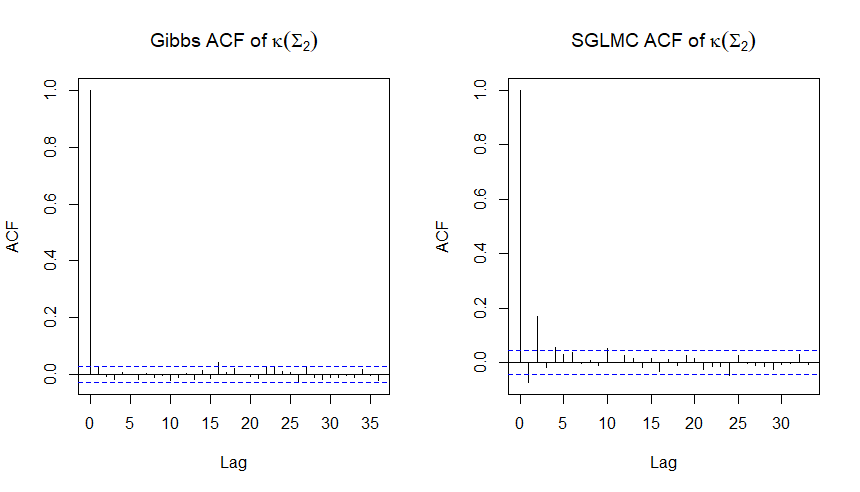} 
    \end{minipage}
    \begin{minipage}{0.45\textwidth}
        \centering
        \includegraphics[width=\textwidth]{images/regularizationcomparisons/alpha.95/eigenvectors1.png} 
    \end{minipage}\hfill
    \begin{minipage}{0.45\textwidth}
        \centering
        \includegraphics[width=\textwidth]{images/regularizationcomparisons/alpha.95/eigenvectors2.png} 
    \end{minipage}
    \caption{Density plots of the traces, determinants for $\Sigma$, their corresponding ACFs of the global $\Sigma$, and the condition number comparisons for $\Sigma_{1}, \Sigma_{2}$ for D-SGLMC, Gibbs ($d_{1} = 4$, $d_{2} = 7$, $\alpha = .5$).}
    \label{fig: D-SGLMC}
\end{figure}

\begin{figure}[ht] 
    \centering
    \begin{minipage}{0.45\textwidth}
        \centering
        \includegraphics[width=\textwidth]{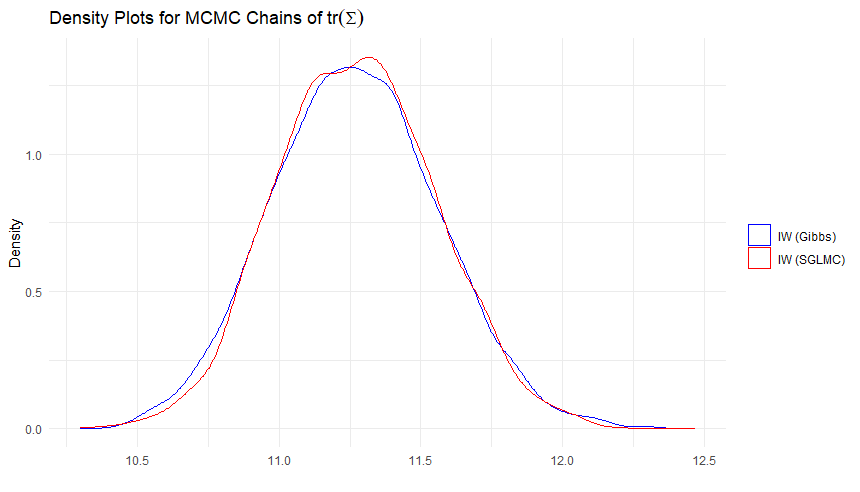} 
    \end{minipage}\hfill
    \begin{minipage}{0.45\textwidth}
        \centering
        \includegraphics[width=\textwidth]{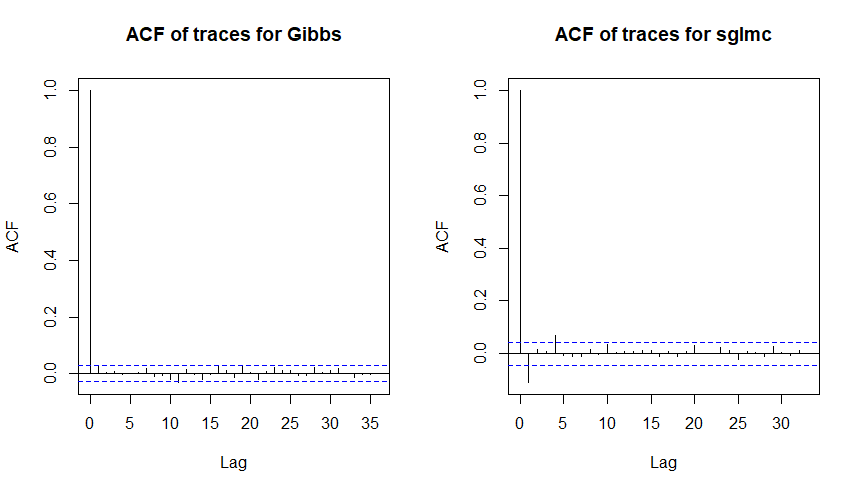} 
    \end{minipage}
           \centering
        \begin{minipage}{0.45\textwidth}
        \centering
        \includegraphics[width=\textwidth]{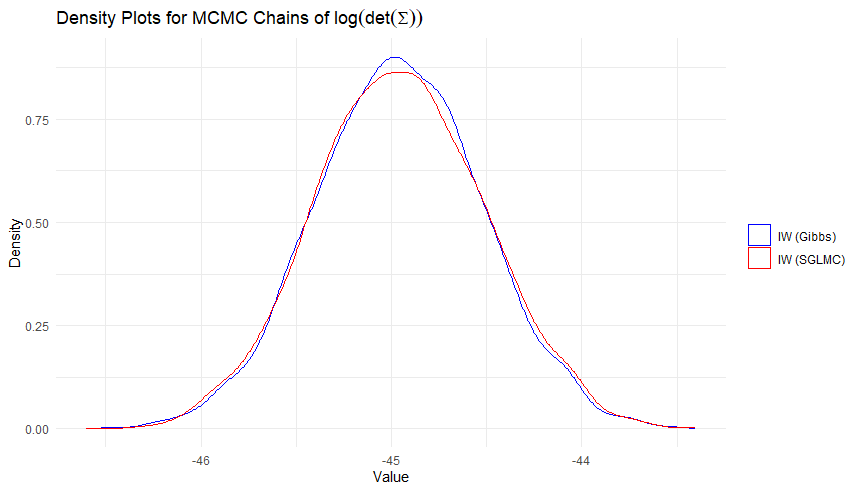} 
    \end{minipage}\hfill
    \begin{minipage}{0.45\textwidth}
        \centering
        \includegraphics[width=\textwidth]{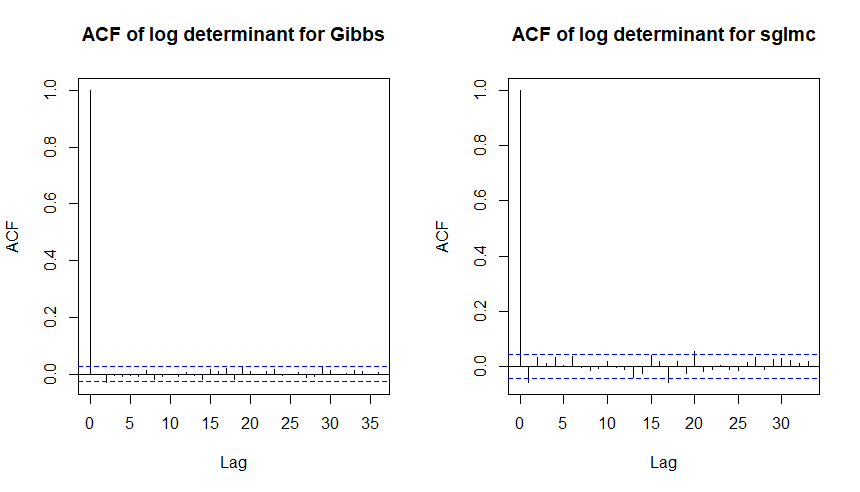} 
    \end{minipage}
        \centering
    \begin{minipage}{0.45\textwidth}
        \centering
        \includegraphics[width=\textwidth]{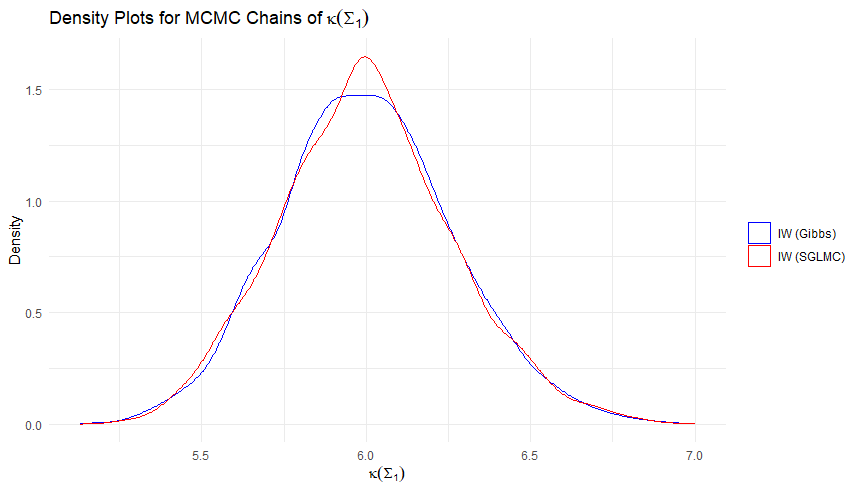} 
    \end{minipage}\hfill
    \begin{minipage}{0.45\textwidth}
        \centering
        \includegraphics[width=\textwidth]{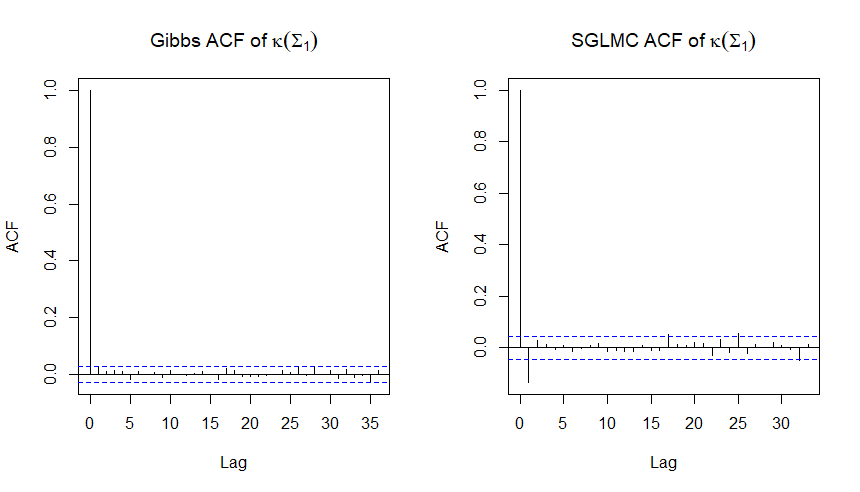} 
    \end{minipage}
        \centering
    \begin{minipage}{0.45\textwidth}
        \centering
        \includegraphics[width=\textwidth]{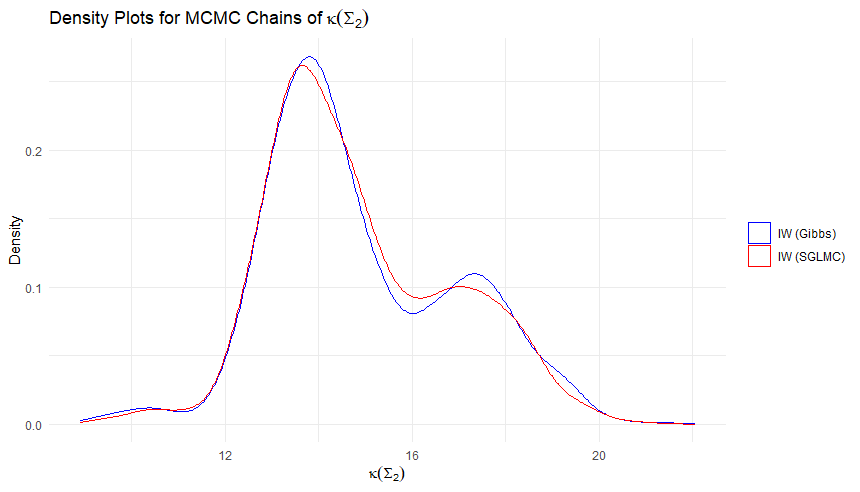} 
    \end{minipage}\hfill
    \begin{minipage}{0.45\textwidth}
        \centering
        \includegraphics[width=\textwidth]{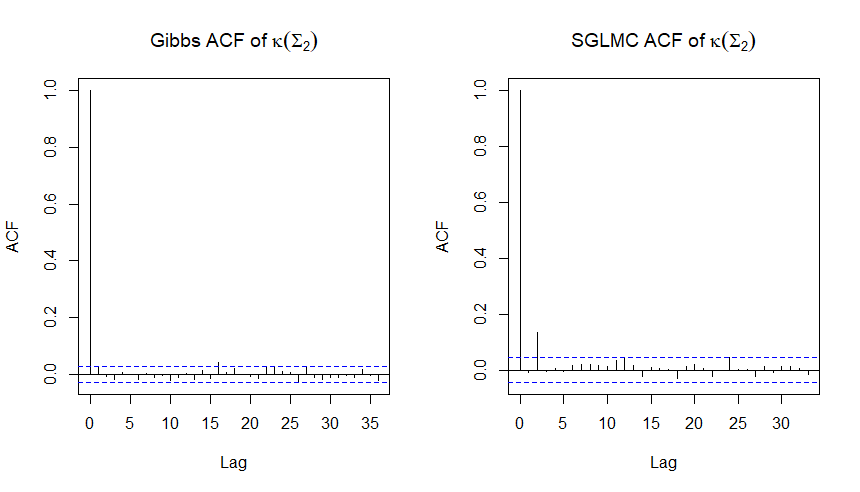} 
    \end{minipage}
            \begin{minipage}{0.45\textwidth}
        \centering
        \includegraphics[width=\textwidth]{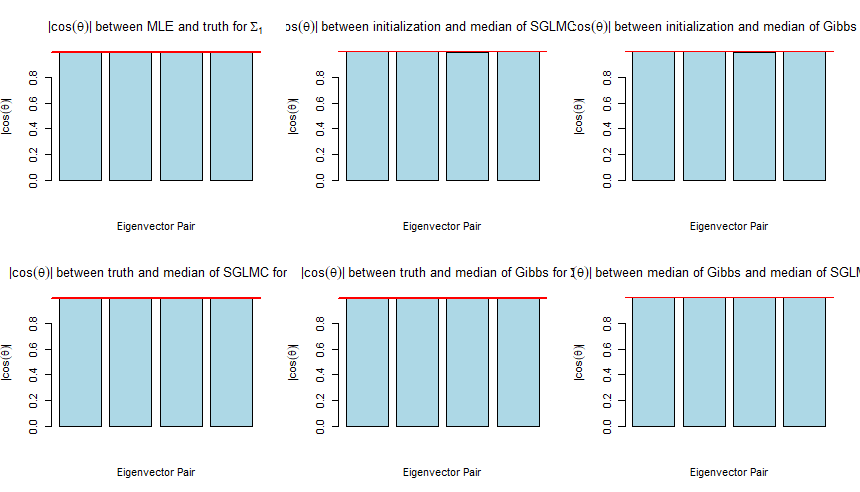} 
    \end{minipage}\hfill
    \begin{minipage}{0.45\textwidth}
        \centering
        \includegraphics[width=\textwidth]{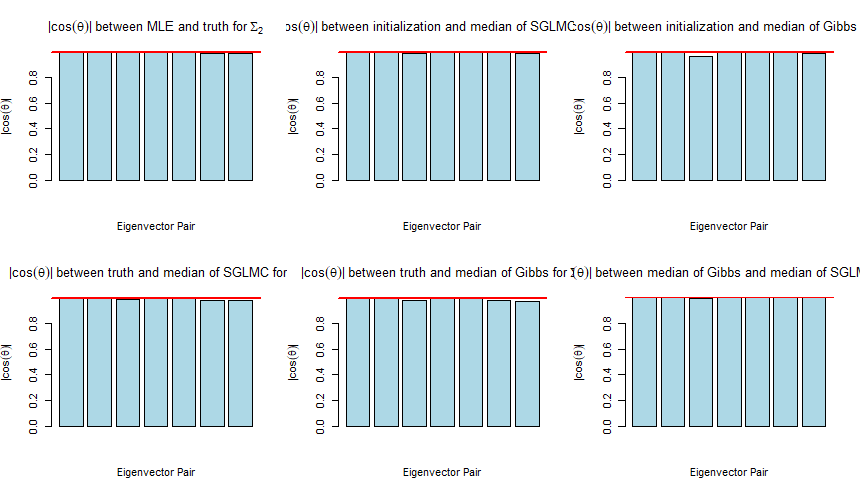} 
    \end{minipage}
    \caption{Density plots of the traces, determinants for $\Sigma$, their corresponding ACFs of the global $\Sigma$, and the condition number comparisons for $\Sigma_{1}, \Sigma_{2}$ for PT-SGLMC, Gibbs ($d_{1} = 4$, $d_{2} = 7$, $\alpha = .5$).}
    \label{fig: PT-SGLMC}
\end{figure}

To make our algorithm more robust to the choice of regularization parameter $\alpha$, as discussed in the previous section, we have two possibilities for further tuning of our algorithm:
\begin{itemize}
    \item An adaptive trajectory termination criteria for $L$
    \item Parallel Tempering
\end{itemize}
In what seemed to be our worst setting from the previous section, with $\alpha = .5$, we give a comparison using a dynamic termination criteria discussed in Section \ref{sec: tuning L} and parallel tempering in Section \ref{sec:Parallel Tempering}. We name these adaptations of SGLMC as D-SGLMC and PT-SGLMC. These comparisons are found in Figures \ref{fig: D-SGLMC} and \ref{fig: PT-SGLMC}, respectively. In each comparison, we used the same data generation and seed from the Experiment of $\alpha = .5$ in Figure \ref{fig: alpha = .5}.
And secondly we consider parallel tempering without adaptive tuning of $L$ under the setting discussed in Section \ref{sec:Parallel Tempering} with $n_c = 5$.
We found the results between these two choices to be competitive in performance.

\subsection{Parallel Tempering} \label{sec:Parallel Tempering}
Parallel tempering aims to alleviate issues with multimodality by running several chains in parallel with different temperatures $T_{1}, \ldots, T_{n_{c}}$ such that $T_{i} \geq 1$ for all $i \in \{1,\ldots, n_{c}\}$ and $T_{n_{c}} = 1$ with
\[
\phi^{c_{i}}(\Sigma_{1}, \Sigma_{2}) = \pi(\Sigma_{1}, \Sigma_{2} \vert Y)^{c_{i}}\sqrt{| G(\Sigma_{1} \times \Sigma_{2}) |}
\]
for $i \in \{1, \ldots, n\}$ for $n$ states being run in parallel, and $c_{n} = 1$ corresponding to the temperature of our target density. Generally, we found $c_{i}$ following a geometrically increasing temperature gradient
\[
c_{i} = c_{1} (\frac{c_{n}}{c_{1}})^\frac{i - 1}{n - 1}
\]
with $c_{1} = .5$ to be suitable for most problems for $n_c = 5$.

At the end of each parallel iteration, a 'swap' between states of adjacent states is proposed and accepted  with probability
\[
\alpha(q_{i}, q_{i + 1}) = \min\left\{1, \frac{E_{i}^{1/T_{i}} E_{i + 1}^{1/T_{i + 1}}}{E_{i}^{1/T_{i+1}} E_{i + 1}^{1/T_{i}}}\right\}
\]
where $E_{i}$ denotes energy of the target density corresponding to chain $i$. Equivalently, this may be stated as
\[
T(q_{i}, q_{i + 1}) = (H(q_{i}) - H(q_{i + 1}))*(\frac{1}{T_{i}} - \frac{1}{T_{i + 1}})
\]
where $H(q_{i})$ is the energy associated to the untempered Hamiltonian configuration associated to state $(q_{i}, p_{i})$. The MH correction then becomes:
\[
\alpha(q_{i}, q_{i + 1}) = \min \{1, T(q_{i}, q_{i + 1})\}.
\]

The goal of tempering is flatten the target distribution, with higher temperatures allowing more flexible movement across improbable regions of the true target density $\pi$. Swapping states to lower temperature densities then encourages HMC to snap back to regions closer to higher probability regions of the true target density as $T_{i} \rightarrow 1$. \\
We used a geometric temperature gradient with $t_s$, $t_l$ to denote the lowest and highest temperatures, respectively,  $T_{i} = t_s*r^{n_{i} - 1}$ such that $r = (\frac{t_s}{t_l})^{n_{chains}}$ and found $\frac{1}{t_{s}} \approx 0.5$ to be suitable in regards to improving effective sample sizes for most dimensions of problems.

\subsection{Dynamic Tuning of $L$} \label{sec: tuning L}
The no U-turn sampler automatically tunes $L$ by terminating at turning point of longest oscillation (TPOLO). This is done by terminating when the momentum is orthogonal to the integrated path length
\[
p_{t}^{T} \int_{0}^{T} p_{\tau} d\tau = 0.
\]
In Euclidean space, this is when $p_{t}^{T} (q_{t} - q_{0}) = 0$. On a Riemannian manifold, it's a bit more nuanced due to the presence of the position specific Riemannian metric which identifies the inner product. Specifically, we would like a method to transport the momentum in along our trajectory so that we can "integrate" our path length along the path. This is challenging, as our gradients don't follow the geodesics from time $t - 1$ to time $t$, so transporting a velocity from time $t - 1$ to time $t$ would follow what is called vector transport instead. This quantity is quite challenging to calculate. 

We can however construct a new geodesic emanating from our current position to the beginning of our trajectory. Let $\Sigma(0)$ and $\Sigma(1)$ be the position at the beginning to the end of a trajectory. The geodesic connecting these two points starting from $\Sigma(1)$ and ending at $\Sigma(0)$ is given by the log and exponential map
\begin{equation}
    \Sigma(t) = exp_{\Sigma(1)}(t log_{\Sigma(0)}(\Sigma(1))).
\end{equation}
That is, starting at $\Sigma(0)$, $log_{\Sigma(1)}(\Sigma(0))$ is a tangent vector at $\Sigma(0)$ such that when exponentiated (integrated) for a time step of $t = 1$ gives $\Sigma(1) = \Sigma_{t}$, with corresponding tangent vector at $t = 1$ given by
\[
V^{*}(1) = \frac{d}{dt} \Sigma(t) \vert_{t = 1}.
\]
Rather than continually transporting velocity vectors along a path, we can directly find a geodesic connecting the end points of our trajectory with a corresponding velocity vector. Because the velocity from our log map follows the geodesic, the inner product between velocities velocities are easily comparable between the beginning and end of our trajectories via
\[
vec^{H}(V)^{T} G(\Sigma_{t}) vec^{H}(V^{*}(0)).
\]
This gives the full termination criteria
\[
vec^{H}(V)^{T} G(\Sigma_{t}) vec^{H}(V^{*}(0)) \geq 0.
\]
Consequently, we terminate when the magnitude of the angle between our current velocity and the tangent vector connecting our current position to our initial position is less than $\frac{\pi}{2}$.

\subsection{Prior Comparisons}
It is known that the inverse-wishart prior tends to overestimate large eigenvalues in the setting where many eigenvalues are near 0. In this section, we demonstrate the influence of the inverse Wishart prior on the estimation of eigenvalues and eigenvectors in the small eigenvalue setting and compare alternative prior choices. We generate $\Sigma_{1} \in \mathcal{P}(4), \quad \Sigma_{2} \in \mathcal{P}(7)$ under 
\[
\Sigma_{1}, \Sigma_{2} \sim IW(d_{i} + 10, \frac{1}{d_{i}} I_{d_{i}}).
\]
Where we shrink the last 4 eigenvalues of $\Sigma_{2}$ by a factor of $.5$.

\begin{figure}[ht] 
    \centering
    \begin{minipage}{0.45\textwidth}
        \centering
        \includegraphics[width=\textwidth]{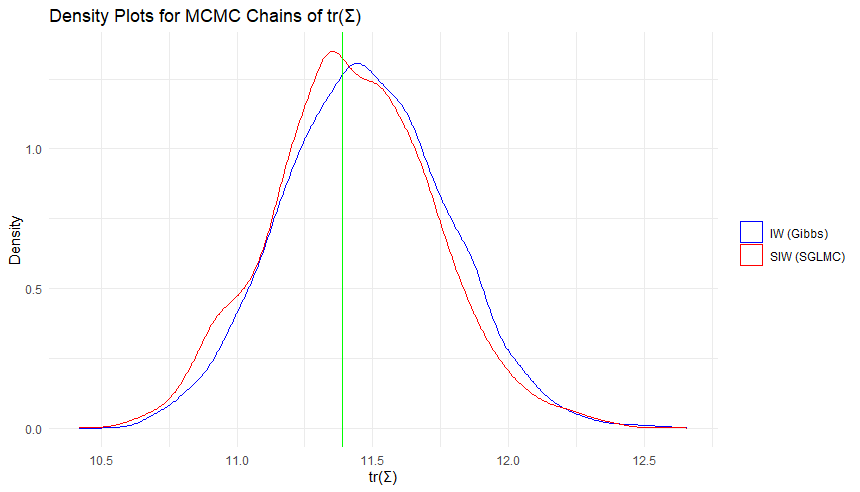} 
    \end{minipage}\hfill
    \begin{minipage}{0.45\textwidth}
        \centering
        \includegraphics[width=\textwidth]{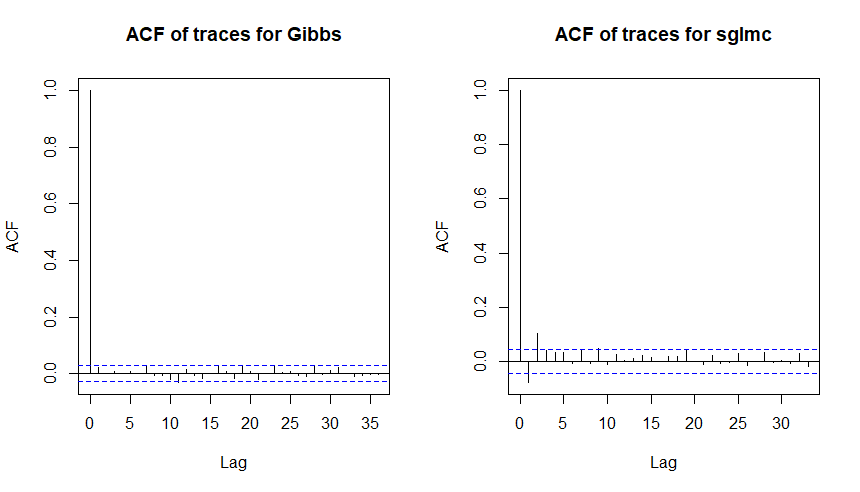} 
    \end{minipage}
           \centering
        \begin{minipage}{0.45\textwidth}
        \centering
        \includegraphics[width=\textwidth]{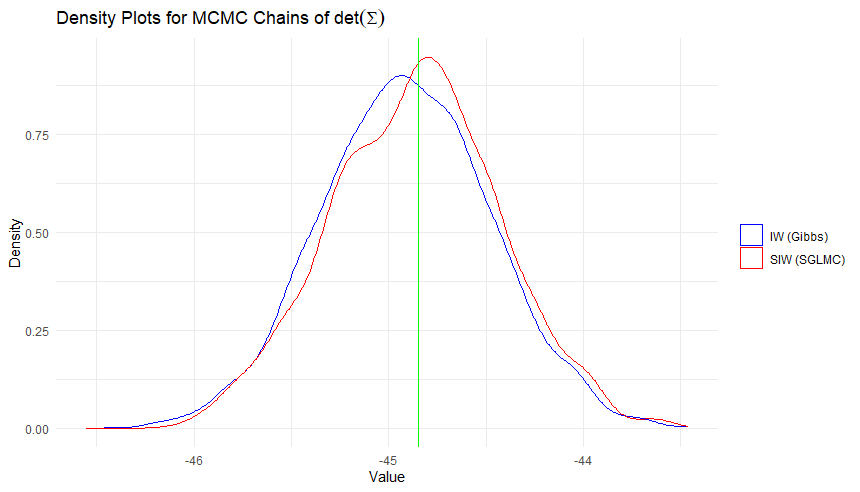} 
    \end{minipage}\hfill
    \begin{minipage}{0.45\textwidth}
        \centering
        \includegraphics[width=\textwidth]{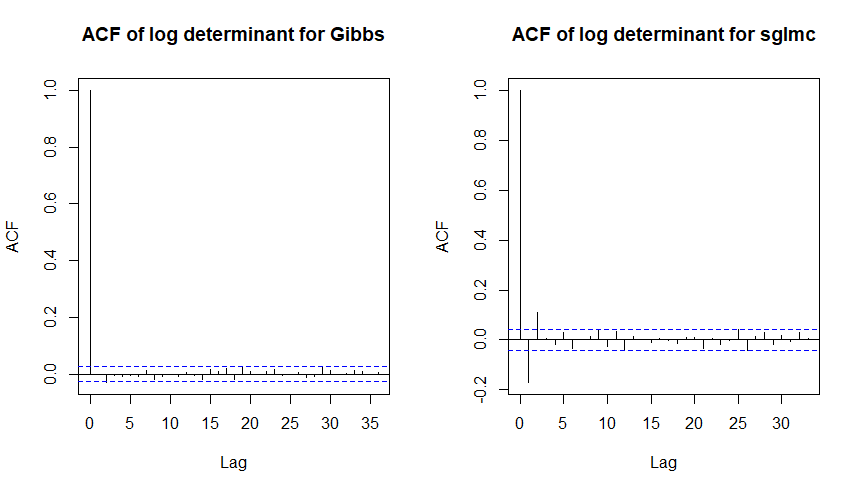} 
    \end{minipage}
        \centering
    \begin{minipage}{0.45\textwidth}
        \centering
        \includegraphics[width=\textwidth]{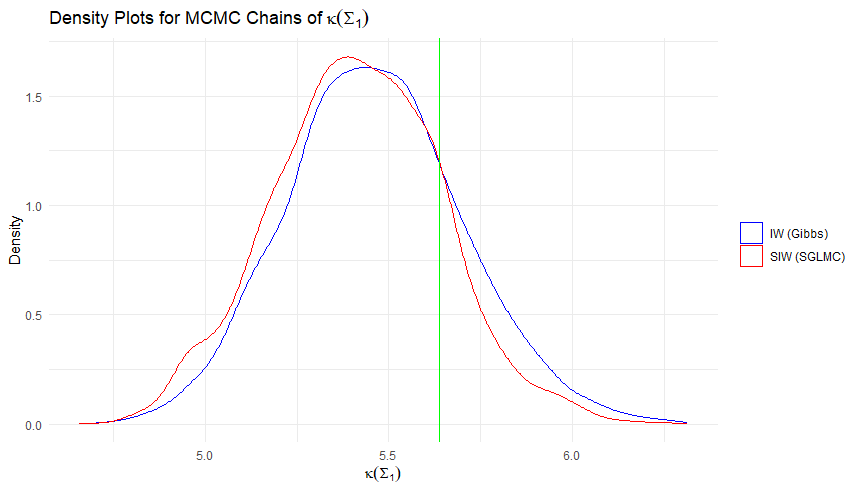} 
    \end{minipage}\hfill
    \begin{minipage}{0.45\textwidth}
        \centering
        \includegraphics[width=\textwidth]{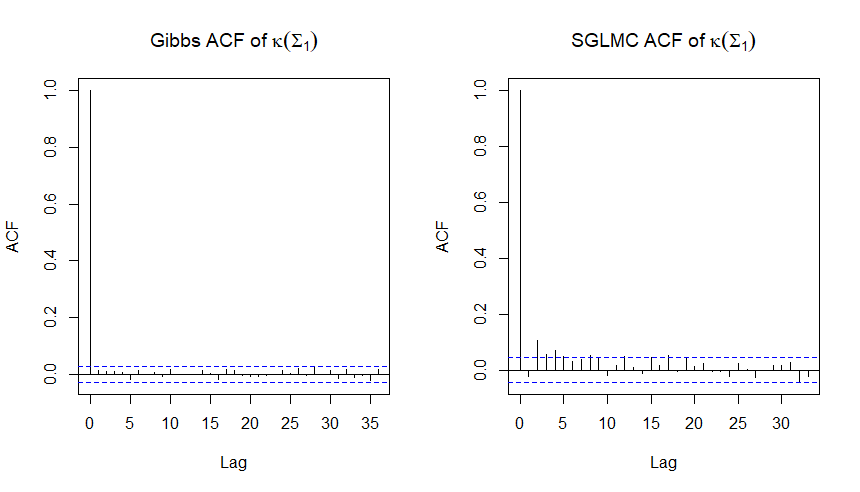} 
    \end{minipage}
        \centering
    \begin{minipage}{0.45\textwidth}
        \centering
        \includegraphics[width=\textwidth]{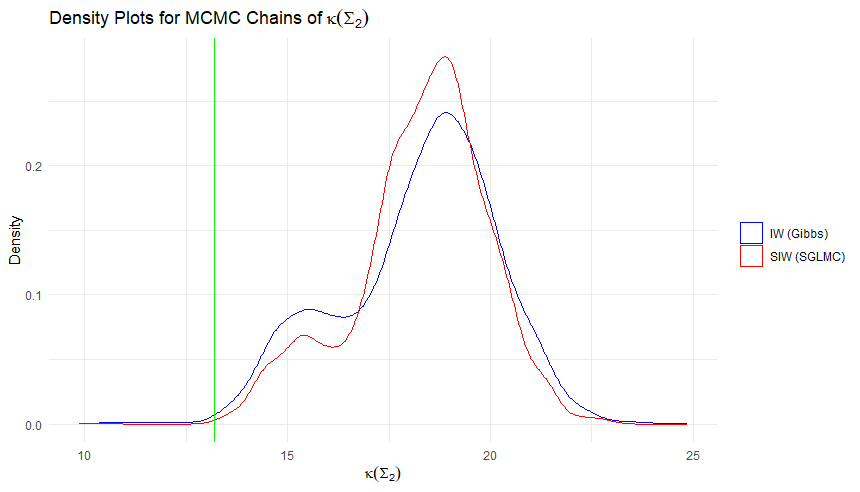} 
    \end{minipage}\hfill
    \begin{minipage}{0.45\textwidth}
        \centering
        \includegraphics[width=\textwidth]{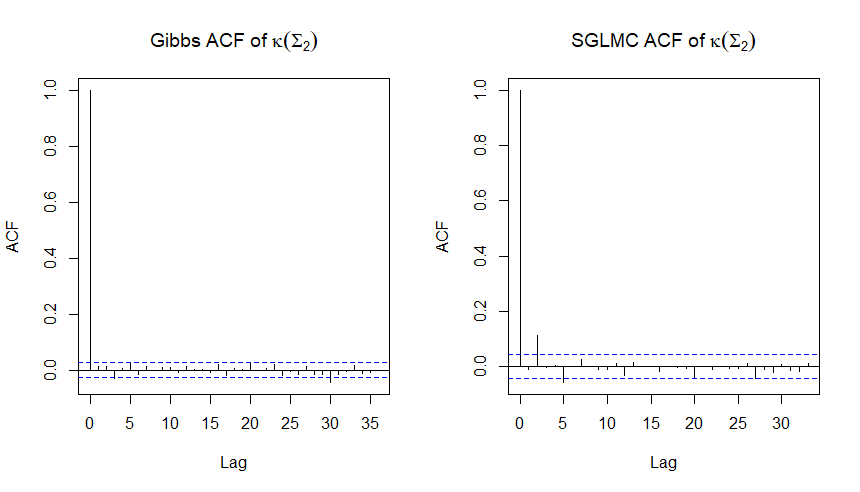} 
    \end{minipage}
            \begin{minipage}{0.45\textwidth}
        \centering
        \includegraphics[width=\textwidth]{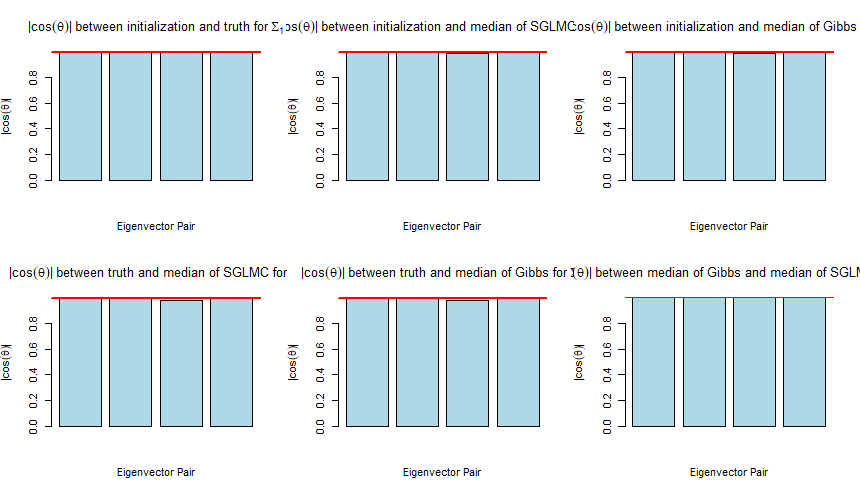} 
    \end{minipage}\hfill
    \begin{minipage}{0.45\textwidth}
        \centering
        \includegraphics[width=\textwidth]{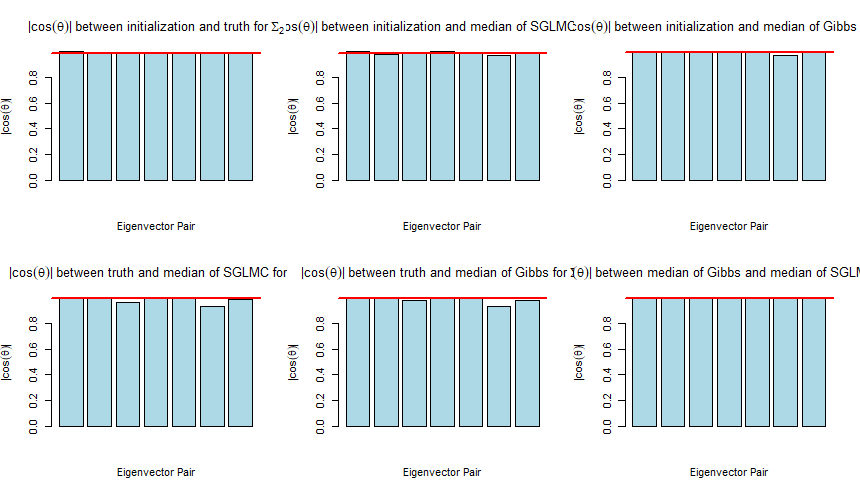} 
    \end{minipage}
    \caption{Density plots of the traces, determinants for $\Sigma$, their corresponding ACFs of the global $\Sigma$, and the condition number comparisons for $\Sigma_{1}, \Sigma_{2}$ for the SIW prior vs Gibbs ($d_{1} = 4$, $d_{2} = 7$, $\alpha = .95$).}
    \label{fig: SIW}
\end{figure}

\begin{figure}[ht] 
    \centering
    \begin{minipage}{0.45\textwidth}
        \centering
        \includegraphics[width=\textwidth]{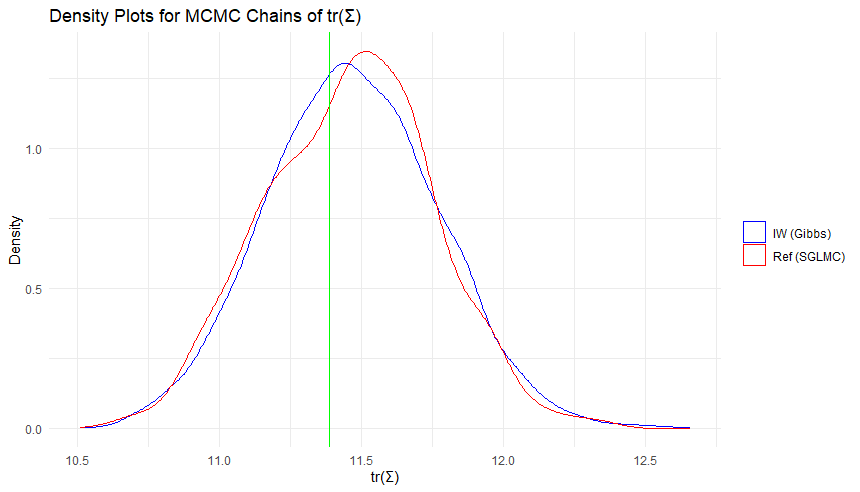} 
    \end{minipage}\hfill
    \begin{minipage}{0.45\textwidth}
        \centering
        \includegraphics[width=\textwidth]{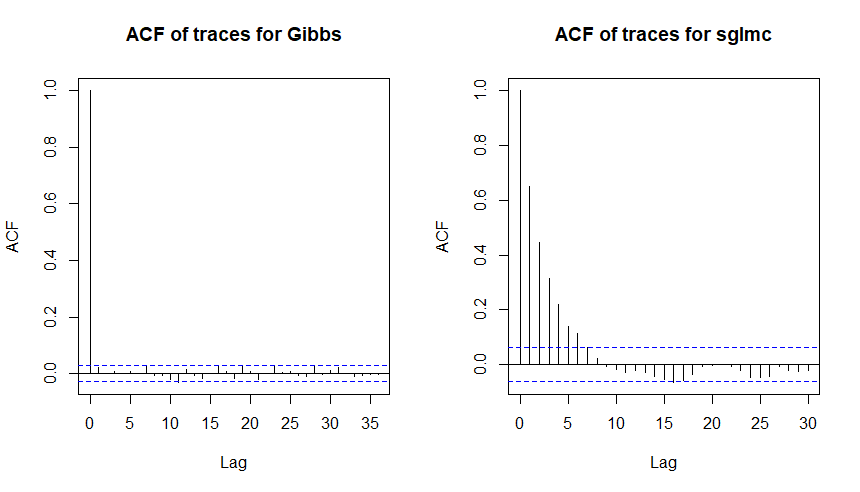} 
    \end{minipage}
           \centering
        \begin{minipage}{0.45\textwidth}
        \centering
        \includegraphics[width=\textwidth]{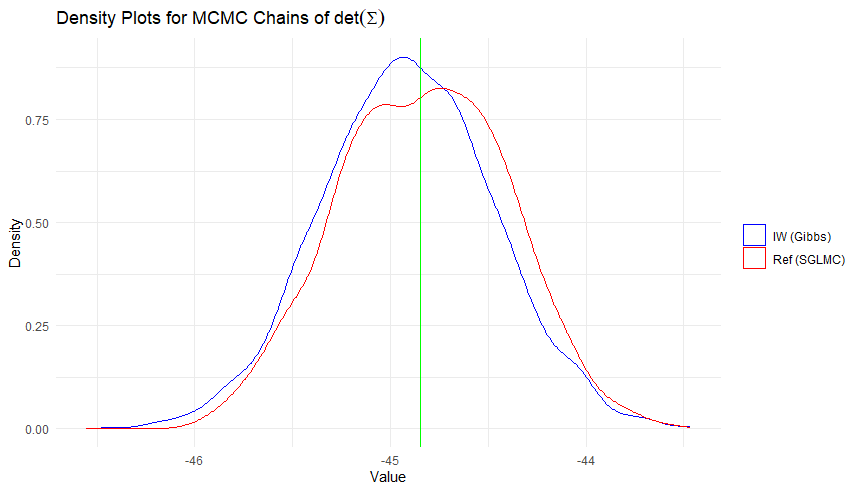} 
    \end{minipage}\hfill
    \begin{minipage}{0.45\textwidth}
        \centering
        \includegraphics[width=\textwidth]{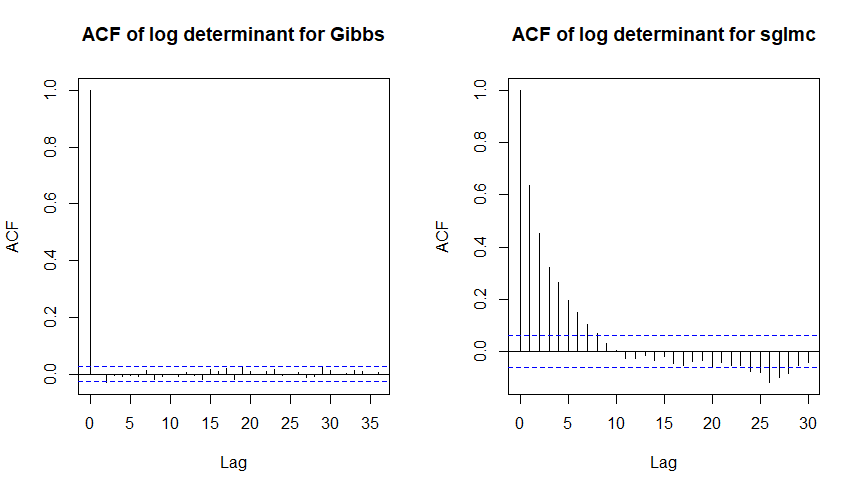} 
    \end{minipage}
        \centering
    \begin{minipage}{0.45\textwidth}
        \centering
        \includegraphics[width=\textwidth]{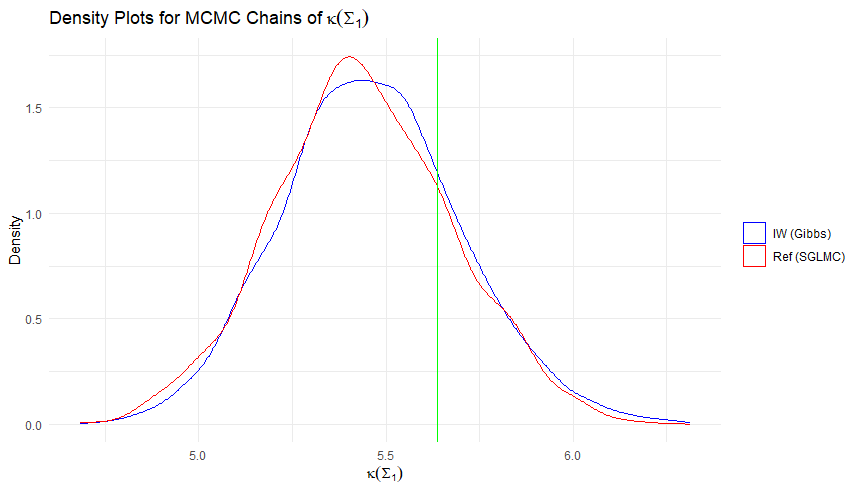} 
    \end{minipage}\hfill
    \begin{minipage}{0.45\textwidth}
        \centering
        \includegraphics[width=\textwidth]{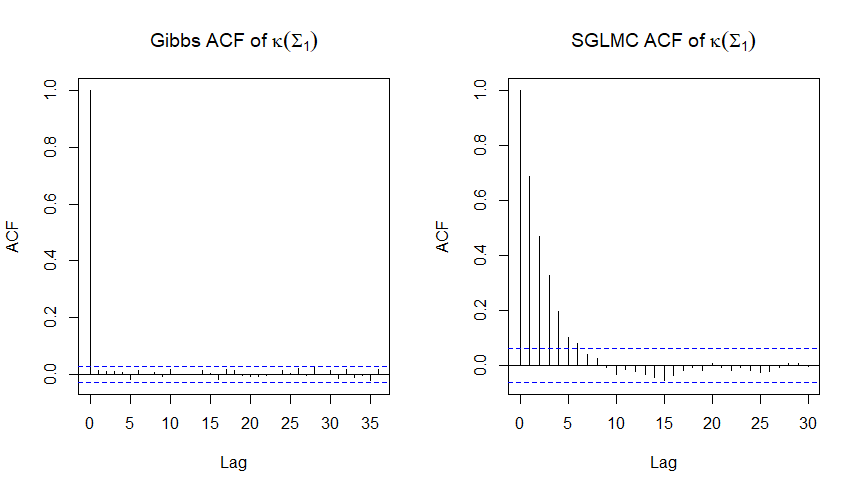} 
    \end{minipage}
        \centering
    \begin{minipage}{0.45\textwidth}
        \centering
        \includegraphics[width=\textwidth]{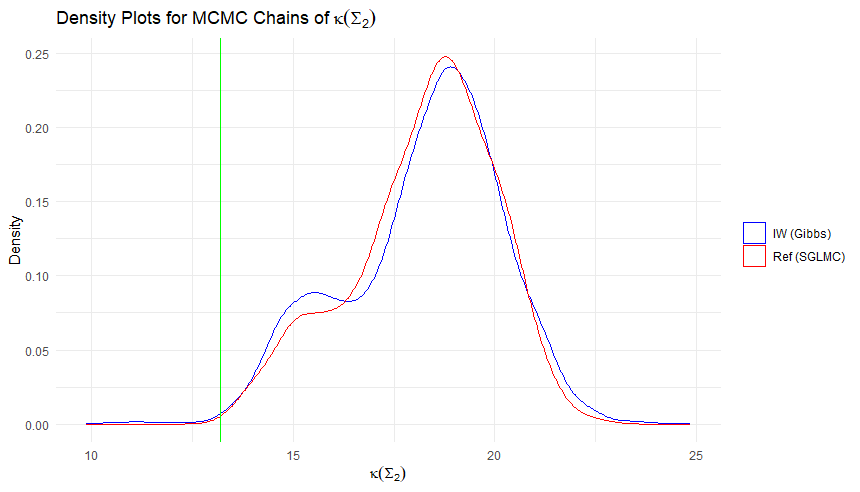} 
    \end{minipage}\hfill
    \begin{minipage}{0.45\textwidth}
        \centering
        \includegraphics[width=\textwidth]{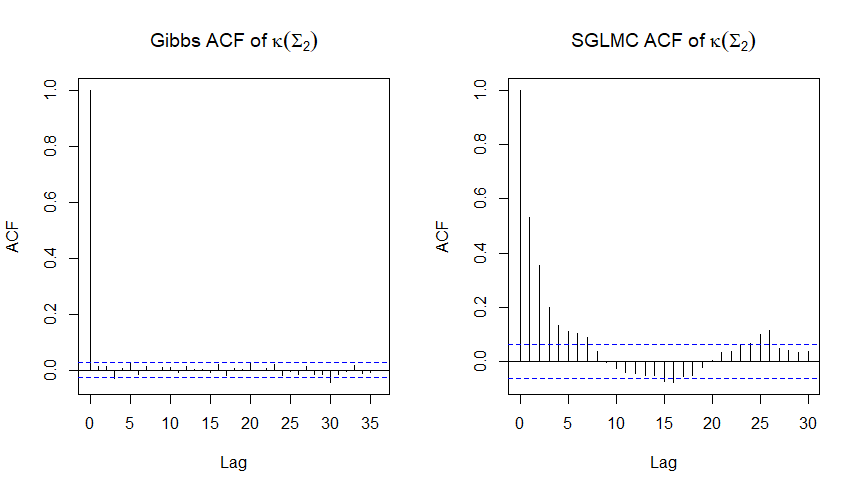} 
    \end{minipage}
            \begin{minipage}{0.45\textwidth}
        \centering
        \includegraphics[width=\textwidth]{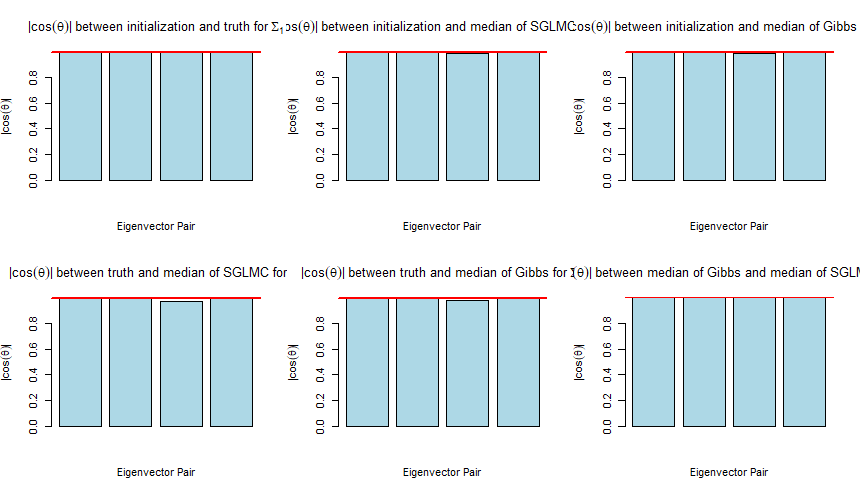} 
    \end{minipage}\hfill
    \begin{minipage}{0.45\textwidth}
        \centering
        \includegraphics[width=\textwidth]{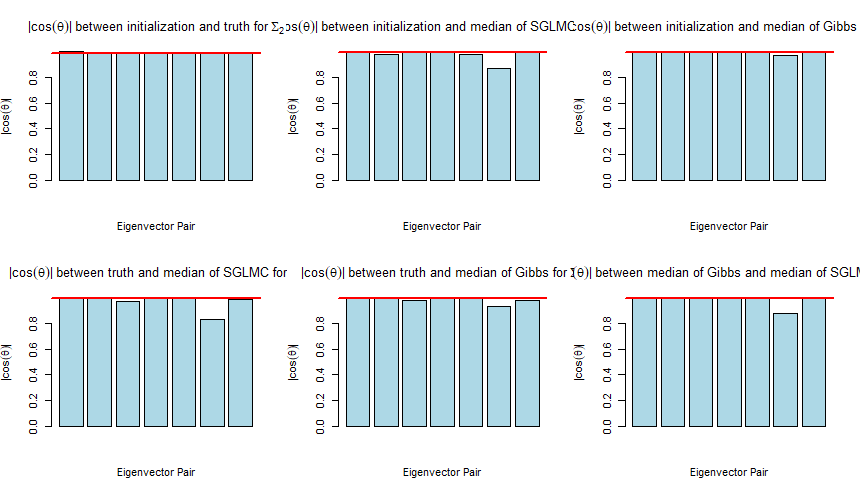} 
    \end{minipage}
    \caption{Density plots of the traces, determinants for $\Sigma$, their corresponding ACFs of the global $\Sigma$, and the condition number comparisons for $\Sigma_{1}, \Sigma_{2}$ for the Reference prior vs Gibbs ($d_{1} = 4$, $d_{2} = 7$, $\alpha = .95$).}
    \label{fig: reference}
\end{figure}

We give comparisons of the Shrinkage Inverse Wishart (SIW) and Reference priors with regularization parameter $\alpha = .95$. In the SIW setting, we use $\alpha_{0} = .7$, where the latter choice was done to ensure numerical stability across prior choices with the same sampling choices made in \ref{sec: Empirical Comparisons}. For the Reference prior, we used $\alpha_{0} = .8$ with the same burn-in and adaptation of Section \ref{sec: Empirical Comparisons}, except the number of samples generated was $1000$. These choices were made to help numerical stability of the reference prior during the MCMC. We used a static choice of $L = 10$ for each comparison.
Comparisons of the SIW and reference prior vs the IW Gibbs samples are found in Figures \ref{fig: SIW} and \ref{fig: reference}, respectively.

We note that each of these prior choices perform similar to the IW prior. However, propriety of the reference prior for this problem was not investigated in this paper, and remains questionable. In our Experiments, we found its stability very sensitive to the choice of $\alpha_{0}$.

